\documentclass[pra,aps,showkeys,showpacs]{revtex4-1}

\usepackage{ucs}
\usepackage[english]{babel}
\usepackage{amsmath}
\usepackage{amssymb,amsfonts}
\usepackage{color}

\usepackage{graphicx}
\usepackage{dcolumn}
\usepackage{bm}

\usepackage{bbold}

\usepackage{subfigure}

\usepackage{tikz}
\usetikzlibrary{shapes.misc}

\usepackage{mathrsfs}

\usepackage{scrpage2}
\usepackage{hyperref}
\usepackage{url}

\newcommand{\qv}{{\mathbf q}}

\newcommand{\bra}[1]{ \left\langle #1 \right | }
\newcommand{\ket}[1]{ \left| #1 \right\rangle }
\newcommand{\ketbra}[2]{ \left| #1 \right\rangle\hspace{-0.13cm}\left\langle #2 \right | }
\newcommand{\ndg}{{\phantom{\dagger}}}
\newcommand{\dg}{\dagger}
\newcommand{\ew}[1]{\pmb{\big\langle} #1 \pmb{\big\rangle}}
\newcommand{\sig}[2]{\sigma_{#1 #2}}
\newcommand{\dt}{\frac{\text{d}}{\text{dt}}}
\newcommand{\avg}[1]{ \langle \hspace{-0.075cm} \langle #1 \rangle \hspace{-0.075cm} \rangle}

\begin{document}

\title{Non-Markovian features in semiconductor quantum optics: Quantifying the role of phonons in experiment and theory}
\author{Alexander Carmele}%
\email{alex@itp.tu-berlin.de}
\affiliation{Technische Universit\"at Berlin, Institut f\"ur Theoretische Physik, Nichtlineare Optik und Quantenelektronik, Hardenbergstra{\ss}e 36, 10623 Berlin, Germany}
\author{Stephan Reitzenstein}
\affiliation{Technische Universit\"at Berlin, Institut f\"ur Festk\"orperphysik, Optoelektronik und Quantenbauelemente, Hardenbergstra{\ss}e 36, 10623 Berlin, Germany}

\begin{abstract}
We discuss phonon-induced non-Markovian and Markovian features in QD-based optics.
We cover lineshapes in linear absorption experiments, phonon-induced
incoherence in the Heitler regime, and memory correlations in two-photon coherences.
To quantitatively and qualitatively understand the underlying physics, 
we present several theoretical models which model the non-Markovian properties of the electron-phonon interaction accurately in different regimes.
Examples are the Heisenberg equation of motion approach, the polaron master equation, and Liouville-propagator techniques in the independent boson limit and beyond via the path-integral method. %
Phenomenological modeling overestimates typically the dephasing due to the finite memory kernel of phonons and we give instructive examples of phonon-mediated coherence such as phonon-dressed anticrossings in Mollow physics, robust quantum state preparation, cavity-feeding and the stabilization of the collapse and revival phenomenon in the strong coupling limit.
\end{abstract}

\maketitle

\section{Introduction} \label{sec:introduction}
Since the seminal demonstration of optically- \cite{michler2000quantum}
and electrically-triggered \cite{yuan2002electrically}
single-photon emission, deterministic generation of entangled photon
pairs \cite{akopian2006entangled,hafenbrak2007triggered}, near-unity indistinguishable photons \cite{santori2002indistinguishable}, 
and strong-coupling to a microcavity \cite{reithmaier2004strong} with semiconductor quantum dots (QDs) \cite{woggon1997optical,bimberg1999quantum} 
acting as active quantum-light emitters, there has been a steadily increasing number of research activities to establish semiconductor systems, in particular QDs as a key element for modern photonic quantum technologies \cite{kiraz2004quantum,michler2009single,michler2017quantum,jahnke2012quantum,shields2010semiconductor,pan,senellart,rastelli,schmidt2007lateral,ding,schlehahn2018}.
Focusing on the goal to implement QDs in quantum sensing, metrology, and quantum cryptography and to establish them as 
``artificial atoms'' and ideal candidates for solid-state quantum bits and scalable quantum information processing 
\cite{yoshie2004vacuum,bose2014all,shields2010semiconductor,senellart_review}, it has become clear that QDs cannot be considered as isolated quantum emitters but interact intrinsically with their semiconductor bulk environment. 
This coupling has usually detrimental impact as it leads to decoherence of the confined exciton acting via superposition states as a qubit \cite{forstner2003phonon,krummheuer2005}.
Interestingly, it can also be of positive impact and facilitate for instance attractive phonon-mediated resonant excitation schemes
\cite{quilter2015phonon,state_preparation_michler,reiter2014role} or phonon-induced quantum coherences \cite{carmele2010antibunching,carmele_cnr}.
A major phonon coupling mechanism stems from the surrounding host material and its lattice vibration \cite{bimberg1999quantum,besombes2001acoustic,borri2001ultralong}.
The associate distortion of the underlying crystal lattice leads inevitably to electron-phonon interaction and subsequently
to substantial new forms of decoherence unknown in atom-molecular optics (AMO). 
Since the phonon reservoir is structured, i.e. the coupling is frequency-dependent, the influence of phonons on the quantum optical properties of QDs is of non-Markovian nature \cite{galland2008non,forstner2003phonon,vagov2007nonmonotonic,madsen2011observation}.
Thus, to quantify and unravel the underlying physics, non-Markovian dissipative processes need to be taken into account.
However, non-Markovianity refuses a time-local formulation, and therefore typical quantum dissipative treatments become unavailable 
\cite{carmichael2009statistical,scully1999quantum} and advanced, or perturbative methods need to be employed 
\cite{breuerbook,gardinerzollerbook,RevModPhys.88.021002,PhysRevLett.112.110401,RevModPhys.89.015001}.
Before we discuss the electron-phonon interaction in more detail, we want briefly comment on how the term 'non-Markovian' and 'Markovian' is used in this review.
We adapt the definition of Markovian and non-Markovian processes as formulated among others in Ref.~\cite{RevModPhys.89.015001,breuerbook}.
A system-reservoir interaction is Markovian if the reservoir correlation time vanishes or can be assumed to be negligible. 
For example, the correlation function of a reservoir consisting of harmonic oscillators with annihilation  (creation) operators $b^{(\dg)}_{\bf q}$ with bosonic commutation relation $[b^{\ndg}_{\bf  q},b^{\dg}_{\bf q^\prime}]=\delta({\bf  q}-{\bf  q^\prime})$ 
can be written as:
\begin{align}
\ew{R(t)R(t^\prime)}
&= \sum_{\qv,\qv^\prime} 
g^{*}_\qv g^{\phantom{*}}_{\qv^\prime}
\left[
e^{-i\omega_qt+i\omega_{q^\prime}t^\prime}
\ew{b^\ndg_\qv b^\dg_{\qv^\prime}} 
+ 
e^{i\omega_qt-i\omega_{q^\prime}t^\prime}
\ew{b^\dg_\qv b^\ndg_{\qv^\prime}} 
\right] \notag \\
&= \sum_{\qv} 
|g^{\phantom{*}}_{\qv}|^2
\big[ (2n^{\phantom{\dg}}_q + 1) \cos[\omega_q(t-t^\prime)]
-i \sin[\omega_q(t-t^\prime)]
\big]
\end{align}
where $R(t)=\sum_{\bf  q} g_{\bf  q} ( b^{\ndg}_{\bf  q} \exp[-i\omega_q t] + b^{\dg}_{\bf  q} \exp[i\omega_q t])$
and an initial thermal equilibrium at $1/\beta=k_BT$ has been assumed with $\ew{b^\dg_\qv b^\ndg_\qv}=[\exp(\beta\hbar\omega_q)-1]^{-1}=n_q$, the Bose-Einstein distribution, and 
$\ew{b^{(\dg)}_\qv b^{(\dg)}_\qv}=0$. 
If now the relaxation time of the reservoir is much smaller than the evolution time of the observed system, i.e. Re$[\ew{R(t)R(t^\prime)}]=\Gamma\delta(t-t^\prime)$,
the system's evolution does not depend on former system-reservoir interaction and the process is Markovian
\cite{RevModPhys.89.015001}.
A specific example of a vanishing correlation time is the  radiative decay into free space, where the vacuum field amplitude is nearly constant in the regime of optical system-reservoir interaction and justifies within the small-bandwidth assumption therefore a Lindblad formulation of the process as shown in standard quantum optics books \cite{carmichael2009statistical,gardinerzollerbook,scully1999quantum,breuerbook}.
Therefore, the dissipative processes can be described with time-independent Lindblad superoperators \cite{carmichael2009statistical}.
A standard example for non-Markovian processes is the electron-phonon interaction in semiconductors \cite{woggon1997optical,michler2003single,forstner2003phonon,krummheuer2005}, where the timescale of the interaction is in the same regime as the corresponding mode structure (ps), and the environment correlation time is finite.
In this case, a backflow of information between environment and system takes place and leads to non-Lorentzian lineshapes (e.g. \cite{stock}), longer Wigner delay times (e.g. \cite{strauss}), reappearance of Rabi oscilllations (e.g. \cite{vagov2007nonmonotonic,ramsay2010phonon}) and allows for robust state preparation protocols (e.g. \cite{kaldewey2017demonstrating,quilter2015phonon,bounouar2015phonon,ardelt2014dissipative,reiter2014role}), phonon-mediated quantum coherences (e.g. \cite{carmele_cnr,iles2016quantum,carmele2010antibunching}) and incoherent excitation processes such as cavity-feeding (e.g. \cite{hughes_incoherent,hohenesterFeeding})
and pulse-area dependent damping and renormalization of Rabi oscillations \cite{ramsay2010phonon,forstner2003phonon}, and excitation-induced dephasing of Mollow triplet sidebands \cite{monniello2013excitation,ulrich2011dephasing,roy2011phonon} as will be discussed in detail below.
Thus, when we refer to non-Markovian processes we mean in general a process which cannot be described with a time-independent, global, Lindblad-based master equation as the paradigm for delta-correlated white noise \cite{breuerbook,gardinerzollerbook,gardinerzoller,carmichael2009statistical,scully1999quantum}.
In the following, we review exemplarily recent work on non-Markovian and Markovian signatures of electron-phonon dynamics in semiconductor
QDs \cite{woggon1997optical,michler2003single}. 
In particular, we focus on InGaAs/GaAs QDs (QDs) with
dominant coupling to longitudinal acoustic (LA) 
and longitudinal optical phonons (LO) via the deformation
or respective Fr\"ohlich coupling element \cite{krummheuer2005}.
For detailed discussion of semiconductor nanostructures,
and the effort of microscopical calculations to quantify 
coupling mechanism in semiconductors, we refer to relevant
books \cite{koch1993QDs,bimberg1999quantum} and reviews \cite{Stier1995QDs,Stier1999QDs,Schliwa2007QDs,chow2013}.
Furthermore, we do also not include explicitly results in the research efforts of higher-dimensional semiconductor nanostructures such as quantum wires, quantum wells or mesoscopic bulk systems \cite{haugkoch,brandes2005coherent_mesoscopic}. 
There is a wide range of exciting phenomena in these systems due to the ubiquitous non-Markovian electron-phonon and electron-electron dynamics.
To name a few, quantum wires show phonon-enabled thermal conductivity \cite{quantumwires} based on a universal quantum of thermal conductance \cite{schwab2000measurement_wire}.
Pronounced non-Markovian decoherence is demonstrated in localized nanotube excitons \cite{galland2008non} and also phonon-assisted Anderson localization phenomena have been investigated \cite{gornyi2005interacting_wires}.
In quantum wells, coherent acoustic oscillations are studied \cite{sun2000coherent_wells}, and non-equilibrium cooling effects bottlenecked by non-Markovian phonon dynamics have been discussed \cite{lugli1987nonequilibrium_cooling_wells}.
Four-wave mixing techniques allow to study and characterize giant excitonic resonances \cite{kim1992giant_excitonic_resonances_wells,gammon1995phonon_broadening_wells}, and via two-dimensional coherent spectroscopy techniques incoherent exciton-phonon Green's function can be probed and extracted in disordered quantum wells \cite{richter2018deconvolution,richter_localization}.
Moreover, optical and electronic two-dimensional spectroscopy has drawn a lot interest recently, as non-Markovian, anomalous lineshapes and relaxation/scattering processes can be characterized and studied in depth.
For example, the study of signatures of spatially correlated noise and non-secular effects \cite{2D_plenio_lim2017signatures} has been performed, the read-out of Rabi oscillations in QDs \cite{2D_wigger2018rabi}, exciton coherence at room temperature \cite{2D_electronic_cassette2015room} investigated, systematic study of dephasing processes including quadratic electron-phonon coupling for elevated temperatures \cite{2D_spec_liu2018vibrational,2D_cundiff_PhysRevB.98.195304} and phonon sidebands in transition metal dichalcogenides have been demonstrated \cite{2D_knorr_PhysRevLett.119.187402}.
Despite the exciting results in higher-dimensional nanostructures and two-dimensional coherent spectroscopy, we focus in the following on a single material platform, semiconductor QDs. 
This allows us to discuss in detail instructive experimental examples in which the phonon interaction is the dominant source of decoherence and theoretical models which are capable to capture their specific details and can partially be treated even analytically.
These models are, however, not limited to the QD case and are used for other material platforms such as quantum wells, quantum wires, or atomic-thin two-dimensional sytems such as graphene as well
\cite{haugkoch,malicbook,kirabook}.
%

\section{Structure of the review and Hamiltonians}
\label{sec:structure}
%
This review covers two different interaction schemes, electron-light $(e-l)$ and electron-phonon $(e-p)$ dynamics, and is basically structured by increasing complexity.
To increase readability, we introduce here the corresponding Hamiltonians which are used throughout the review.
The electron-light interaction is treated either semiclassically $H_{e-l}^{sc}$,
quantized in a mode continuum $H_{e-l}^{con}$, or within a cavity or single mode description $H_{e-l}^{cav}$.
The Hamiltonians read:
\begin{align}
H_{e-l}^{sc} 
&= 
\hbar \Delta \sig{2}{2} +
\hbar 
\Omega(t)
\left(
\sig{2}{1}
+
\sig{1}{2}
\right)
\label{eq:H_sc}, \\      
H_{e-l}^{cav} 
&=
\hbar \omega_{21} \sig{2}{2} 
+
\hbar\omega_0 c^\dg c 
+ 
\hbar g \left(c^\dg\sig{1}{2} +\sig{2}{1}c\right)
\label{eq:H_cav} \\
H_{e-l}^{con}
&=
\hbar \omega_{21} \sig{2}{2} 
+\hbar
\int d\omega \left( \omega \ c^\dg_\omega c^\ndg_\omega + g \
c^\dg_\omega \sigma_{12} + g \sigma_{21} c^\ndg_\omega \right)
\label{eq:H_con},
\end{align}
where the rotating-wave approximation has been applied, and
$\Omega(t)$ is the Rabi frequency, including the external driving field $E(t)$ a $\omega_L$ and the dipole moment of the QD ${\bf d}_{12}={\bf d}_{21}$, assumed in the following as real number, between the conduction $2$ and valence band $1$ with a band gap energy of $\hbar\omega_{21}$.
In the semiclassical case $(H_{e-l}^{sc})$, the Hamiltonian is written in the rotating frame of the laser frequency, leading to a detuning of $\Delta=\omega_{21}-\omega_L$. 
The operator $\sigma_{ij}$ flips the state $\ket{j}$ to 
$\ket{i}$, whereas $c^{(\dg)},c^{(\dg)}_{\omega}$ annihilates (creates) a photon in the corresponding mode with bosonic commutation relation $[c^\ndg_\omega,c^\dg_{\omega^\prime}]=\delta(\omega-\omega^\prime)$.
The interaction strength between photons and electron is denoted $g$, assuming a Wigner-Weisskopf-like coupling with an approximate constant vacuum field amplitude \cite{scully1999quantum,carmichael2009statistical}.
The second interaction we consider is the electron-phonon interaction, either in a semiclassical limit $H_{e-p}^{sc}$
or for longitudinal acoustical $H_{e-p}^{LA}$ or 
longitudinal optical phonons $H_{e-p}^{LO}$.
The Hamiltonians read:
\begin{align}
H_{e-p}^{sc} &= \hbar F(t) \sigma_{22}
\label{eq:H_st}, \\
H_{e-p}^{LA} &=
\hbar
\sum_{\bf q}
\omega_{\bf q}
b^\dg_{\bf q}
b^\ndg_{\bf q}
+
\hbar
\sig{2}{2}
\sum_{\bf q} 
g^{\bf q}_{12} \left[ b^\dg_{\bf q} + b^\ndg_{\bf q} \right]
\label{eq:H_la}, \\
H_{e-p}^{LO} &=
\hbar
\omega_{LO}
\sum_{\bf q}
b^\dg_{\bf q}
b^\ndg_{\bf q}
+
\hbar
\sig{2}{2}
\sum_{\bf q} 
f^{\bf q}_{12} \left[ b^\dg_{\bf q} + b^\ndg_{\bf q} \right],
\label{eq:H_lo}
\end{align}
where $F(t)$ denotes a stochastic force acting on the QD,
and $g^{\bf q}_{12}$ the electron-longitudinal acoustic and $f^{\bf q}_{12}$ the electron-longitudinal optical phonon coupling element.
Throughout the review, the standard GaAs phonon bulk parameter are used.
For example, in case of an approximative spherical geometry 
the acoustical phonon coupling element $g^{\bf q}_{12}=g^{{\bf q}}_{11}-g^{{\bf q}}_{22}$ reads, where 
\begin{align}
g^{{\bf q}}_{ii}
=
\sqrt{\frac{\hbar q}{2 \rho c_s 
V}}D_{i}e^{-\frac{q^{2}\hbar}{4 m_{i}\omega_{i}}},
\end{align}
and typical parameters for InGaAs/GaAs QDs are given with sound velocity of GaAs $c_\text{s}=0.00511\text{nm/fs}$, deformation potentials $D_{1}=-5.38\text{eV}, D_{2}=-11.68\text{eV}$, effective masses: $ m_{2} = 0.063$, $m_{1} = 0.45$, confinement energies $\hbar\omega_{2}=0.040\text{eV}$, $\hbar\omega_1=0.02\text{eV}$ and 
mass density of GaAs $\rho=5370\text{kg/m}^3$ 
\cite{krummheuer2005,carmele_cnr,hohenester,hohenesterFeeding,forstner2003phonon,koch1993QDs}.
For the longitudinal optical phonons, the Fr\"ohlich coupling applies and reads:
\begin{align}
f^{{\bf q}}_{ii}
=
-i
\sqrt{\frac{e^2 \hbar \omega_{LO}}{2 V \epsilon_0 \epsilon^\prime}} \frac{1}{q} e^{-\frac{q^{2}\hbar}{4 m_{i}\omega_{i}}},    
\end{align}
with $1/\epsilon^\prime = 1/\epsilon_\infty - 1/\epsilon_s=0.0119347$ with the static dielectric constant $\epsilon_s=12.53$ and the high frequency dielectric constant $\epsilon_\infty=10.9$, and the reziprocal dielectric constant $1/\epsilon_0=18.1 e^2$/(eV nm).
The longitudinal optical phonon frequency is in GaAs $\hbar\omega_{LO}=36.4$meV.
This review is structured in five parts.
After the introduction [Sec.~\ref{sec:introduction}] 
and this structure section  [Sec.~\ref{sec:structure}],
we discuss non-equilibrium phonon dynamics in the 
semiclassical regime in Sec.~\ref{sec:semiclassical}.
Here, the light field is treated classically 
with $H^{sc}_{e-l}$ and acts as an excitation source and a way to probe the system's dynamics.  
In this section. we discuss the electron-phonon and electron-light interaction in different regimes, from weak coupling and weak driving to strong coupling and strong driving, and important theoretical tools are presented and explained in detail.
In Sec.~\ref{sec:quantized}, we include also quantized electron-light interaction in cavity quantum electrodynamics
$H^{cav}_{e-l}$ or for a mode continuum $H^{con}_{e-l}$
and show how phonons contribute to quantum optical phenomena, rendering the field of semiconductor quantum optics exciting and novel.
This section is structured from the single- and two-photon regime to the many-photon dynamics, including non-Markovian phonons as the main source for decoherence.
We then conclude the review in Sec.~\ref{sec:conclusion}
and give a short outlook on future directions.
Please refer to the given Sections for a detailed discussion of the presented material.  
%

\section{Non-equilibrium phonon dynamics in semiclassical light-matter interaction}
\label{sec:semiclassical}
%
Experimental data on quantum emitters in a solid-state environment, as discussed in the introduction, show the necessity of non-Markovian decoherence models.
Modelling the required intrinsic memory of the electron-phonon interaction is theoretically a demanding but also very rewarding task as it provides the link between experiment data and its non-Markovian description. 
Analyzing decoherence mechanisms in solid-state devices is therefore crucial to understand the underlying physics and to optimize their functionality, in particular when it comes to quantum properties. 
It is also an important part in the pursuit to realize reliable protocols in future quantum communication networks which are based
on coherent light-matter interfaces and the exchange of quantum information via single photons and entangled photon pairs with high indistinguishability \cite{briegel_quantum_repeater,kimble_quantum_internet}.
Here, light-matter interaction between two-level emitters and the coherent light-fields of an external laser is a well-established approach to investigate the time evolution of the polarization density within a medium \cite{mukamelbook,axt_mukamel}. 
This time evolution can be governed by semiclassical, Maxwell-type of dynamics or takes place deep into the quantum regime of single-emitter-light interaction. 
In this section, we discuss non-equilibrium phonon dynamics in the semiclassical regime.
The achievements in fabrication of high-quality QDs acting as close-to-ideal two-level emitters in the solid-state 
\cite{senellart_review,warburton_review} opened the possibility to investigate the light-matter interaction in different regimes.
This section is structured from the weak electron-phonon coupling and weak driving limit (linear) to the strong electron-phonon coupling and strong driving regime (Mollow regime).
We start in Sec.~\ref{subsec:absorption} with the discussion of the impact of lattice distortion processes in absorption experiments \cite{mukamelbook,krummheuer2005,matthiesen_subnatural}. 
In this linear regime, the QD emission can be expressed via the  absorption coefficient $\alpha(\omega)$ and calculated exactly via the independent boson model \cite{may2008charge,breuerbook}. 
Phonon-induced non-Lorentzian lineshapes are unraveled in the absorption coefficient and a clear deviation from Gaussian white noise correlation is observed and successfully explained with wavefunctions based on 
8 band ${\bf k}\cdot {\bf p}$ theory and with inclusion of non-Markovian phonon effects \cite{stock}.
We continue in Sec.~\ref{subsec:wigner_delay} with the weak-driving or Heitler regime in which we investigate the emission and scattering of a single QD under pulsed excitation
\cite{strauss}.
This regime allows us to explore an intriguing quantum optical effect namely the Wigner time delay known from atomic physics \cite{bourgain2013direct,leuchsWignerDelay,wignerDelayAtomic}. 
This delay originates form the phase-shift between the exciting and emitted field due to the finite dwelling time in which the QD absorbs excitation from the time-dependent driving laser-field and re-emits this excitation. 
Phonon-induced incoherent processes explain maximal achievable Wigner delay times $ \tau_W(\Delta)$ in the non-trivial detuning dependence between the laser and the QD transition \cite{strauss} which cannot be described by conventional optical Bloch equations \cite{leuchsWignerDelay}.
Here, we discuss semiconductor Bloch equations 
with phonon contributions via the cluster-expansion approach \cite{kirabook,kirareview} which allows perturbatively to take into account non-equilibrium and non-Markovian phonon dynamics.
We also show that for low-temperatures and weak electron-phonon coupling, the perturbative cluster-expansion solution agrees well with the solution from the exact independent boson model, given in Sec.~\ref{subsec:absorption}.
In the weak electron-phonon coupling but strong driving limit [Sec.~\ref{subsec:heisenberg}], non-Markovian features are already captured in the perturbative cluster expansion approach \cite{cicek1969,forstner2003phonon,krummheuer2002,PhysRevLett.103.087407,kirabook,PhysRevLett.100.027401}.
In this limit, ultralong dephasing times can be explained \cite{borri2001ultralong} and the recurrence of Rabi oscillations for large pulse areas \cite{forstner2003phonon,vagov2007nonmonotonic,kaldewey2017demonstrating} are seen.
This method allows a fast and good approximation to quantify 
influences of the microscopic properties of the nanostructure
under investigation \cite{reiter,malicbook}.
As an example that already second-order perturbation theory addresses signatures typically not present in Lindblad (Markovian) master equation simulations, we present in detail damped Rabi-oscillations via longitudinal acoustic phonons. 
Within the same theoretical framework but in fourth-order perturbation theory, phonon-assisted state preparation protocols are discussed. 
Those protocols are robust against the underlying geometry of the nanostructure, however the wavefront of the outgoing acoustic excitations differ strongly for different geometries \cite{reiter,kaldewey2017demonstrating,reiter2014role}.
Beyond the weak electron-phonon coupling but still in the perturbative regime [Sec.~\ref{subsec:master_equation}], master equation approaches within a dressed-state basis are 
typically employed \cite{breuerbook,carmichael2009statistical,wilsonPMEQ}.
We first discuss the Markovian limit \cite{laussy,PhysRevB.84.195313,delvalle_phd,PhysRevLett.105.233601} in Sec.~\ref{subsec:master_equation}
and give an analytical solution for the power spectrum of a resonantly and optically driven QD subjected to Markovian pure 
dephasing \cite{kreinberg2018quantum}.
The Markovian limit, however, cannot capture incoherent excitation processes via phonon-feeding \cite{hughes_incoherent}.
To model such processes correctly, a Polaron master equation is a feasible model to include as much information as possible 
from the non-equilibrium phonon dynamics in second-order perturbation theory \cite{wilsonPMEQ,nazirPMEQ,mansonPMEQ,hohenesterFeeding}, and
shows already very good agreement with experimental data \cite{hughes_incoherent}.
We derive the polaron master equation explicitly and show that for the weak coupling limit in secular approximation, the cluster-expansion solution and polaron master equation dynamics agree well.
We conclude this section with exact solutions of the electron-phonon interaction.
In case of a dispersionless phonon mode, such as longitudinal optical (LO) phonons in the Einstein approximation, an inductive equation of motion scheme leads to numerically exact solutions, cf.~Sec.~\ref{subsec:inductive_eom_approach}.
As an example, we show phonon-assisted anticrossings in the Mollow regime which give experimental access to the Huang-Rhys factor.
This shows that the equation of motion approach based on the semiconductor Bloch equations \cite{haugkoch,kirabook,kirareview} 
is not limited to the weak coupling regime in principle. 
If the dispersion relation is nearly constant, as for LO phonons, an inductive equation of motion approach
can be employed to take into account higher-order phonon processes \cite{carmele2010antibunching,kabuss2011inductive,axt1999coherent}.
In this case, the equation of motion approaches becomes a numerically exact method and allows to compute LO-phonon assisted Mollow triplet spectra.
Due to the strong-driving limit, anticrossings between polaronic and polaritonic dressed-states occur and the electron-LO phonon coupling strength can be spectrally obtained due to the emerging splitting \cite{kabuss}.
In case of linear, or non-constant dispersion, this method is possible but technically demanding to evaluate and therefore not feasible anymore.
In the limit of time-dependent pulses and/or strong coupling to phonons with non-constant dispersion relation [Sec.~\ref{subsec:path_integral}], perturbative approaches and master equation models do not capture the phonon-induced dynamics accurately anymore \cite{clustervspathintegral}. 
Given the non-Markovian and therefore entangled system-bath dynamics, density matrix renormalization group techniques 
\cite{schollwock2011density,caldeira1983path} or exact diagonalization become the only choice.
They allow for numerical expensive but also exact treatments.
Examples are matrix product state evolution techniques \cite{vidal_mps,phonon_vidal,strathearn2018efficient,droenner2018two}
or the real-time path integral method \cite{makri1998quantum,vagov2007nonmonotonic} in which
the time evolution is discretized and the dissipative
quantum kinetics becomes solvable due to
the finite memory of the dissipative kernel, here of the acoustic phonons.
Within the path-integral method technique, excellent agreement between theory and experiment has been demonstrated for state-preparation protocols
\cite{quilter2015phonon}, and the description of phonon-induced dephasing of coupled QD-microcavity systems in the strong coupling regime of cavity quantum electrodynamics \cite{cygorek2017nonlinear,axt_hopfmann_prb}.
%

\subsection{Non-Lorentzian lineshapes (Independent boson model)}
\label{subsec:absorption}
%
For weak coherent pumping, a system is well-described in the linear response regime \cite{mukamelbook}. 
Via a probe field, the absorption of the nanostructure, here, for example an ensemble of QDs is quantified via the Beer-Lambert law \cite{jackson2012classical}.
The absorption coefficient reveals resonances of the emitter but, more importantly, the interaction with the environment is probed also via the detuning dependence of the absorption.
The corresponding lineshape gives access to important information about the kind of coupling (e.g. Fr\"ohlich, deformation, or piezoelectric interaction between electrons and phonons) and the linewidth allows one to conclude about the effective electron-phonon coupling strength \cite{jahnke2012quantum,bimberg1999quantum,woggon1997optical,koch1993QDs}.
A characteristic quantity to connect theory to experiment is the absorption coefficient which can be expressed via the susceptibility $\alpha(r,\omega)=\text{Im}[\chi(r,\omega)]$. 
In the semiclassical description we assume an incoming electromagnetic field $ {\bf E}(r,t) $ described via Maxwell's equations including material contributions.
The dynamics of the electric component is described via the 
homogeneous equation: $ \nabla\times {\bf E} = -\partial_t {\bf B} $,
and the inhomogeneous $ \nabla\times {\bf B} =\mu_0 {\bf j} + \partial_t {\bf E}/c^2 $ with the speed of light in the vacuum $ c_0=(\epsilon_0\mu_0)^{-1/2}$.
Averaging over microscopic degrees of freedom below the optical wavelengths, assuming an electrically neutral sample without a macroscopic current and neglecting magnetic contributions, we yield for the averaged current: $ {\bf j}(r,t) \equiv \partial_t {\bf P}(r,t)$ \cite{mukamelbook,malicbook,jackson2012classical}.
Deriving the wave equation in frequency domain for the electric field, we obtain:
\begin{align}
\left[
\Delta + \frac{\omega^2}{c_0^2}
\right]
{\bf E}(r,\omega)
=&
-\mu_0 \omega^2 {\bf P}(r,\omega),
\end{align}
where we assume a purely transversal wave and we neglect in the following out of notation convenience the background material refractive index.
If the sample is isotropic, homogeneous and linear, the polarization is related to the electric field:
$ {\bf P}(r,\omega)=\epsilon_0 \chi(r,\omega) {\bf E}(r,\omega) $
and we derive the solution of the field intensity 
(propagating in $z$-direction):
\begin{align}
I(z,\omega) 
=& |{\bf E}(r,\omega)|^2 
= |E_0|^2 \exp[i(k-k^*)z]
= |E_0|^2 \exp[-2\text{Im}[k]z],
\end{align}
with $ k=k_0\sqrt{n(r,\omega)} $, the refraction index of the investigated sample $ n(r,\omega)=\sqrt{1+\chi(r,\omega)} $ and $ k_0=\omega/c_0$.
The imaginary part of the wave number is directly proportional to the imaginary part of the susceptibility.
The susceptibility in the linear regime is connected to the dipole density, which is microscopically given via the 
transition amplitudes multiplied by their dipole momentum and the average dipole density in the sample with emitter density $n_0$:
\begin{align}
{\bf P}(r,\omega) =& n_0 \sum_{ij} {\bf d}_{ij}(r) \rho_{ij}(\omega)
\end{align}
with $ \rho_{ij}(t)=\langle i| \rho(t) | j \rangle $.
Therefore, the density matrix equation gives access to the linear and, also, nonlinear response of the system via solving the Liouville-von 
Neumann equation $ i\hbar\dot\rho(t)=[H,\rho]$.
In the following, a sample consisting of QDs in the  two-level approximation is assumed which is probed via an incoming (linear polarized) field with an amplitude parallel to the dipole moment of the two-level emitters \cite{allen1987optical}. 
The associated Hamiltonian $H_{e-l}^{sc}$ is given in Eq.~\eqref{eq:H_sc}.
The non-coherent contributions are considered by the decoherence inducing Hamiltonian $ H^{st}_\text{e-p}(t)$ 
given in Eq.~\eqref{eq:H_st} with a c-valued stochastic force F(t). 
The dynamics of the transition amplitudes read in general:
\begin{align}
\partial_t \rho_{12} =& 
i\Delta \rho_{12}
-i\Omega(t) \rho_{22}
+i\Omega(t) \rho_{11}
-\frac{i}{\hbar}\bra{1}[H^{st}_{e-l}(t),\rho(t)]\ket{2}.
\end{align}
In the linear regime (no population change is induced), we set $ \rho_{11}(t)\approx\rho_{11}(0)=1$ and a time-independent weak driving field $ \Omega(t)=\Omega_0$.
Using a phenomenological decoherence model, we choose a stochastic force contribution to the energy splitting between ground $ \ket{1} $ and excited state $ \ket{2} $, i.e. $ \omega_{21}+F(t)$  \cite{gardiner2009stochastic,gardinerzollerbook}.
The formal solution of the transition dynamics in the linear regime reads then with $\xi(t)=\int_0^t dt^\prime F(t^\prime)$:
\begin{align}
\rho_{12}(t)=& 
e^{i\Delta t+i\xi(t)}
\left(
\rho_{12}(0) 
-
i\Omega_0\rho_{11}(0)
\int^t_0 dt^\prime
e^{-i\Delta t^\prime-i\xi(t^\prime)}
\right)
\end{align}
Due to the decoherence-inducing contributions $ \xi(t)$, the equation must be averaged \cite{gardiner2009stochastic}.
Assuming the white noise limit, we average with a Gaussian probability distribution, taking only two-body correlations into account \cite{gardiner2009stochastic}.
In this limit, we characterize the correlation function as  $\avg{F(t)F(t^\prime)}=2\gamma\delta(t-t^\prime)$ with $ \avg{F(t)}=0$ and obtain
$\avg{e^{\pm i\xi(t^\prime)}}=\exp[-\gamma t]$.
Given the dephasing, the solution of the transition dynamics is obtained after a simple integration:
\begin{align}
\avg{\rho_{12}(t)}=& 
e^{i\Delta t-\gamma t}
\left(
\rho_{12}(0) 
-
i\Omega_0\rho_{11}(0)
\frac{e^{-i\Delta t+\gamma t}-1}{-i\Delta +\gamma}
\right)
\end{align}
Differentiating with respect to time and taking the 
Fourier transform into account, we derive an expression for the polarization density in the linear regime and projected along the direction of the dipole:
\begin{align}
P(r,\omega) 
=& n_0 d_{12} [ \avg{\rho_{12}(\omega)} + \avg{\rho_{21}(\omega)}] 
\notag
\\ 
\label{eq:sc_absorption}
=& n_0 d_{12}
\left[
\frac{\Omega_0\rho_{11}(0)}{\omega-\Delta-i\gamma}
-
\frac{\Omega_0\rho_{11}(0)}{\omega+\Delta-i\gamma}
\right].
\end{align}
We assume now a linearly polarized electrical field with a real amplitude parallel to the dipole: $ \Omega_0=d_{21} E_0/\hbar$.
Ignoring the off-resonant parts of the susceptibility, we find with $\rho_{11}(0)=1$:
\begin{align}
\chi(r,\omega) =&
\frac{n_0 |d_{12}|^2}{\epsilon_0 \hbar} \frac{i\gamma}{(\omega-\omega_{21})^2+\gamma^2},
\end{align}
which connects to the absorption of the incoming electric field via
the Beer-Lambert law: $ I(z,\omega)=|E_0|^2 \exp(-2\text{Im}[\chi]z)$.
The higher the dipole density, the larger the individual dipole moments and the closer the incoming wave is resonant with the transitions, the stronger is the absorption of the incoming wave into the sample.
The aforementioned absorptive behavior is experimentally accessible via luminescence signals.
In the limit of Fermi's golden rule and after subtracting a constant offset, the absorption is proportional to the luminescence signal. 
This is assumption is valid up to second-order in the time-evolution operator and within the classical Condon approximation for semiclassical light-matter interaction
\cite{mukamelbook}.
These requirements are nevertheless fulfilled in most of the cases and absorption and luminescence lineshapes are mirror images of each other. 
In the phenomenological model, a Lorentzian lineshape is obtained, cf.~Eq.~\ref{eq:sc_absorption}.
\begin{figure}
\includegraphics[width=0.75\columnwidth]{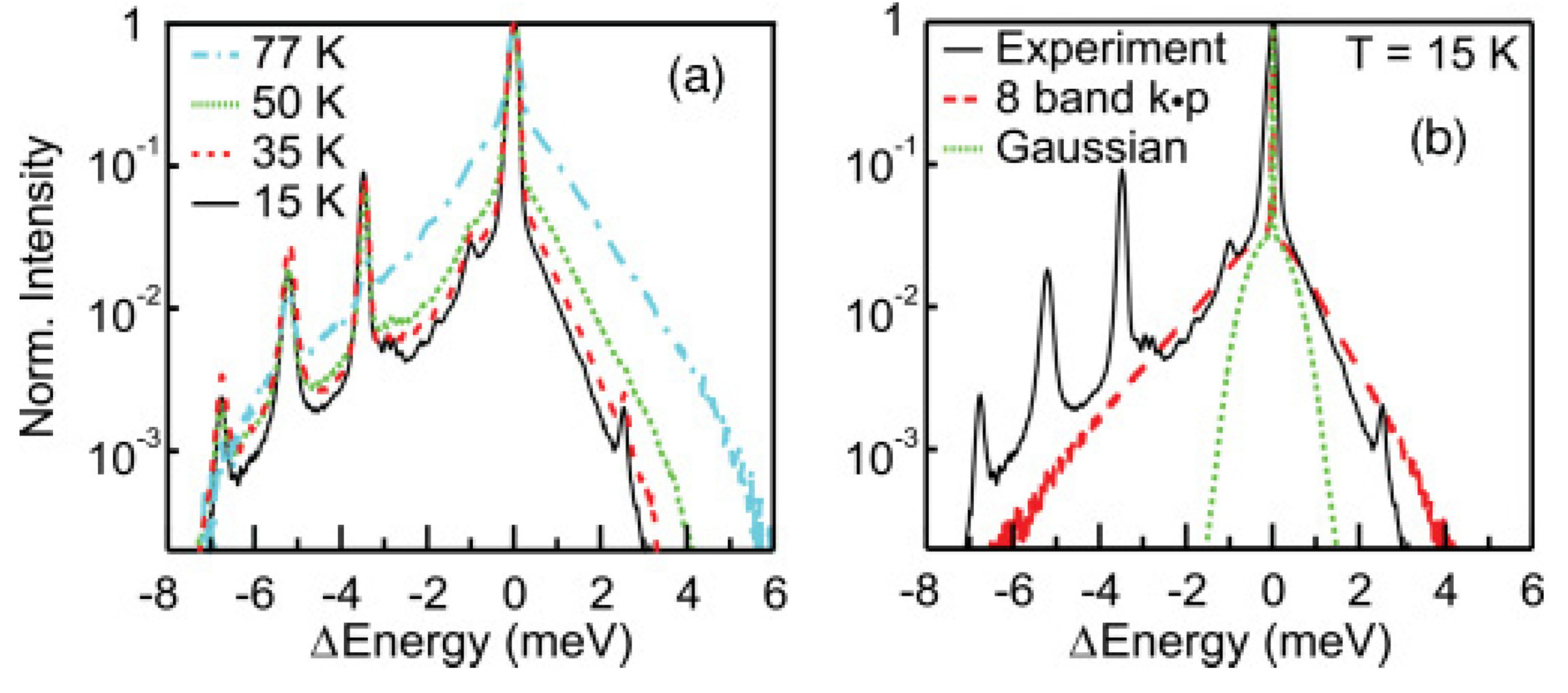}
\caption{Semi-logarithmic plot of the electroluminescence intensity of a single QD. (a) With increasing temperature the line broadening increases due to stronger acoustic phonon interaction. (b) Comparison of the measured line-shape with two calculated spectra, using alternatively a simple Gaussian (green dotted curve) or a realistic 8 band ${\bf k }\cdot{\bf p}$ wave function (red dashed curve). Reprinted figure with permission from \cite{stock}. \copyright2011 by the American Physical Society.}
\label{fig:lineshape_la}
\end{figure}
However, comparing the experimental luminescence signal of a single QD non-resonantly driven via current injection in the p-i-n diode structure, cf.~Fig.~\ref{fig:lineshape_la}(a) and (b), strong deviations from a Lorentzian oscillator model are visible. 
The line broadening has a temperature-dependent shoulder, which cannot be exlained with a white noise model as outlined above. 
In contrast, the asymmetric lineshape is a direct consequence of the underlying non-Markovian dynamics between the QD and lattice vibrations \cite{QD_lum_besombes2001acoustic,QD_lum_jakubczyk2016impact,krummheuer2005,krummheuer2002,langbein_fourwave_mixing,2D_cundiff_PhysRevB.98.195304,2D_wigger2018rabi}.
To model the microscopic interaction between the weakly driven QD and its semiconductor host matrix more accurately, we choose the quantized electron-phonon Hamiltonian $ H_{e-p}^{la}(t)$ given in Eq.~\eqref{eq:H_la} already in the interaction picture with the free evolution governed by $ H_p = \hbar \sum_{\bf q} \omega_{\bf q} b^\dg_{\bf q} b^\ndg_{\bf q}$ \cite{krummheuer2002,krummheuer2005,madelung2012introduction}.
To solve the quantum mechanical decoherence model, we have to take into account the quantum mechanical character of the decoherence Hamiltonian.
The full solution of the density matrix evolution reads in the Liouville space:
\begin{align}
\rho(t)=
T\left\lbrace \exp[\int_0^t L(t_1) dt_1] \rho(0) \right\rbrace
\end{align}
with time-ordering operator T and Liouvillean superoperator $L(t)\rho=(-i/\hbar)[H(t),\rho]$.
To derive the systems dynamics, we trace out the phonon degrees of freedom:
\begin{align}
\text{Tr}_B\left\lbrace\rho(t)\right\rbrace=\rho_s(t)=
T_s\left\lbrace\text{Tr}_B\left\lbrace T \exp[\int_0^t L(t_1) dt_1] \rho(0)\right\rbrace \right\rbrace,
\label{eq:Liouville_full_solution}
\end{align}
where the time-ordering operators ensure that the corresponding von-Neumann equation is still solved.
To evaluate this solution, typically a path-integral method
is employed, cf.~Sec.~\ref{subsec:path_integral}.
Here, we assume a vanishing light-matter coupling and stay in the linear regime. 
In this case, Eq.~\eqref{eq:Liouville_full_solution} can be solved analytically for $H(t)=H_{e-p}^{LA}(t)$ \cite{breuerbook,krummheuer2005}.
%

%
%
%
%
%

%
For weak excitation, the luminescence spectrum and the absorption coefficient coincide \cite{krummheuer2005,mukamelbook,pullerits1995temperature} and give access to the lineshape and coherence times of the QD ensemble.
To solve the quantum mechanical model in the linear regime, 
one traces out first the electronic degree of freedom:
\begin{align}
\dt\bra{2}\rho(t)\ket{1}  
&= -i\bra{2}[H_{e-p}^{LA}(t),\rho(t)]\ket{1} \\ \notag
&= -i\sum_{\bf q} 
g^{\bf q}_{12} \left[ b^\dg_{\bf q}(t) + b^\ndg_{\bf q}(t) \right]
\bra{2}\rho(t)\ket{1}.  
\end{align}
This equation can be formally solved via the Dyson series with $\bra{2}\rho(t)\ket{1}=P_B(t)$:
\begin{align}
P_B(t)  \notag
&=
T\left\lbrace 
\exp\left[-i\int_0^t dt_1 
H_{e-p}^{LA}(t_1)/\hbar
\right] P(0) 
 \right\rbrace 
\end{align}
Assuming discrete time steps $\Delta t$, we can write
\begin{align}
P_B(N\Delta t)  \notag
&=
T\left\lbrace 
\exp\left[-i \sum_{n=1}^N 
H_n 
\right] P(0) 
\right\rbrace \\
&=
\exp\left[-i H_N\right]
\cdots 
\exp\left[-i H_2 \right]
\exp\left[-i H_1\right]
P(0) 
\end{align}
after performing a Suzuki-Trotter decomposition and using
the abbreviation
\begin{align}
H_N :=
\int_{(N-1)\Delta t}^{N\Delta t}
dt \ H_{e-p}^{LA}(t) /\hbar. 
\end{align}
The time-order is now trivially fulfilled and therefore, the
time-ordering operator is omitted. 
The necessary condition for the analytical solution in this 
case is that the commutator of two Hamiltonians is a c-value and not an operator anymore:
\begin{align}
\left[H_N,H_{N-1}\right]   
= 
-2i
\int_{(N-1)\Delta t}^{N\Delta t} dt_1 
\int_{(N-2)\Delta t}^{(N-1)\Delta t} dt_2
\sum_{\bf q} \left|g^{\bf q}_{12}\right|^2 
\sin[\omega_q(t_1-t_2)].
\end{align}
This result means that the commutator between of the Hamiltonian with the commutator between the Hamiltonians vanishes $[[H_i+H_j,[H_i,H_j]]=0$, and therefore the Baker-Campbell-Hausdorff formula can be employed $\exp[A]\exp[B]=\exp[A+B]\exp[[A,B]/2]$ to obtain in the limit of $\Delta t\rightarrow 0$:
\begin{align}
P_B(t)  \notag
&=
\exp\left[-i\int_0^t dt_1 
H_{e-p}^{LA}(t_1)/\hbar
\right] P(0)  \\ 
&\phantom{=}
\exp\left[\frac{1}{2\hbar^2}\int_0^t dt_1
\int_0^{t_1} dt_2
[H_{e-p}^{LA}(t_1),H_{e-p}^{LA}(t_2)]
\right].
\end{align}
Since there is no time-ordering necessary, one can trace out 
the phonons and the independent boson model is given via 
$P(t):=Tr_B[P_B(t)]$
\begin{align}
P(t) =& 
\label{eq:ibm_solution}
\exp
\bigg[
i\omega_{21}t+i\text{Im}[\phi_I(t)]-\text{Re}[\phi_I(t)]
\bigg],
\\
\label{eq:ibm_phonon_correlation}
\phi_I(t) &=
\sum_{\bf q}
\frac{|g^{\bf q}_{12}|^2}{\omega_q^2}
\left[
(2n_q+1)(1-\cos[\omega_qt])
+
i
\omega_q t
-i
\sin(\omega_qt)
\right]
.
\end{align}
Obviously, the time-order is fully taken account of in the imaginary phase whereas the dephasing stems alone from the part
left untouched by the time-ordering.
Therefore, neglecting the time-order completely will yield the wrong result only if phase-dependent quantities are studied.
Taking the Fourier transform, we yield $\rho_{12}(\omega) $ and $\rho_{21}(\omega)$ and the susceptibility includes then non-Markovian effects due to the quantum character of the structured bosonic reservoir.
The semiconductor host matrix is considered in the coupling elements, as well as the temperature-dependence mainly governed by the phonon frequencies from the dispersion relation, which in principle can be calculated ab initio.
In Fig.~\ref{fig:lineshape_la}(b), a microscopic wavefunction is used to model the correct, experimentally observed lineshape \cite{stock}. 
To conclude this section, a non-Markovian dephasing model was applied to describe the emission spectra of a single QD embedded in an electrically pumped device which is based on a p-i-n diode design and an oxide aperture to spatially restrict the current flow. 
Despite the complexity of the device design and the electrical excitation scheme the agreement between theory and experiment is striking. 
This together with the possibility to include microscopic parameters into the model for the description of experimental 
results proves the strength of the non-Markovian dephasing model.
Noteworthy, the lineshape of the calculated spectra depends only on the electron-phonon coupling matrix elements. 
Moreover, the sensitivity of the choice of electronic wave function is clearly visible, comparing a Gaussian wave function with a 8 band ${\bf k }\cdot{\bf p}$  theory \cite{Schliwa2007QDs,Stier1999QDs}.
This is an important feature of the advanced model and highlights its usefulness for further optimization and technological fine tuning of single-QD quantum devices.  

%
\subsection{Phonon-enhanced Wigner time delay (Cluster expansion)}
\label{subsec:wigner_delay}
\begin{figure}
\includegraphics[width=0.5\columnwidth]{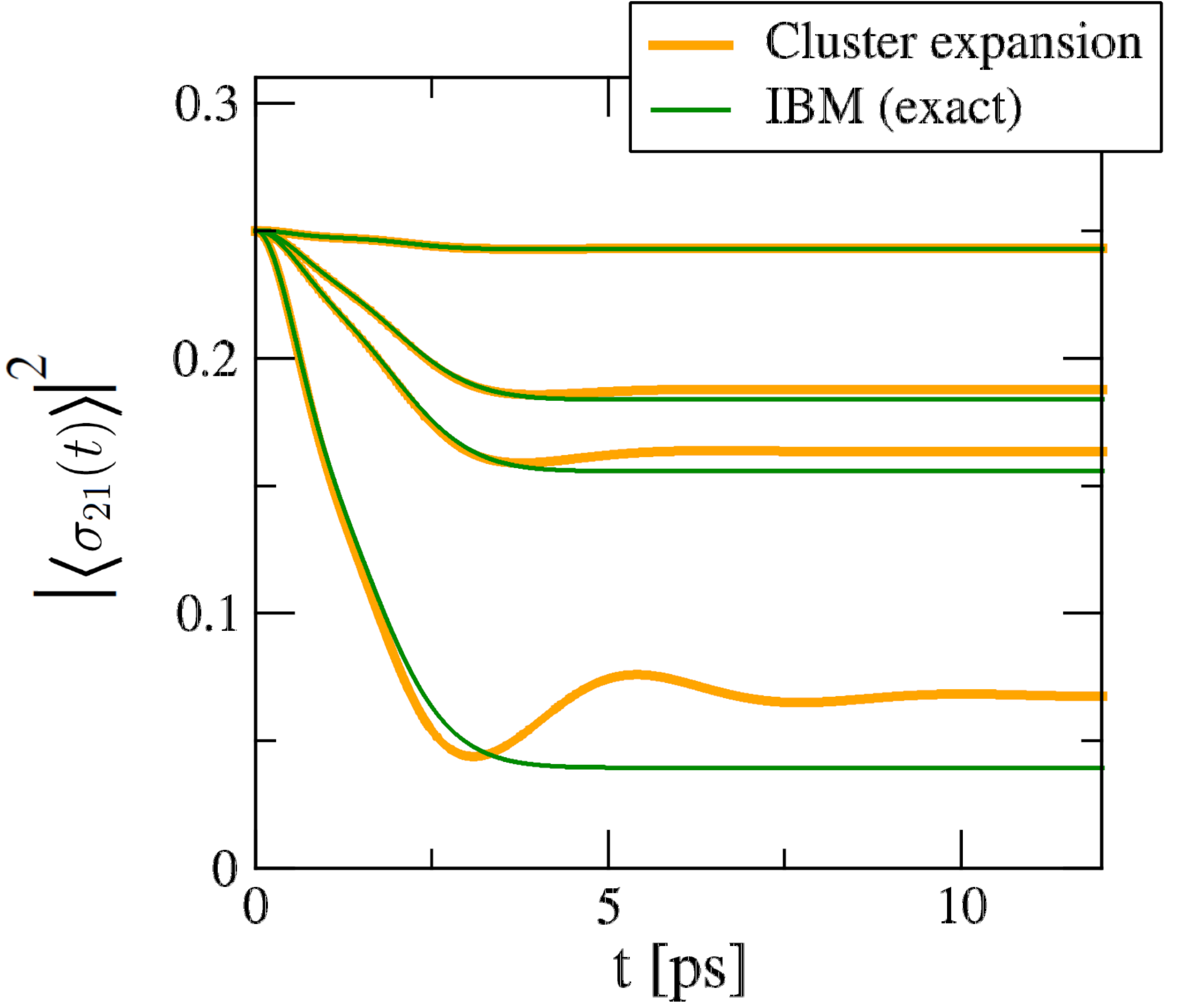}
\caption{Comparison for different temperatures (from top to bottom $4, 50, 77, 100$K) between the independent boson solution from Eq.~\eqref{eq:ibm_solution} (green line) and the second-order cluster expansion solution from Eq.~\eqref{eq:cluster_expansion_polarization} with $\Omega(t)=\delta(t)/2$ (orange line). For low temperatures and typical semiconductor parameter as given in the text, the solutions agree well up to around $60$~K.}
\label{fig:cluster_vs_ibm}
\end{figure}

%
Beyond the linear regime but still in the weak excitation limit, the Heitler regime includes nonlinear contributions as it studies the excited state dynamics $\rho_{22}(t)=\ew{\sigma_{22}(t)}$ as a figure of merit in strong contrast to the simpler linear regime which addresses $ \rho_{12}(t)=\ew{\sigma_{21}(t)}$.
In the Heitler regime, the incoming light is mainly coherently scattered, i.e. proportional to $ |\rho_{12}(t)|^2 $, but in dependence on the pulse length and pulse area, a part of the incoming laser excitation is absorbed $ \rho_{11}(t)<1 $, converted into electronic excitation $\rho_{22}(t)>0$ before being emitted back via incoherent scattering.
The process of re-emission needs a finite time between the incoming pulse and outgoing emission, and the delay between the maximum of the excitation pulse and emission pulse is called Wigner delay or dwelling time.
The Wigner delay is ultimately limited by the coherence time $1/T_2=1/(2T_1)+1/T_2^*$ \cite{leuchsWignerDelay} where $T_1$ denotes the radiative lifetime and $T_2^*$ is the coherence timescale.
If no or negligible decoherence is present ($T_2^*\rightarrow \infty$) such as in certain atomic systems \cite{bourgain2013direct,wignerDelayAtomic}, 
the maximum Wigner delay reads $ 2T_1 $ as the signal stems from radiative relaxation, and every relaxation induces a partial decoherence between the electronic levels.
Since the Wigner delay is limited by the decoherence process in case of solid-state emitters, it is an interesting figure of merit to quantify the effective $T_2^*$ time.
In particular when probing the spectral response, the detuning dependence of the Wigner delay reveals clearly a non-Markovian and phonon-coupling-dependent feature as we discuss in the following
\cite{strauss}.
\begin{figure}
\includegraphics[width=0.5\columnwidth]{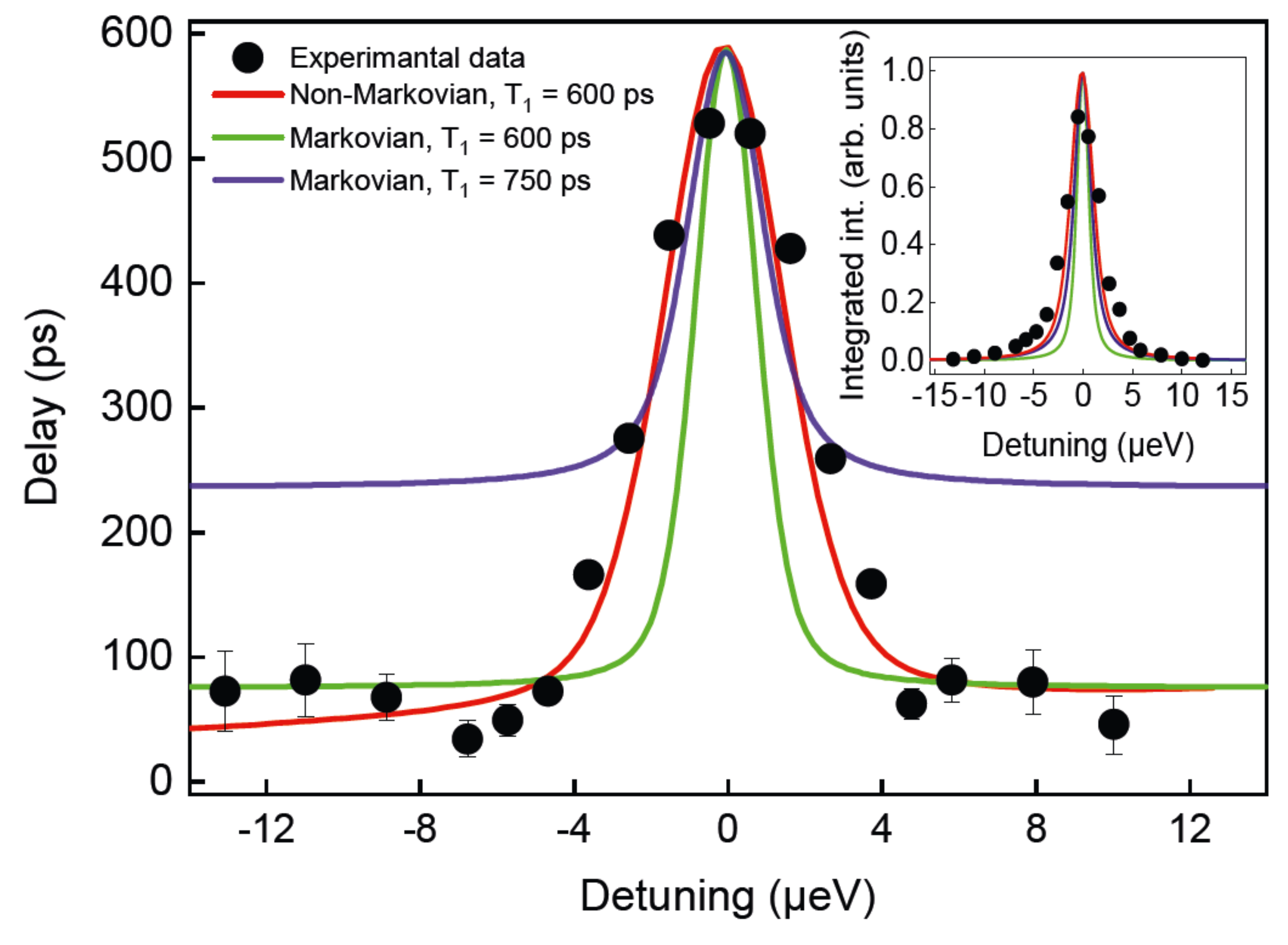}
\caption{Time delay as a function of spectral detuning between the laser and the TLS. The red line and black dots correspond respectively to the non-Markovian simulations and the experiment as discussed in the main text. Green and blue lines show the simulation obtained via the Markovian approximation. Inset: Integrated intensity of the scattered pulses as a function of detuning. \copyright2019 American Physical Society, reprinted from \cite{strauss}.}
\label{fig:wigner_delay}
\end{figure}
Since the pulse induces population dynamics of the electronical system, we need to solve the optical Bloch equations. Limiting our analysis of the electronic system to a two-level system, which is in good agreement with the experimental situation of studying InGaAs QD with close to ideal quantum properties at low temperature, we restrict the dynamics first to the Lindblad master equation case with 
\begin{align}
\dot \rho =& -\frac{i}{\hbar}[H_{e-l}^{sc}(t),\rho] 
+ \frac{\Gamma}{2}
\mathcal{D}[\sig{1}{2}]\rho
+
\frac{\gamma_p}{2}
\mathcal{D}[\sig{2}{2}]\rho.
\label{eq:lindblad_meq}
\end{align}
Dissipative contributions are taken into account via the 
Lindblad formalism and can be derived via second-order perturbation theory based on $H_{e-p}^{st}$ for the pure dephasing $\gamma_p$ and $H_{e-l}^{con}$ for the radiative decay $\Gamma$, cf.~Sec.~\ref{subsec:master_equation} for an exemplarily derivation of the pure dephasing Lindblad.
The dynamics of the system reads:
\begin{align}
\dot \rho_{22} =& -\Gamma\rho_{22} + 2\text{Im}[\Omega(t)\rho_{12}] \\
\dot \rho_{12} =& (i\Delta-\Gamma/2-\gamma_p)\rho_{12} 
-i \Omega(t) \left(2\rho_{22}-1\right),
\end{align}
where we use for the excitation pulse 
$\Omega(t) =  \Omega_L \sqrt{\frac{\pi}{(2\tau^2)}} \exp\left[-\frac{(t-t_0)^2}{2\tau^2}\right]$ with the amplitude $ \Omega_L$
and $ \Delta=\omega_{21}-\omega_L$.
The Heitler regime is given in the low excitation limit, 
when no Rabi oscillations will be induced and the coherence 
$ \rho_{12} $ is driven weakly and is mainly governed by the detuning and dephasing dynamics $ \Delta,\gamma=\Gamma/2+\gamma_p$.
Therefore, we yield after assuming quasi-steady dynamics ($\dot\rho_{12}\approx0)$ $\rho_{12}\approx \Omega(t) \left(2\rho_{22}-1\right) (\Delta-i\gamma)/(\Delta^2+\gamma^2)$ and the excitation density reads: $ 2\rho_{22} \approx [1+(\Delta^2+\gamma^2)\Gamma/(4\gamma\Omega^2)]^{-1} $.
Since the detection is proportional to the coherence time of the electronic signal, the phase shift of the transition dynamics is given in the adiabatic limit.
To describe the Wigner time delay in the presence of decoherence, we are not interested in the amplitude but in the phase which is proportional to $\phi(\omega)=\text{atan}[\Delta/\gamma]+\pi/2$ of the field amplitude $\Delta-i\gamma$.
The derivative of the phase with respect to the frequency allows us to find the Wigner delay in the steady-state limit:
\begin{align}
\tau_W=
\frac{d \phi}{d\omega} =&
\frac{1}{\gamma+\Delta^2/\gamma}
\end{align}
which for $ \Delta=0,\gamma_p=0 $ yields the maximum of the Wigner delay, namely $ \tau^\text{max}_W=2/\Gamma=2T_1$.
However, if we compare the Markovian theory (green and blue lines) with experimental data (black dots) in Fig.~\ref{fig:wigner_delay}, a Lindblad-based pure dephasing mechanism (green, blue lines) reproduces either the small or comparably large detuning limit but misses the asymmetry between positive and negative detunings. 
Importantly, it is not valid in both limits for the same phenomenological $\gamma_p$ value while the radiative decay $\Gamma$ and the pulse $\Omega(t)$ is fixed.
In contrast to the phenomenological decoherence model, the non-Markovian model, cf. Fig.~\ref{fig:wigner_delay}(red line) based on the semiconductor Bloch equation approach reproduces the experimental data for both limits and also exhibits the slight asymmetry between positive and negative detunings with respect to the laser frequency
\cite{strauss}.
We will now discuss how to obtain the non-Markovian system response.
In the weak coupling limit, the experimental data is modeled by semiconductor Bloch equations in the Heisenberg picture $-i\hbar \dt A = [H^{sc}_{e-l}+H_{e-p}^{la},A]$ from Eq.~\eqref{eq:H_sc} and \eqref{eq:H_la} with corresponding non-equilibrium phonon-contributions  \cite{kirabook,kirareview,carmele2010antibunching,haugkoch}
and $A$ a quantum mechanical operator such as $\sigma_{22}$. 
In this context it is important to note that lattice vibrations in semiconductor nanostructures give rise to new effects not encountered in typical atomic quantum optics \cite{cohen2011advances,gardiner2015quantum}.
These features stem from the non-, i.e. sub- or super Ohmian spectral density of the semiconductor electron-phonon interaction \cite{krummheuer2002,axt_mukamel,woggon1997optical,Stier1995QDs}. 
Corresponding Lindblad-based master equation treatments are derived via Markovian-, Born-, and secular approximation and neglect frequency-depending system-bath interaction strengths.
Furthermore, master equation approaches fail in this case due to the time-dependent pulse which enforces a time-reordering procedures \cite{mccutcheon2010quantum}.
Here, we model the dynamics via a Born-factorization approach which is valid up to a temperature of $60~$K and in the weak-driving limit, cf. Ref.~\cite{clustervspathintegral}, relevant for the description of the data presented in Fig.~\ref{fig:wigner_delay}.
When solving the semiconductor Bloch equation in the Heisenberg picture the one-electron assumption is considered $\ew{\sigma_{11}}=1-\ew{\sig{2}{2}}$.
The phonon dynamics is treated non-Markovianly in the bath assumption limit, i.e. second-order Born factorization and the derived set of equations of motion 
reads (including a Markovian radiative decay constant $ \Gamma $) with $ \ew{A(t)}=\text{Tr}[\rho(0)A(t)]$:
\begin{align}
\label{eq:cluster_sig22}
\frac{d}{dt} \ew{\sig{2}{2}} =&
-2\Gamma \ew{\sig{2}{2}}
+ 2 \text{Im}
\left[
\Omega(t) 
\ew{\sig{1}{2}}
\right], \\
\label{eq:cluster_expansion_polarization}
\frac{d}{dt} \ew{\sig{1}{2}} =&
-(\Gamma+i\Delta) \ew{\sig{1}{2}}
-i \Omega(t)
\left( 2 \ew{\sig{2}{2}} - 1
\right) 
-i \sum_{\bf q} 
g^{\bf q}_{12} \ew{b^\ndg_{\bf q} \sig{1}{2}} 
+ g^{{\bf q}*}_{12} \ew{b^\dg_{\bf q} \sig{1}{2}},
 \\
\label{eq:coherence_bndg}
\frac{d}{dt} \ew{b^\ndg_{\bf q} \sig{1}{2}} =&
-(\Gamma+i\Delta+i\omega_q) \ew{b^\ndg_{\bf q} \sig{2}{2}}
-i \Omega(t)
\left( 2 \ew{b^\ndg_{\bf q} \sig{2}{2}} - \ew{b^\ndg_{\bf q}}
\right) 
-i g^{{\bf q}*}_{12} \ew{b^\dg_{\bf q} b^\ndg_{\bf q}} \ew{\sig{1}{2}},
\\
\label{eq:coherence_bdg}
\frac{d}{dt} \ew{b^\dg_{\bf q} \sig{1}{2}} =&
-(\Gamma+i\Delta-i\omega_q) \ew{b^\dg_{\bf q} \sig{1}{2}}
-i \Omega(t)
\left( 2 \ew{b^\ndg_{\bf q} \sig{2}{2}}^* - \ew{b^\ndg_{\bf q}}^*
\right) 
-i g^{{\bf q}}_{12} \ew{b^\ndg_{\bf q}b^\dg_{\bf q}} \ew{\sig{1}{2}}, \\
\frac{d}{dt} \ew{b^\ndg_{\bf q} \sig{2}{2}} =&
-(2\Gamma+i\omega_q) \ew{b^\ndg_{\bf q} \sig{2}{2}}
-i \Omega(t)
\left( 
\ew{b^\ndg_{\bf q} \sig{1}{2}} 
- 
\ew{b^\dg_{\bf q} \sig{1}{2}}^*
\right) 
+i g^{{\bf q}*}_{vc} \ew{\sig{2}{2}}, \\
\label{eq:cluster_bndg}
\frac{d}{dt} \ew{b^\ndg_{\bf q}} =&
-i\omega_q \ew{b^\ndg_{\bf q}}
-i g^{{\bf q}*}_{12} \ew{\sig{2}{2}} .
\end{align}
The phonon occupation number are given via the Bose-Einstein distribution:
\begin{align}
\ew{b^\dg_{\bf q} b^\ndg_{\bf q}}  =
\left[ 
\exp(\hbar\omega_q/(k_BT)) - 1
\right]^{-1}.
\end{align}
Before discussing the Heitler regime, we show the validity of the second-order truncation in the weak-coupling limit. 
In Fig.~\ref{fig:cluster_vs_ibm}, we compare the solution of the independent boson model from Eq.~\eqref{eq:ibm_solution} (green, solid line) with the perturbative solution in second-order (orange, dotted-line) for different temperatures (from top to bottom, $4, 50, 77, 100$K.
Assuming an initial delta-pulse $\Omega(t)=\delta(t)/2$, the dynamics of the set of equations [Eq.~\eqref{eq:cluster_sig22}-\eqref{eq:cluster_bndg}] reduces to three equations [Eq.~\eqref{eq:cluster_expansion_polarization},\eqref{eq:coherence_bndg}, and \eqref{eq:coherence_bdg}] and can also be solved analytically \cite{krummheuer2005,krummheuer2002}.
The resulting dynamics agrees well with the analytically exact solution up to $60$K.
If the cluster-expansion solution is expanded up to the fourth-order, the agreement is even better and holds up to $150$K \cite{clustervspathintegral}.
Having shown the validity of the perturbative approach, we can discuss the role of phonons in the Heitler regime.
In the calculations, the radiative decay time $\Gamma=650\text{ps}^{-1}$ is taken from the experiment, and we choose a phonon-coupling strength so that we reproduce the maximum of the Wigner delay on resonance at $ \Delta=0$ for the experimentally given pulse width of $ \tau=1.05$ns.
Via the detuning dependence, the Wigner delay probes the amount of phonon-induced incoherence in the signal and is a good figure of merit to unravel and to monitor the electron-phonon interaction \cite{kabuss,hughes_incoherent}.
Clearly, the non-Markovian theory interpolates between both limits (large and small detuning), cf.~Fig.~\ref{fig:wigner_delay}(red, solid line) and reproduces also the small temperature-dependent asymmetry due to the preference of the system to emit rather than to absorb phonons at low temperatures.
This can be explained with Eq.~\eqref{eq:coherence_bndg} and \eqref{eq:coherence_bdg}, as in the low temperature limit mainly $\ew{b^\dg_{\bf q} \sig{1}{2}}$ contributes due to spontaneous phonon emission rather than induced absorption and stimulated emission.
Since the phonon frequency enters with a different sign into the dynamics of both of the phonon-assisted coherences, the pure dephasing is blue-detuned effectively larger which leads subsequently to an efficiently larger Wigner delay.
Applying this theory is therefore a key to describe an exciting optical effect known up until now only from atomic physics.
Together with experimental data on the Wigner time delay of a single two-level emitter represented by a high-quality QD it provides important access to the underlying decoherence processes and associated timescales which will be of  importance to tailor such quantum emitters for applications in quantum technology which require a high degree of coherence in the generation, transfer and interfacing of single photon in solid-state quantum devices.   
%

\subsection{Phonon-assisted damping of Rabi oscillations and state preparation (Cluster expansion)}
\label{subsec:heisenberg}
%
\begin{figure}[t!]
\centering
\includegraphics[width=7cm]{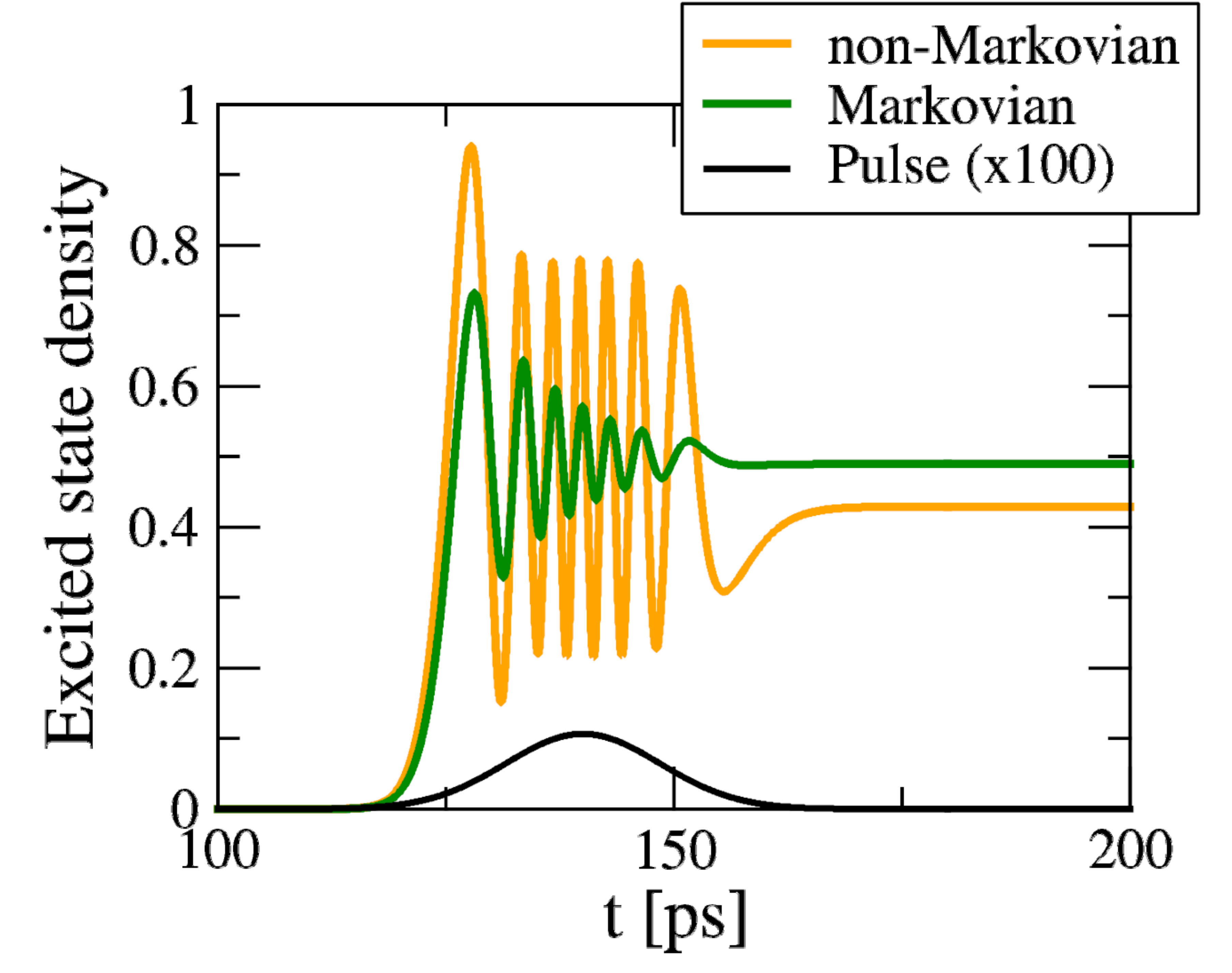}
\caption{Rabi oscillations of a QD under time-dependent driving (Gaussian pulse, black line). Non-Markovian electron-phonon interaction (orange line) leads to a renormalization of the Rabi frequency and recovers partial coherence after the first ps in contrast to Markovian damping (green line).}
\label{fig:heisenberg_rabi}
\end{figure}
%
As have been discussed in the previous section, in the weak coupling regime for the electron-phonon coupling, quantified here via $g^{\bf q}_{12}$, a factorization approach already mirrors qualitatively and quantitatively the system's dynamics very accurately.
A typical factorization approach for time-dependent and time-independent pulses is the Born factorization, typically within the Heisenberg picture of the operator of interest $A$, e.g. $\ew{A b^\dg_{\bf q} b^\ndg_{\bf q}}\approx\ew{A}\ew{b^\dg_{\bf q} b^\ndg_{\bf q}}$ \cite{gardinerzollerbook,breuerbook}.
To go beyond the Born-limit, a systematic correlation expansion approach can be applied \cite{haugkoch,kirabook,kirareview,chow2013,julia_raman}.
Here, the factorization is accompanied with a correction and the dynamics of this correction is taken into account up to arbitrary high orders: $\ew{A b^\dg_{\bf q} b^\ndg_{\bf q}}=\ew{A}\ew{b^\dg_{\bf q} b^\ndg_{\bf q}}+\delta\ew{A b^\dg_{\bf q} b^\ndg_{\bf q}}$ \cite{carmele2009photon,kirabook}.
For the weak coupling limit and non-entangled system-reservoir dynamics, this cluster expansion approach gives reliable quantitative agreement for weakly-correlated many-body systems \cite{kirabook}.
Already on a second-order level, the quantum kinetic dephasing dynamics of optically induced nonlinearities in GaAs QDs can be calculated accurately and in agreement with experiments for arbitrary pulse strengths. 
Evaluating the set of equations of motion [Eq.~\eqref{eq:cluster_sig22}-\eqref{eq:cluster_bndg}] beyond the Heitler regime, Rabi oscillations occur, cf.~Fig.~\ref{fig:heisenberg_rabi}.
Here, acoustical phonons renormalize the Rabi energy and result in damping that depends strongly on the input pulse strength, which is not included in a Markovian treatment \cite{forstner2003phonon,QD_rot_PhysRevLett.104.017402,QD_rot_PhysRevB.84.125304,QD_rot_PhysRevB.69.193302,QD_rot_PhysRevB.73.035302}.
This can be seen by comparing the dynamics induced by a Markovian, time-independent dephasing (green line) with the non-Markovian, phonon-induced dephasing (orange line) in the excited state density, cf.~Fig.~\ref{fig:heisenberg_rabi}. 
The Markovian dephasing dynamics overestimates the influence of the phonon strongly and acts continuously with the same damping strength.
In contrast, acoustical phonons attack initially the coherences strongly but saturate after few ps which leads to a constant Rabi oscillation amplitude.
This shows that Markovian an non-Markovian treatment of decoherence processes lead to qualitative different behavior and are difficult to compare on the same footing.
Note also, that the results can be obtained in a rate equation approach using corresponding dressed states \cite{PhysRevB.95.125308}.

\begin{figure}[t!]
\centering
\includegraphics[width=8cm]{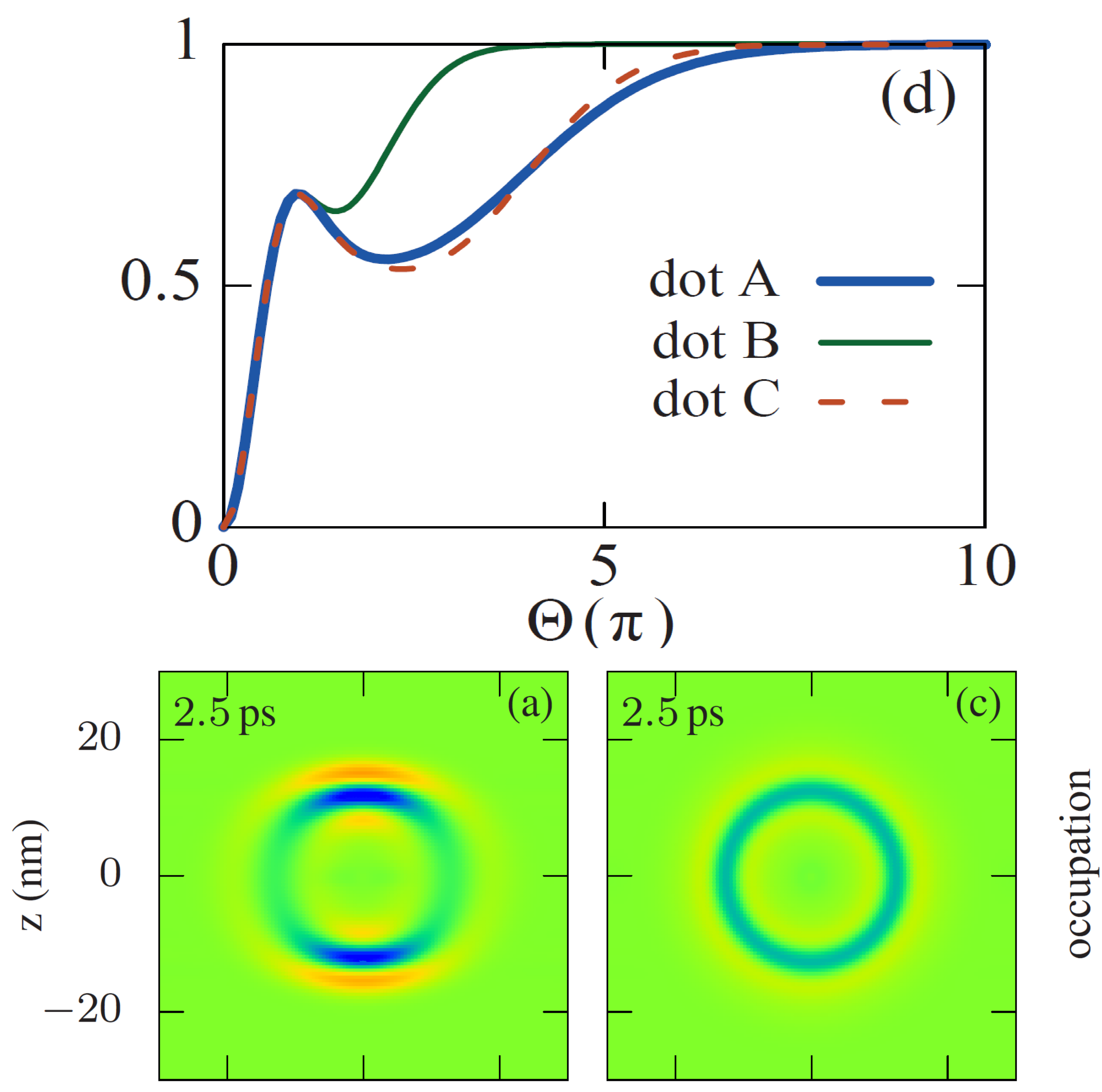}
\caption{(d) Final occupation of the excited QD state for a strongly lens-shaped QD (A) and two spherical shaped QDs (B,C). Despite different geometries QD A and C show quantitative and qualitative the same dependence on the pulse area. (a) and (c) the relative volume change after an excitation with a $2\pi$ pulse for (a) a lens-shaped and (c) a spherical shaped QD. The outgoing phonon waves differ strongly in directionality.
Reprinted figure with permission from \cite{reiter}. \copyright2017 by the American Physical Society.}
\label{fig:heisenberg_geometry}
\end{figure}
%

Another example is all-optical state preparation \cite{reiter2014role}.
The cluster-expansion techniques allows to investigate the impact of different geometries of the QD in state-preparation protocols. 
For example the electron-phonon coupling of a spherical QD and a more realistic lens-shaped QD can explicitly be compared \cite{reiter}.
Interestingly, the numerical analysis shows that the QD (electronic) dynamics is hardly influenced on the actual nanostructure geometry, cf.~Fig.~\ref{fig:heisenberg_geometry}(d),
and depends mainly on the smallest dimension which governs the electron-phonon interaction. 
For instance, comparing the state preparation dynamics for different geometries (lens-shaped QD A, spherical B,C with different radii) of the QD's excited state, the qualitative dependence on the pulse area is the same.
In this regard it is noteworthy that the strongly lens-shaped QD A has nearly exact the same pulse area dependence as the spherical QD C despite different spectral densities for the electron-phonon interaction strength.
This result allows one to map the electronic dynamics of lens-shaped QDs with spectral densities derived via a spherical geometry when studying the phonon influence on the electronic system. 
In contrast to the electronic kinetics which are mainly governed by the exciting laser field, the actual nanostructure geometry has a very strong impact on the spatio-temporal properties of the phonon dynamics, cf. Fig.~\ref{fig:heisenberg_geometry}(a) and (c). 
An example is given, where for a lens-shaped QD (a) the phonon emission is strongly concentrated along the direction of the smallest axis of the QD which is important for a phonon-mediated coupling of different QD \cite{naumann2016solid,droenner2017collective}.
Therefore, the QD shape plays an important role in determining the properties of the created phonons and possible application for phonon lasing and sensing with solid-state nanostructures 
\cite{julia_phonon_laser,rabl_phonon_laser,reiter_bayer_phonon_laser}. 
Also, it has been shown that adding a high chirp rate to ultrashort laser pulses, the QD can be decoupled from the phononic environment and thus a reappearance of rapid adiabatic passage can be established \cite{kaldewey2017demonstrating}.
%

\subsection{Phonon-assisted incoherent excitation processes (Polaron master equation)}
\label{subsec:master_equation}
\begin{figure}[t!]
\centering
\includegraphics[width=5cm]{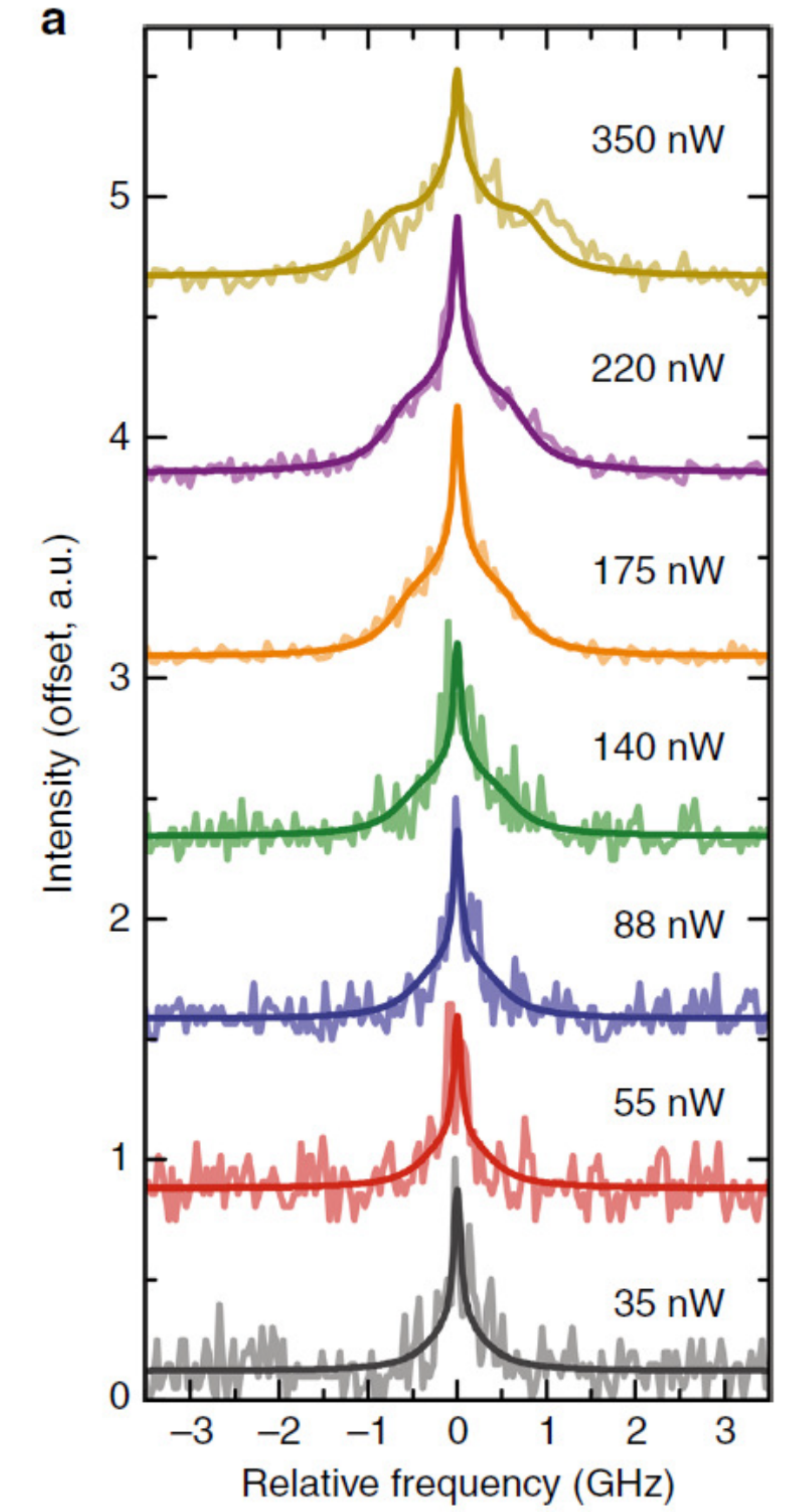}
\caption{Excitation dependent resonance fluorescence emission spectra and photon autocorrelation function under cw excitation by a state-of-the-art high-$\beta$ QD-microlaser.  Increasing excitation power lead towards a Mollow-triplet like emission spectrum. Reprinted figure from \cite{kreinberg2018quantum}.}
\label{fig:markovian_mollow}
\end{figure}

%
%
The Heisenberg equation of motion method is in general valid 
in the weak coupling limit. 
This is not the case for the strong coupling and high temperature limit, where most perturbative approaches fail. 
However, in case of a time-independent pump, the Hamiltonian in Eq.~\eqref{eq:H_sc} can be rewritten in the polaron frame, and a more convenient non-Lindblad type of master equation can be derived \cite{mccutcheon2010quantum,hughes_incoherent}.
The main goal of the polaron transformation is to trace out the degrees of freedom of the phonon reservoir in a second-order Born theory but to keep as much information as possible about the electron-phonon interaction.
In second-order perturbation theory, the reduced density matrix $ \text{Tr}_B\left\lbrace\rho(t)\right\rbrace=\rho_s(t)$ in the interaction picture and in second-order Born approximation reads:
\begin{align}
\dot\rho_s(t) 
=&
-\frac{i}{\hbar}
[H^{sc}_{e-l},\rho_s(t)]
-\frac{i}{\hbar}
\text{Tr}_B
\left\lbrace 
[H_I,\rho^I_s(t)\otimes\rho_B]
\right\rbrace \\ \notag
&
-\frac{1}{\hbar^2}
\int^t_0 d\tau 
\text{Tr}_B
\left\lbrace 
[H_I,[H_I(-\tau),\rho_s(t-\tau)\otimes\rho_B]]
\right\rbrace \label{eq:second_order_meq}. 
\end{align}
As a first approach, one can choose a stochastic force via $H^{st}_{e-p}$.
In this case, the interaction Hamiltonian reads $H_I(\tau)=\hbar\sigma_{22}(\tau)F(t+\tau)$ and the trace over the bath degrees of freedom is a Gaussian average 
$\text{Tr}_B\lbrace \dots \ \rho_B\rbrace \to \avg{ \dots}$.
As the electronic operators are not affected by the statistical average and c-values commute, we obtain:
\begin{align}
\avg{[H_I,[H_I(-\tau),\rho_s(t-\tau)]]}
= \hbar^2 (\gamma_p/2) \delta(\tau) 
[\sigma_{22},[\sigma_{22}(-\tau),\rho_s(t-\tau)]]
\end{align}
where we used the white noise correlation for the stochastic force $\avg{F(t_1)F(t_2)}=\gamma_p\delta(t_1-t_2)/2$ and $\avg{F(t)}=0$.
In this limit, the master equation reads as given in Eq.~\eqref{eq:lindblad_meq}:
\begin{align}
\dt\rho_s(t) 
=&
-i
[\Delta\sig{2}{2}+\Omega(\sig{1}{2}+\sig{2}{1}),\rho_s(t)]
+\gamma_p \left[\sig{2}{2}\rho_s(t)\sig{2}{2}-\rho_s(t)\right]
\label{eq:mollow_pure_dephasing_meq}. 
\end{align}
Such a Markovian limit is valid in many experimental setups and driving scenarios, as long as the dynamics of the chosen observable evolves on a much smaller timescale than the environmental correlation times.
For example Fig.~\ref{fig:markovian_mollow}(a) depicts excitation dependent resonance fluorescence emission spectra for different excitation which were obtained in a unique experiment using a high-$\beta$ QD microlaser as cw pump~\cite{kreinberg2018quantum}.
%
%
Acting as excitation the electrically-driven high-$\beta$ microlaser drives with its output field a semiconductor QD.
This QD acts as a two-level system within an advanced fiber-coupled experimental setup based on two cryostats which host the QD-microlaser and single QD sample, respectively. 
The experimental data on the resonantly driven QD is obtained at cyrogenic temperatures of $7$K to minimize electron and hole escape from the QDs and phonon-induced decoherence $\gamma_p$.
However, the emerging Mollow spectra for increasing excitation powers show driving-induced  pure dephasing which we model quantitatively by our model Eq.~\eqref{eq:mollow_pure_dephasing_meq}.
Indeed, the Markovian theory (solid lines) reproduces well the experimental data for optimum $ T_2 $ values around $ 500$ps.
The Mollow spectrum, derived from the power spectrum formula $S(\omega)$ reads:
\begin{align}
S(\omega) =& \notag 
\lim_{t \to \infty}
\text{Re}
\left[
\int_0^\infty d\tau \ew{\sigma_{21}(t)\sigma_{12}(t+\tau)}
e^{-i\omega\tau}
\right],
\\
S(\omega) =& \notag
\frac{\Gamma_R^2}{4\Omega^2+\Gamma_P\Gamma_R}\pi\delta(\omega-\omega_0)
+\frac{\Gamma_P}{(\omega-\omega_0)^2+\Gamma_P^2} \\
&+
\frac{1}{2}
\frac{\Gamma_+Z_C+(\omega-\omega_0+\Omega_R)Z_S}
{\Gamma_+^2+(\omega-\omega_0+\Omega_R)^2}
+
\frac{1}{2}
\frac{\Gamma_+Z_C-(\omega-\omega_0-\Omega_R)Z_S}
{\Gamma_+^2+(\omega-\omega_0-\Omega_R)^2},
\end{align}
where the following abbreviations are used:
$ \Gamma_R=2\Gamma $, $ \Gamma_P=\Gamma+\gamma_p$, 
and $ \Gamma_\pm=(\Gamma_R\pm\Gamma_P)/2 $, 
$ \Omega_R=\sqrt{4\Omega^2-\Gamma_-^2}$ and
the oscillator strength functions are introduced:
\begin{align}
Z_C =& 1-\frac{\Gamma^2_R}{4\Omega^2+\Gamma_R\Gamma_P} \\
Z_S =& \frac{\Gamma_R}{\Omega_R}
\left(2-\frac{\Gamma_+}{4\Omega^2+\Gamma_R\Gamma_P}
\right).
\end{align}
Although a Markovian model may apply to special situations, more insight into the decoherence processes is feasible to overcome temperature and single-photon repetition-rate limitations.
As shown above, Markovian, i.e. global, frequency- and time-independent dephasing dynamics do not always model the dynamics correctly, and it is necessary to go beyond the Born-Markovian and secular limit of the master equation. 
To include as much information as possible in second-order perturbation theory, we transform the Hamiltonian for a time-independent Rabi frequency $ \Omega(t)=\Omega_0 $ in 
Eq.~\eqref{eq:H_sc} into the polaron frame
\cite{wilsonPMEQ,nazirPMEQ,mansonPMEQ,hohenesterFeeding}.
We use the electron-phonon interaction Hamiltonian and apply a unitary transform, i.e. diagonalization of the phonon part of the Hamiltonian, via: $ U_p=\exp[\sig{2}{2}(R^\dg-R)]$ and $ R=\sum_{\bf q} g^{\bf q}_{12} b^\ndg_{\bf q}/\omega_q$.
The Hamiltonian in the polaron frame can be calculated via $ \exp[x]y\exp[-x]=\sum_{n=0}^\infty [x,y]_n/n!$ and $ [x,y]_0=y, [x,y]_n=[x,[x,y]_{n-1}]$:
\begin{align}
U^\ndg_p \sig{2}{2} U^\dg_p 
=&
\sig{2}{2}, \\
U^\ndg_p (\sig{1}{2}+\sig{2}{1}) U^\dg_p 
=
&\cosh[R^\dg-R](\sig{1}{2}+\sig{2}{1}) \\ \notag
+
&\sinh[R^\dg-R](\sig{1}{2}-\sig{2}{1}), \\
U^\ndg_p b^\dg_{\bf q} b^\ndg_{\bf q} U^\dg_p 
=&
b^\dg_{\bf q}b^\ndg_{\bf q}
-\sig{2}{2}
\frac{g^{\bf q}_{12}}{\omega_q}
(b^\dg_{\bf q}+b^\ndg_{\bf q})
+
\sig{2}{2}
\left(
\frac{g^{\bf q}_{12}}{\omega_q}
\right)^2, \\
U^\ndg_p \sig{2}{2} (b^\dg_{\bf q}+b^\ndg_{\bf q}) U^\dg_p 
=&
\sig{2}{2} (b^\dg_{\bf q}+b^\ndg_{\bf q})
-2\sig{2}{2}\frac{g^{\bf q}_{12}}{\omega_q}.
\end{align}
The transformed Hamiltonian is split into an interaction Hamiltonian: 
\begin{align}
H_I/\hbar =&
(\sig{1}{2}+\sig{2}{1})(\Omega\cosh[R-R^\dg]-\bar\Omega) \\ \notag
&
+\Omega(\sig{2}{1}-\sig{1}{2})\sinh[R-R^\dg],
\end{align}
with $ \bar\Omega=\Omega\ew{\exp[R^\dg-R]}=\Omega\exp[-\phi(0)/2]$ where
$ \phi(t) $ denotes the phonon correlation of the diagonalized electron phonon-interaction:
\begin{align}
\phi(t) =& \sum_{\bf q} \left|g^{\bf q}_{12}\right/\omega_q|^2 \left(
\coth\left[\frac{\hbar\omega_q}{2k_bT}\right]
\cos(\omega_q t) -i \sin(\omega_q t) 
\right).
\label{eq:phonon_correlation}
\end{align}
The corresponding Hamiltonian of the free evolution reads:
\begin{align}
H_0/\hbar
=&
\Delta\sig{2}{2}
+\bar \Omega (\sig{1}{2}+\sig{2}{1})
+\sum_{\bf q} \omega_q b^\dg_{\bf q} b^\ndg_{\bf q}, 
\end{align}
with $ \Delta=\omega_{21}-\omega_{L}-\sum_{\bf q} \frac{|g^{\bf q}_{12}|^2}{\omega_q}$.
Additionally the Franck-Condon renormalization $ \bar\Omega(\sig{1}{2}+\sig{1}{2})$ is introduced to the free evolution and subtracted from the interaction Hamiltonian 
to ensure in second-order perturbation theory in $H_I(t)$ a vanishing first-order contribution:
$ \text{Tr}\left\lbrace 
[H_I(t),\rho^I_s(t)\otimes\rho_B]
\right\rbrace=0$,
and the resulting non-Markovian master equation reads:
\begin{align}
\dt\rho_s(t) 
=&
-i
[\Delta\sig{2}{2}+\bar\Omega X_+,\rho_s(t)] \\ \notag
&-
\frac{\bar\Omega^2}{4}
\sum_{i=+,-}
\int_0^t
d\tau
\left\lbrace
G_i(\tau)[X_i,X_i(-\tau)\rho_s(t)] + \text{h.a.}
\right\rbrace,
\end{align}
with $ X_+ = \sig{1}{2}+\sig{2}{1}=X_+^\dg$
and $ X_- =i (\sig{2}{1}-\sig{1}{2})=X_-^\dg$ to allow for a compact formulation of the master equation.
The corresponding phonon Green's functions read
$
G_+(t)=
\cosh[\phi(t)]
-1$, 
and
$
G_-(t)=
\sinh[\phi(t)]
$.
Interestingly, this trace-preserving master equation (after setting $\rho(t-\tau)=\rho(t)$) simulates the complex decoherence dynamics in more detail for time-independent system dynamics, and poses a feasible alternative to time-convolutionless techniques
\cite{breuerbook,richter2010time}.
In the limit of a vanishing coupling element, one recovers the system dynamics without any contributions from the environment, as
$ G_\pm(t) \equiv 0 $ and $ \phi(t)=0 $ it follows $ \Omega_R =\Omega $.
The time dynamics of $X_i(\tau)$ is given via the electronic part of $H_0$.
The general solution for a detuned, driven two-level system reads with $\eta=\sqrt{4\Omega^2+\Delta^2}$:
\begin{align}
X_-(t) &= 
\frac{2\Omega}{\eta}\sin(\eta t) X^0_z
+
\cos(\eta t) X^0_- 
-
\frac{\Delta}{\eta} \sin(\eta t) X^0_+, \\
X_+(t) &= 
\frac{2\Omega\Delta}{\eta^2}[1-\cos(\eta t)] X^0_z
+
\frac{X^0_+}{\eta^2}
\left[
4\Omega^2+\Delta^2 \cos(\eta t)
\right]
+
\frac{\Delta}{\eta}
\sin(\eta t)
X^0_-, \\
X_z(t) &=
\left[ \cos(\eta t)-\frac{\Delta^2}{\eta^2}[1-\cos(\eta t]\right]X^0_z
+ \frac{2\Omega\Delta}{\eta^2} [1-\cos(\eta t)] X^0_+
-\frac{2\Omega}{\eta} \sin(\eta t) X^0_-,
\end{align}
where $X_i^0=X_i(0)$ are the initial values and $X_z=\sigma_{22}-\sigma_{11}$ is given for completeness.
\begin{figure}[t!]
\centering
\includegraphics[width=12cm]{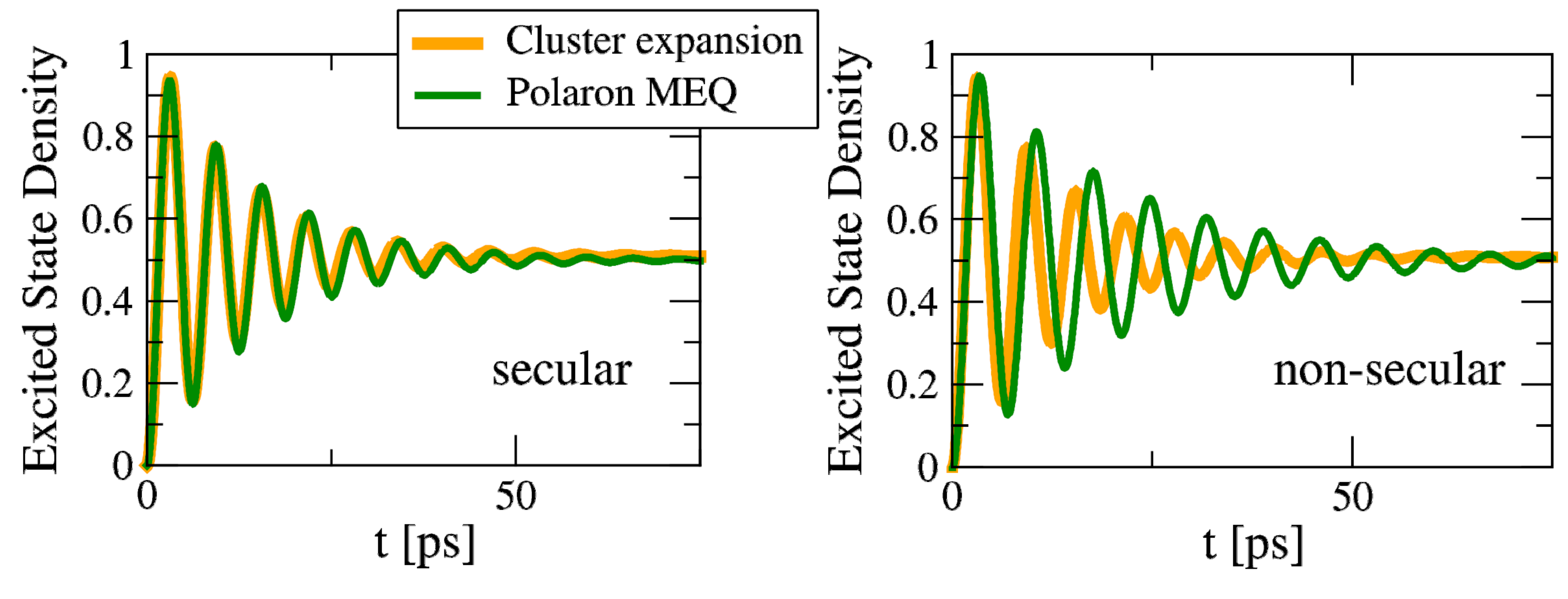}
\caption{Comparison between the polaron master equation and cluster expansion solution in the second-order and weak driving limit ($\hbar\Omega=50\mu$eV). The solutions agree well without non-secular terms and Rabi energy renormalization (left) for temperatures up to $T=77$K.}
\label{fig:cluster_vs_polaron}
\end{figure}
Another interesting limiting case assumes 
$X_i(\tau)\approx X_i(0)$, i.e. the system's dynamics is too slow and the environmental time correlation dominate the dissipative part of the master equation.
As $X_+^2=\mathbb{1}=X_i^2$ and $X_i(t)\approx X^0_i$, we can write
\begin{align}
\sum_{i=+,-}
&\left\lbrace
G_i(\tau)[X_i,X_i(-\tau)\rho_s(t)] + \text{h.a.}
\right\rbrace \\ \notag
&\approx  2 
\sum_{i=+,-}
\text{Re}[G_i(\tau)] 
\left[ 
\rho_s(t) -  X_i \rho_s(t) X_i
\right] \\ \notag
&=
-
\text{Re}[G_+(\tau)+G_-(\tau)] 
\left[ 
2 \sigma_{12} \rho_s(t)\sigma_{21}
+
2 \sigma_{21} \rho_s(t)\sigma_{12} 
- 2\rho_s(t)
\right] \\ \notag
&\phantom{=}
-
\text{Re}[G_-(\tau)-G_+(\tau)] 
\left[ 
2 \sigma_{12} \rho_s(t)\sigma_{12}
+
2 \sigma_{21} \rho_s(t)\sigma_{21} 
\right],
\end{align}
which leads to the following master equation:
\begin{align}
\label{eq:polaron_weak_driving}
\dt\rho_s(t) 
=&
-i
[\Delta\sig{2}{2}+\bar\Omega X_+,\rho_s(t)] +\Gamma_+(t) 
\left(
\mathcal{D}[\sigma_{21}]
+
\mathcal{D}[\sigma_{12}]
\right) 
\rho_s(t)
\\ \notag
\phantom{=}&
+\Gamma_-(t)
\left(
\sigma_{12} \rho_s(t)\sigma_{12}
+
\sigma_{21} \rho_s(t)\sigma_{21}
\right)
\end{align}
with the time-dependent damping and dephasing 
coefficients:
\begin{align}
\Gamma_+(t)
&= 
\frac{\bar\Omega^2}{4}
\text{Re}
\left[
\int_0^t d\tau
(
e^{\phi(\tau)} 
-1)
\right] \\
\Gamma_-(t)
&= 
\frac{\bar\Omega^2}{4}
\text{Re}
\left[
\int_0^t d\tau
(
e^{-\phi(\tau)} 
-1)
\right].
\end{align}
In Fig.~\ref{fig:cluster_vs_polaron}, we plot the solutions obtained from the polaron master equation in  Eq.~\eqref{eq:polaron_weak_driving} in the weak driving limit $\Omega=50\mu$eV and the solution of the cluster expansion of second-order from the set of equations in Eq.~\eqref{eq:cluster_sig22}-\eqref{eq:cluster_bndg} for standard GaAs material parameter and $T=77$K. 
In the secular approximation $\Gamma_-(t)=0$ and without the Rabi energy renormalization $\bar\Omega=\Omega$ in $H_0$ (left), the solutions agree well. 
Deviations with non-secular terms and the renormalized Rabi energy become apparent even in the weak coupling, weak driving limit.
However, the solution have been obtained in the very
weak and resonant driving scenario.
The advantage of the polaron master equation is the possibility to take into account dressed state dynamics
in the case, when the time dependence of $X_i(\tau)$
becomes non-negligible.
%

\begin{figure}
\centering
\includegraphics[width=10cm]{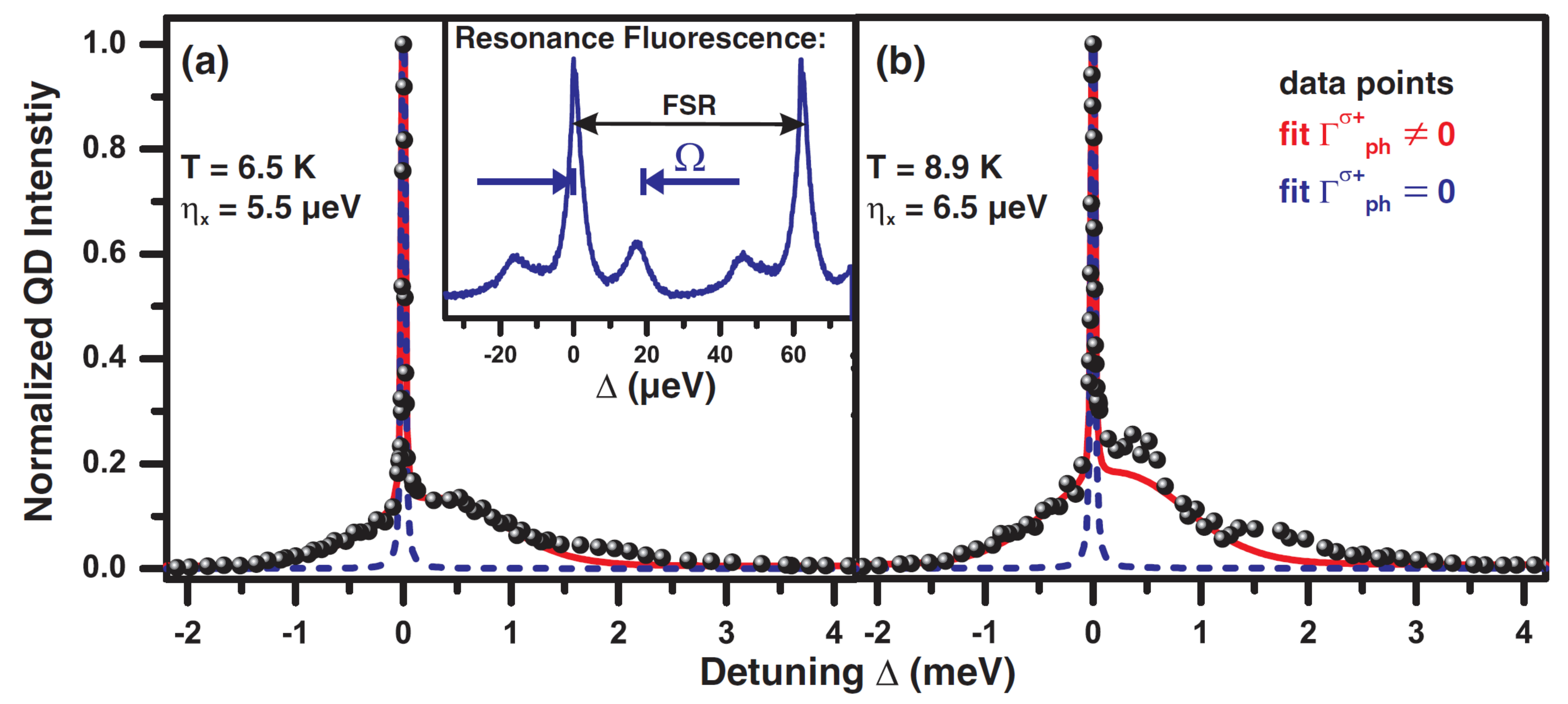}
\caption{Integrated QD intensity derived from a frequency scan (black dots, experiment) and theory with (red) and without (blue) phonon-included processes on the basis of the polaron master equation for different temperatures (a) and (b).
Reprinted figure with permission from \cite{hughes_incoherent}. \copyright2012 by the American Physical Society.}
\label{fig:polaron_meq}
\end{figure}
%
An interesting result from such a scenario is plotted in Fig.~\ref{fig:polaron_meq}. 
The experimental data points are obtained from self-assembled In(Ga,As)/GaAs QDs grown by metal-organic vapor-phase epitaxy. 
The corresponding sample includes a single layer of QDs centered in a 1-$\lambda$-thick planar GaAs cavity surrounded by alternating $\lambda/4$ periods of AlAs/GaAs as 4 top and 20 bottom distributed Bragg reflectors to enhance the photon extraction efficiency.
A narrow band ($∼500$ kHz) tunable Ti:Sapphire cw laser served as an excitation source and the integrated intensity of an incoherently driven QD is compared with theory~\cite{hughes_incoherent}.
This detailed study of a phonon-assisted incoherent excitation mechanism of single QDs allows one to explore the underlying acoustic phonon bath dynamics and shows very good agreement with the polaron master equation theory. 
Please note that the experimental data points (black spheres) are only well-produced, if the full phonon-induced quantum kinetics is taken into account such as in a polaron master equation model (red) with a constant external driving field and emerging additional incoherent excitation channels.
Therefore, the QD coupling to the phonon reservoir does not only introduce pure dephasing and an enhancement of radiative decay processes, but also new and in Markovian models not included incoherent excitation/scattering channels if spectral detuning becomes important.
The applied phonon-assisted incoherent excitation provide a unique excitation mechanism of a semiconductor QD and can be employed as an effective new tool to map the characteristic features of the phonon bath present in such a solid-state quantum-emitter system. 
Additional, it serves as an interesting quasi-resonant excitation scheme for the triggered generation of single photons with high indistinguishability \cite{state_preparation_michler}.
%

\subsection{Phonon-anticrossings in the Mollow regime (Inductive equations of motion)}
\label{subsec:inductive_eom_approach}

\begin{figure}[t!]
\centering
\includegraphics[width=12cm]{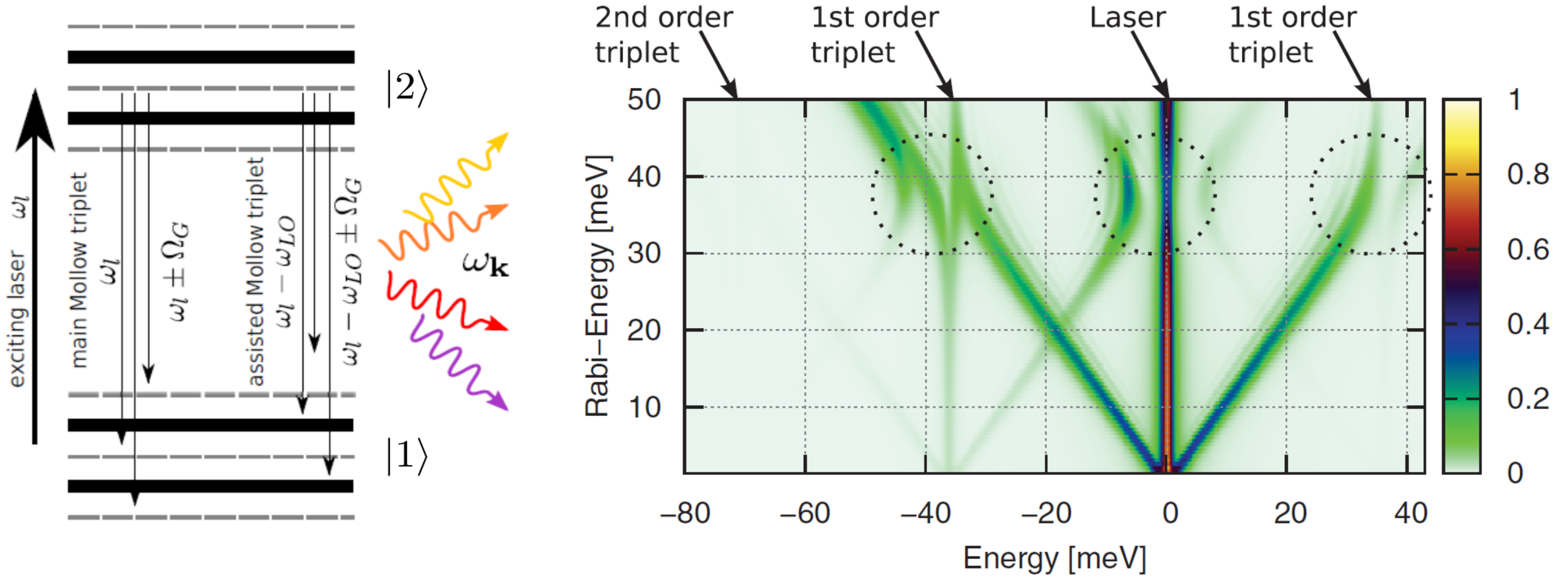}
\caption{Scheme and power spectrum of a two-level system (e.g. QD) interacting coherently with strong external laser field. The emission consists of triplets centered at the the driving laser frequency (Mollow triplet) and at the Raman frequencies (only Stokes contribution shown). For driving strength of the order of the LO phonon energy, additional anticrossings occur. 
\copyright
2011
American Physical Society, reprinted from
\cite{kabuss}.
}
\label{fig:mollow_lo}
\end{figure}

In addition to the perturbative models above, we
address now numerically exact solutions.
We start with the Heisenberg equation of motion approach which can be used to calculate non-equilibrium phonon dynamics up to arbitrary order if an inductive equation of motion method is employed.
For example, for longitudinal optical (LO) phonons the dispersion is constant in the Einstein approximation: $ \omega(q)\equiv\omega_{LO}$
\cite{axt1999coherent,krummheuer2005,stauber}.
This allows for an exact treatment of the electron-LO phonon interaction \cite{stauber,axt1999coherent,carmele2010antibunching,kabuss2011inductive}.
Due to the constant dispersion, we may write the Hamiltonian in Eq.~\eqref{eq:H_lo} in the interaction picture as 
$ H^{lo}_{e-p}(t)=\sum_{\bf q} f^{\bf q}_{12} [b^\ndg_{\bf q}(t)+b^\dg_{\bf q}(t)]= B e^{-i\omega_{LO}t} +B^\dg e^{i\omega_{LO}t} $.
This leads to the commutation relation $[B(t),B^\dg(t)]=|f|^2$ with 
$|f|^2=\sum_{\bf q} |f_{12}^{\bf q}|^2$, which corresponds to the renormalized harmonic oscillator picture.
This collective operator leads to a numerically exact solvable set of equations.
Employing the Heisenberg equation of motion for a non-explicit time-dependent operator: $ -i\hbar\dt A = [H^{sc}_{e-l}+H^{lo}_{e-p},A] $ from Eq.~\eqref{eq:H_sc} and \eqref{eq:H_lo}, 
a set of differential equations, defining $ \ew{A^{n,m}}=\ew{A B^{\dg n}B^{m}}$, can be derived.
As an example, we give the full set of equations of motion for a driven-QD with LO-phonon interaction:
\begin{align}
\dt \ew{\sig{1}{2}^{n,m}} 
=
&-i (\omega_{21}-(n-m)\omega_{LO}-i\Gamma/2)\ew{\sig{1}{2}^{n,m}}  
\\ \notag
&-i\Omega(t) 
\left(2 \ew{\sig{2}{2}^{n,m}} 
-
\ew{\mathbb{1}^{n,m}} 
\right) \\ \notag
&-i \sum_{\bf q} f^{\bf q}_{12} 
\left(\ew{\sig{1}{2}^{n+1,m}}
+
\ew{\sig{1}{2}^{n,m+1}}
\right)
\\ \notag
&-i (1-\delta_{n,0})
n |f|^2 \sum_{\bf q} f^{\bf q}_{12} \ew{\sig{1}{2}^{n-1,m}},
\end{align}
\begin{align}
\dt \ew{\sig{2}{2}^{n,m}} 
=
&-(\Gamma-i(n-m)\omega_{LO})\ew{\sig{2}{2}^{n,m}}  \\
\notag
&+2\text{Im}
\left[
\Omega(t) 
\ew{\sig{1}{2}^{n,m}} 
\right]
\\
\notag
&-i \sum_{\bf q} f^{\bf q}_{12} 
\left(\ew{\sig{2}{2}^{n+1,m}}
+
\ew{\sig{2}{2}^{n,m+1}}
\right)
\\
\notag
&+i (1-\delta_{n,0})
n |f|^2 \sum_{\bf q} g^{\bf q}_{12} \ew{\sig{2}{2}^{n-1,m}} \\
&
\notag
-i (1-\delta_{m,0})
m |f|^2 \sum_{\bf q} f^{\bf q}_{12} \ew{\sig{2}{2}^{n,m-1}} , \end{align}
\begin{align}
\dt \ew{\mathbb{1}^{n,m}} 
=
&i(n-m)\omega_{LO}\ew{\mathbb{1}^{n,m}}  \\ 
\notag
&+i (1-\delta_{n,0})
n |f|^2 \sum_{\bf q} f^{\bf q}_{12} \ew{\sig{1}{2}^{n-1,m}} \\ \notag
&-i (1-\delta_{m,0})
m |f|^2 \sum_{\bf q} g^{\bf q}_{12} \ew{\sig{1}{2}^{n,m-1}}. 
\end{align}
This set of equation is complete and allows one to solve the dynamics of the coupled phonon-emitter system and calculate the associated emission spectra, cf.Fig.~\ref{fig:mollow_lo}.
Note that with given initial conditions and for a fixed time interval, hierarchies only up to a certain $N$ in the phonon interaction contribute.
This maximum $ N $ is tested until convergence is reached, depending on the time interval and coupling strength of the system.
This inductive equation of motion method has been applied, e.g. to calculate the luminescence spectrum of a strongly-driven QD with LO phonon satellite peaks \cite{kabuss}. 
Due to the present electron-LO phonon interaction, the typical Mollow triplet is changed and additional side peaks appear at the frequency of the LO phonon satellite peaks $36$meV.
In Fig.~\ref{fig:mollow_lo}, an emission scheme is depicted showing that besides the strong drive induced Mollow sideband also LO phonon emission and absorption occurs and introduces Raman-like features into the spectrum under coherent excitation.
Another important aspect is the fact that the phonon emission and absorption is strongly temperature-dependent.
Here, the temperature is included in the initial conditions for the  expectation values: $\ew{B^{\dg n}(0)B^{m}(0)}
=\delta_{nm} n! \left[ \exp(\beta\hbar\omega_{LO}) -1 \right]^{-1}$
and $ 1/\beta=k_B T$.
For low temperatures, due to the LO-phonon frequency in the range of e.g. $ \hbar\omega_{LO}=36$meV, spontaneous
emission dominates, leading to strong Stokes contributions.
Given the set of equations of motions, the dynamics of the coupled laser-QD system can be calculated and due to the included Markovian radiative decay constant $ \Gamma $, a steady-state is inevitably reached.
Via a two-time correlation function, using the quantum regression theorem \cite{gardinerzollerbook,carmichael2009statistical,scully1999quantum}, the power spectrum $S(\omega)$ is obtained.
In Fig.~\ref{fig:mollow_lo}, the power spectrum of a QD under cw-excitation $ \Omega(t)=\text{const}$ is plotted for different Rabi energies $\Omega$.
Clearly, for low-excitation amplitudes, the Mollow triplet can be identified and an additional LO-phonon satellite peak triplet becomes visible, where both scale linearly with the Rabi energy.
Interestingly, if the excitation strength becomes comparable to the LO-phonon energy, anticrossings occur and both, the electronic and phononic triplet interfere, forming a new Eigenstate \cite{kabuss}. 
This has the interesting practical consequence that the splitting allows one to obtain independently and spectrally resolved the electron-phonon interaction strength without relying on the delicate measurement of the relative peak heights \cite{stock,PhysRevLett.83.4654,PhysRevB.64.241305} to estimate the Huang-Rhys factor $ F_{HR}=\sum_{\bf q} |f_{12}^{\bf q}|^2/\omega^2_{LO}$.
%

\subsection{Phonon-assisted population inversion (Real-time path integral)}
\label{subsec:path_integral}
Due to the intrinsic non-Markovian nature of the electron-phonon dynamics, perturbation approaches such as reviewed above break down either in the high temperature or strong coupling limit. 
With exception of phonon modes with constant dispersion, in which the Heisenberg inductive equations of motion is numerically exact, only the real-time path integral method is capable to treat the coupling of QDs to a continuum of acoustic phonons exactly for time-dependent excitation scenarios.
This method relies on slicing the time evolution into discrete steps and is similar to recently widely employed time-evolving block decimation or matrix-product state methods \cite{phonon_vidal,vidal_mps,droenner2018two,makri1998quantum,droenner2019stabilizing}.
These models take into account not only the non-Markovian features of the system dynamics but also the growing degree of entanglement between system and reservoir states due to excitation exchange and information backflow.

The real-time path integral method is derived from the Caldeira-Leggett model \cite{caldeira1983path} and has been reviewed widely within the well-known spin-boson problem \cite{RevModPhys.59.1,makri1998quantum,vagov2007nonmonotonic}. 
The application to the QD-phonon kinetics has been successfully employed for exciton \cite{vagov2007nonmonotonic} and biexciton systems \cite{bounouar2015phonon}. 
Interestingly, the real-time path integral method has also successfully been applied to describe phonon-induced dephasing in a coherently coupled QD-microcavity systems in the regime of cavity quantum electrodynamics (cQED) with efficient numerical protocols \cite{axt_hopfmann_prb,cygorek2017nonlinear}.  
Starting point is the general solution of the Liouville-von Neumann equation:
\begin{align}
\rho(t)=
T\left\lbrace
\exp[-i\int_0^t H(t^\prime) dt^\prime] 
\right\rbrace
\rho(0)
\left[
T
\left\lbrace
\exp[i\int_0^t H(t^\prime) dt^\prime] 
\right\rbrace
\right]^\dg,
\end{align}
in contrast to Sec.~\ref{subsec:absorption}, we take light-matter interaction into account: $H(t)=H^{sc}_{e-l}(t)+H_{e-p}^{LA}$.
Choosing a final QD state we are interested in $\ketbra{i_N}{i^\prime_N}$ at time $t=N \Delta t$, and summing over all possible initial system and phonon and final phonon states in the coherent state representation, we yield:
\begin{align}
\bra{i_N}\rho^s(t)\ket{i_N^\prime}
= \iiint &\frac{d^2z_N}{\pi}\frac{d^2z_0}{\pi}\frac{d^2z_0^\prime}{\pi}
\bra{z_N,i_N}U(t,0)\ket{z_0,i_0}\\ \notag
&\bra{z_0,i_0}\rho(0)\ket{z^\prime_0,i^\prime_0}
\bra{z^\prime_0,i^\prime_0}U^\dg(t,0)\ket{z_N,i^\prime_N}
\end{align}
We proceed as before when we derived the independent boson model in 
Eq.~\eqref{eq:ibm_solution} and trace out the system degrees of freedom first. 
We use a stroboscopic evolution in time slices $\Delta t$, 
applying the Suzuki-Trotter decomposition and trivially fulfilling the time-order.
Then, we insert between the time slices (N-1)-times the system identity $\mathbb{1}=\sum_{i_n=1,2} \ketbra{i_n}{i_n}$ and yield following matrix elements:
\begin{align}
\bra{i_n}\exp[-iH(n)]\ket{i_{n-1}}
=
\bra{i_n}\exp[-iH_{e-l}^{sc}(n)]\ket{i_{n-1}}
\exp[-iH_{e-p}^{LA}(n,i_{n-1})],
\end{align}
where we assumed $\Delta t$ to be small that the electron-phonon and light-matter Hamiltonian commute to first order, and we exploit the fact that electron-phonon interaction is diagonal in the system states.
The light-matter interaction reads explicitly (for resonant 
excitation):
\begin{align}
&\bra{i_n}\exp[-iH_{e-l}^{sc}(n)]\ket{i_{n-1}}
= M^{i_n}_{i_{n-1}}
\\ \notag
&=\bra{i_n}
\begin{bmatrix}
\cos[f_n] & -i \sin[f_n] \\
 -i\sin[f_n] & \cos[f_n]
\end{bmatrix}
\ket{i_{n-1}},
\end{align}
reducing the system dynamics to c-values per time slice
and $f_n=\int_{(n-1)\Delta t}^{n\Delta t} \Omega(t) dt$.
Now, the electron-phonon interaction is reduced to a Gaussian
problem and the Feynman-Vernon influence functional can be 
derived.
It is convenient to use the coherent-state representation,
inserting (N-1)-times the phonon base identity, the evolution reads:
\begin{align}
\bra{z_N,i_N}U(t,0)\ket{z_0,i_0}
=\prod_{n=1}^N M_{i_n-1}^{i_n}
\int \mathcal{D}Z T[e^S]
\end{align}
with $S$ as the phonon action and formally as a path integral with corresponding trajectories for Gaussian degrees of freedom.
Taking the extremum, solving for the classical 
equation of motion for the phonon degrees of freedom, and tracing
out the phonon reservoir, one yields the reduced density matrix 
evolution:
\begin{align}
\label{eq:path_integral_evolution}
\bra{i_N}\rho^s(t)\ket{i_N^\prime}
=
\prod_{n=1}^N M_{i_n-1}^{i_n} \left( M_{i^\prime_n-1}^{i^\prime_n}\right)^*
\prod_{m=1}^n
e^{-S_{nm}} 
\bra{i_0}\rho^s(0)\ket{i^\prime_0}
\end{align}
where the impact of the phonon reservoir is included via:
\begin{align}
S_{nm}
=
\int_{{n-1}\Delta t}^{n\Delta t}
dt_1
\int_{{m-1}\Delta t}^{m\Delta t}
dt_2
\left(i_n - i_n^\prime \right)
\left[\phi(t_1-t_2)\ i_m - \phi^*(t_1-t_2) \ i_m^\prime \right]
\end{align}
with the phonon correlation $\phi(t)$ given in Eq.~\eqref{eq:phonon_correlation}.
Note, the solution assumes a real phonon coupling element and a two-level system in which the phonons couple to the excited state only. 
We recover the solution of the independent boson model
when $\Omega(t)\equiv0$ and $i_0 \neq i_0^\prime$, 
\begin{align}
\bra{2}\rho^s_{IBM}(N\Delta t)\ket{1}
=
\prod_{n=1}^N
\prod_{m=1}^n
e^{-S_{nm}} 
\bra{2}\rho^s(0)\ket{1}
\end{align}
leading numerically to solution given in Sec.~\ref{subsec:absorption} in Eq.~\eqref{eq:ibm_solution}.
%

\begin{figure}
\centering
\includegraphics[width=12cm]{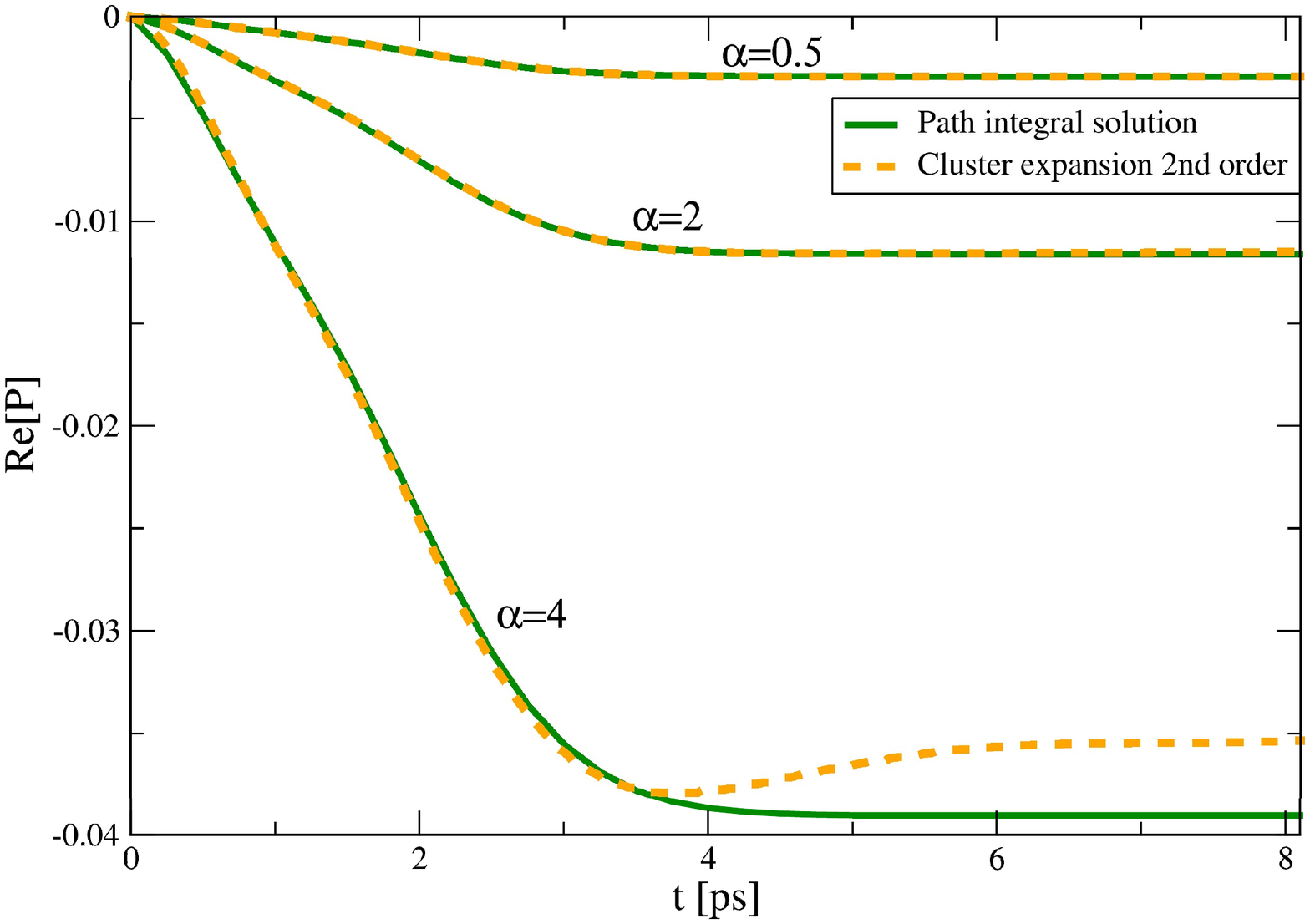}
\caption{ Comparison of the real time path-integral solution with
the cluster expansion time-trace for cw excitation.
Clearly, for weak coupling ($\alpha < 2$), the cluster expansion solution
of second-order (orange) captures the correct steady-state of the
real part of the microscopic coherence $P=\ew{12}$. However, for stronger
coupling and elevated temperatures deviations arise, and only the
path-integral solution (green) predicts the correct experimental
values.}
\label{fig:path_vs_cluster}
\end{figure}

%
In principle, the problem is solved with Eq.~\eqref{eq:path_integral_evolution}.
However, the evaluation is numerically still expensive and can be reduced strongly by taking into account an "on fly" screening \cite{vagov2007nonmonotonic,vagov2006high,vagov2011dynamics} due to the fact that for longitudinal acoustic phonon coupling (or other superohmic environments) the memory falls off rapidly.
Furthermore, boundary conditions need to be imposed to recover known exact solutions \cite{vagov2007nonmonotonic,vagov2006high,vagov2011dynamics}.
This exact method can also be employed to benchmark perturbative approaches.
In Fig.~\ref{fig:path_vs_cluster} and in Ref.\cite{clustervspathintegral}, the cluster expansion solution of the second-order (orange line) is compared with
the numerically exact solution from the real-time path
integral method (green line). As a figure of merit, the real
part of the microscopic coherence is calculated for a QD
exposed to a weak cw driving. Due to the electron-phonon
interaction, the Eigenstate of the system change and a
real part of the coherence emerges in dependence on the
driving and the electron-phonon coupling strength. This
real part saturates fast and captures intrinsical phonon-
induced effects.
%
\begin{figure}
\centering
\includegraphics[width=10cm]{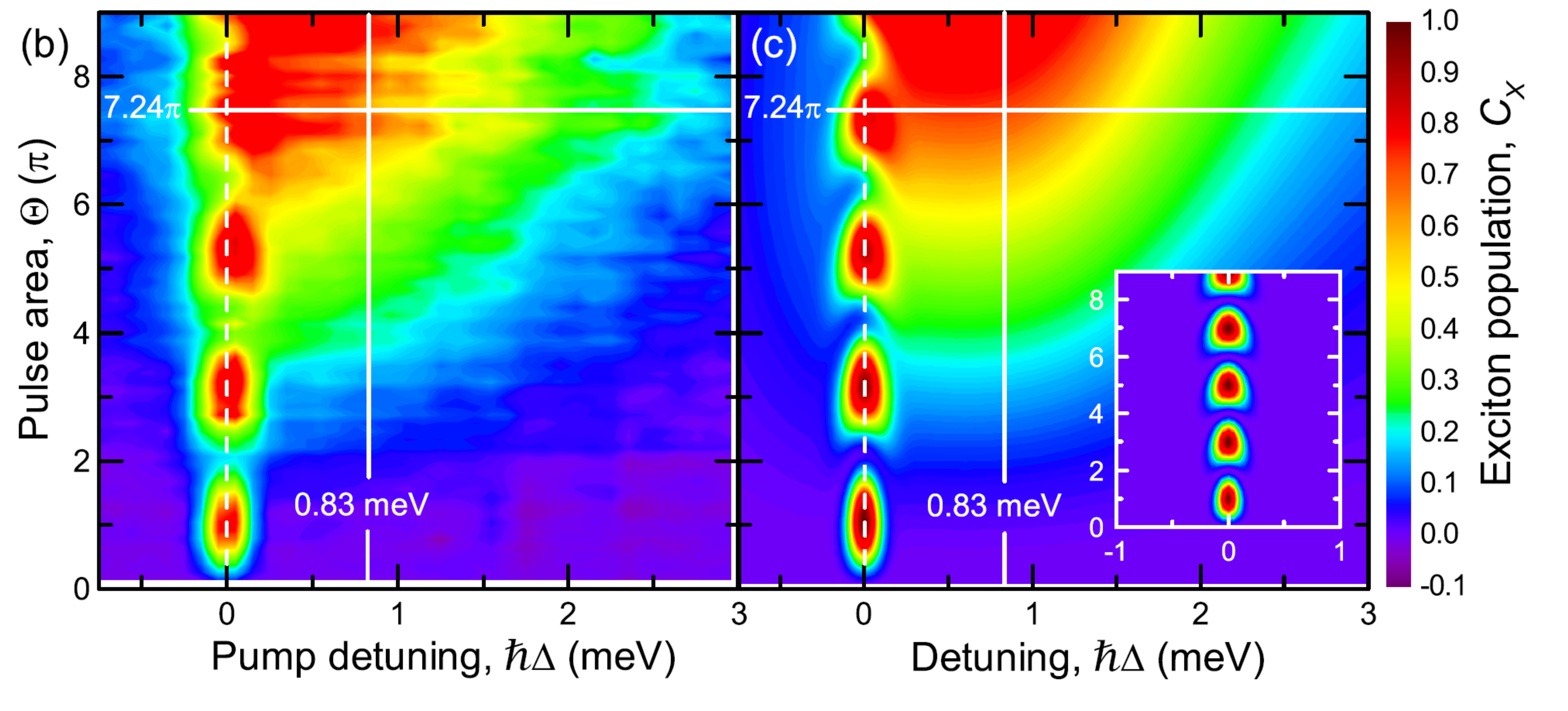}
\caption{Experimentally obtained (left, b) exciton population versus the pulse area and laser detuning. The path-integral solution is given in (c,right) and in very good agreement with the measured signal. The inset (c) shows the calculated values without electron-phonon interaction.
Reprinted figure with permission from \cite{quilter2015phonon}.
\copyright2015
by the American Physical Society.
}
\label{fig:path_integral}
\end{figure}
%

%
Full demonstration of the reliability of the path-integral method has been achieved in population inversion protocols where the laser pulses are tuned within the neutral exciton phonon sideband \cite{quilter2015phonon}. 
This unconventional method achieves the inversion by rapid thermalization of the optically dressed states for which incoherent phonon-assisted relaxation processes are necessary.
In Fig.~\ref{fig:path_integral}, the exciton population $\text{Tr}_B[\bra{2}\rho(t)\ket{2}]$ is plotted (experiment b; theory c). 
The experimental data have been obtained from a layer of InGaAs/GaAs QDs embedded in the intrinsic region of an n-i-Schottky diode structure at $4.2$K, where the measured 
photocurrent is a figure of merit of the final occupation of the exciton state.
We observe a very good overall agreement between experiment and theory and all qualitative features are reproduced and confirmed by microscopic input parameters.
The population inversion arises due to the incoherent phonon-induced relaxation between optically dressed states and the occurring phonon scattering becomes an enabling 
factor in the high driving limit. 
%

\section{Non-equilibrium phonon dynamics in quantized light-matter interaction}
\label{sec:quantized}
\begin{figure}
\includegraphics[width=6.5cm]{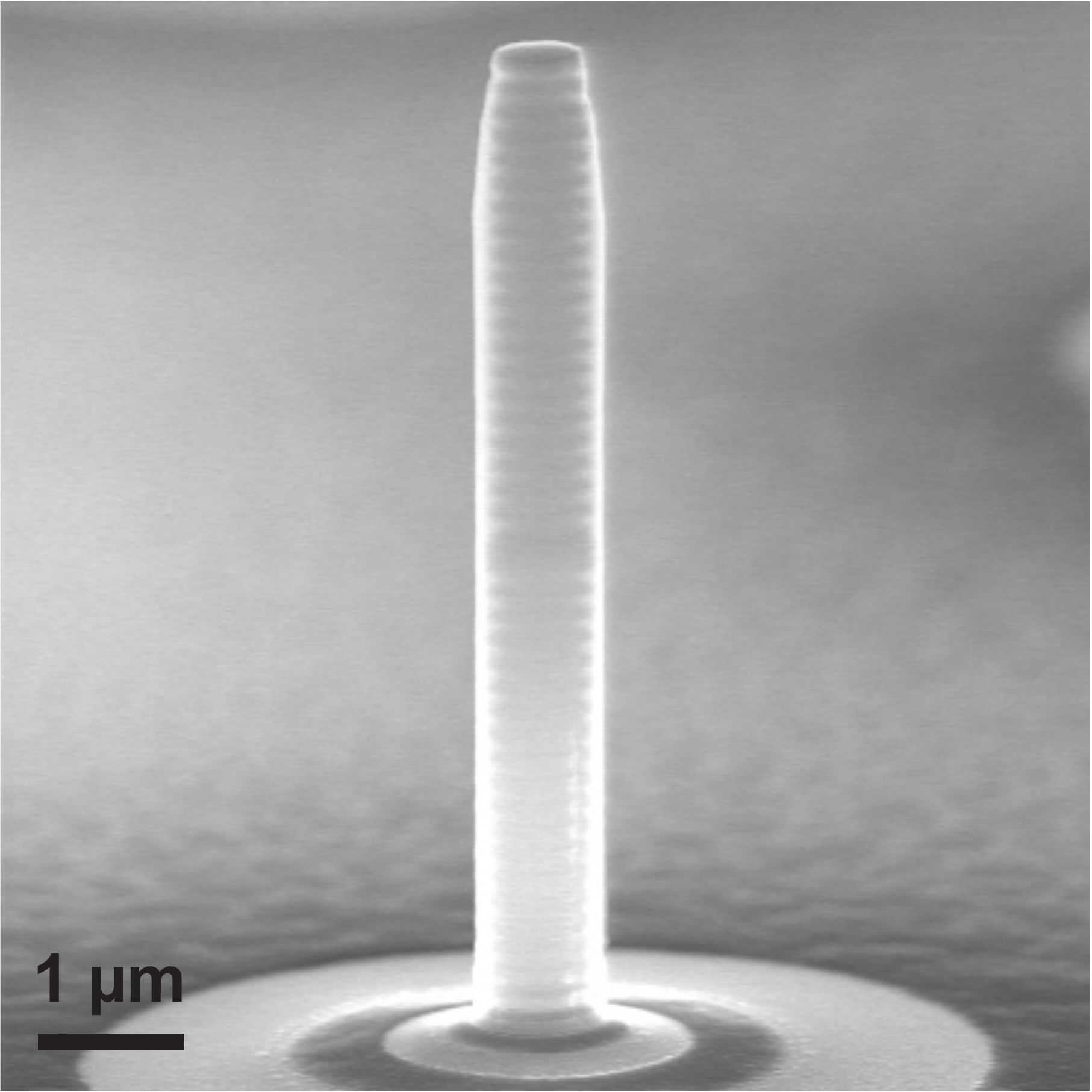} 
\caption{Scanning electron micrograph of a pillar with a diameter of about 0.8 mm. A combination of electron-beam lithography and reactive dry etching allows one to obtain micropillars with close to vertical and defect-free sidewalls.
Figure reprinted from \cite{reithmaier2004strong}.
}
\label{fig:sem_micropillar}
\end{figure}

Quantum optical experiments based on solid-state quantum emitter platforms have made significant advances over the past decade. 
Manifestly “atom-like” properties have been demonstrated for semiconductor QDs and further progress is expected from coupling those semiconductor nanostructures to microcavities as development of in-situ lithography techniques advances rapidly  \cite{dousse,gschrey2013situ}.
Related progress in the fabrication of QD-microcavities pave the way to use the Purcell effect in the optical regime to build efficient and deterministic sources of single photons with tunable photon statistics \cite{sotier2009femtosecond,flagg2009resonantly}.
Exploiting such phenomena QDs coupled to a cavity mode become even more atom-like \cite{reithmaier2004strong,laussy2008strong,kasprzak2010up,kavokin2017microcavities,delvalle_phd,lodahl2015}.
Many well-known features of AOM physics have already been demonstrated on solid-state based platform such as Rabi splitting \cite{reithmaier2004strong} or spectral Jaynes-Cummings ladder signatures \cite{schneebeli2008,kasprzak2010up,hopfmann2017transition}.
In contrast to atoms, solid-state environments are also tailorable and provide the possibility to position nanostructures on demand such as QDs permanently in a high-$\mathcal{Q}$ microcavity \cite{yoshie2004vacuum}, cf.~Fig.~\ref{fig:sem_micropillar}. 
Independent on natural given atomic features, the size and geometry of the QD are controlled properties, e.g., the coupling strength and the confinement energies \cite{Stier1999QDs}. 
Those systems are scalable and perform as material platforms for future technological applications, including single-photon emitters \cite{lounis2005single,michler2000quantum,PhysRevLett.116.020401,somaschi2016near}. 
However compared to atomic systems, quantum emission features in semiconductor nanostructures are ultimately accompanied with decoherence and scattering in the semiconductor environment \cite{iles2017phonon,savona}, as also we have reviewed in Sec.~\ref{sec:semiclassical}(A-F).
%

In this section, we selectively review phonon-induced phenomena in cQED based on QDs as active medium and where the quantized nature of the light field enters in the description of the dynamics.
The following examples are ordered from the single- via the two- to the many-photon regime.
First [Sec.~\ref{subsec:cQED_phonon_decoherence}], we extend the theoretical toolbox by introducing the time-convolutionless method to model the Jaynes-Cummings physics in the single-excitation limit via $H^{cav}_{e-l}$ in Eq.~\eqref{eq:H_cav} and show that acoustic phonons introduce dephasing but also a Rabi frequency renormalization via $H^{LA}_{e-p}$ \cite{mork,axt_hopfmann_prb}.
In this single-photon regime, two examples for the detrimental aspects of the semiconductor environment in QD-based cQED are given, e.g. disadvantageous impact of phonons on the goal to reach the strong coupling regime \cite{mork}. 
Further, and also in the single-photon limit, phonon-mediated cavity feeding is discussed, an interesting effect that strongly deviates from the idealized ``artificial atom''-model typically applied  \cite{hohenester,PhysRevB.83.165313,winger2009explanation,hughes2011influence,carmele2013highly}. 
In the two-photon limit [Sec.~\ref{subsec:HOM}],
we discuss the emission of photon pairs into free space, based on $H^{con}_{e-l}$.
In this quantum optically nonlinear regime, we discuss the effect of colored noise in $H^{st}_{e-p}$ on the visibility of subsequently emitted photons in a Hong-Ou-Mandel setup.
This colored noise can be interpreted as semiclassically treated phonon influence as in Sec.~\ref{subsec:absorption} when the susceptibility has been derived.
We demonstrated that changing the pulse-separation in a two-pulse sequence allows one to monitor environmental correlations and unravels the intrinsic memory kernel \cite{thoma}.
Overall, the semiconductor environment is inevitably involved in excitation processes and renders coherent excitation processes partially incoherent which are nevertheless crucial for a quantitative understanding of experimental results. 
However, as mentioned already above, the same environment which has detrimental effects on the coherence may also in certain scenarios become advantageous and supports quantum optical properties. 
As an example of phonon-assisted coherence increase, we discuss in Sec.~\ref{subsec:cQED_phonon_coherence} the collapse and revival phenomenon known from AMO physics based on the Jaynes-Cummings model Hamiltonian $H^{cav}_{e-l}$.
Here, we employ the inductive equations of motion approach from Sec.~\ref{subsec:inductive_eom_approach} to model a QD strongly coupled to microcavity mode which is initialized in a coherent state.
This coherent state leads to the collapse and revival phenomenon \cite{rempe,eberly1980,scully1999quantum}.
Typically, for low photon numbers the collapse and revival patterns vanish fast into an irregular oscillatory behavior but a QD coupled to an acoustic phonon reservoir exhibits a stabilization via $H^{LA}_{e-p}$ and the collapse and revival phenomenon survives much longer in the presence of phonon-mediated coherence support \cite{carmele_cnr}.  
Therefore, we show before concluding this review, that
phonon non-equilibrium dynamics may even supported and enhanced quantum coherences.

\subsection{Phonon-induced decoherence in QD-cQED}
\label{subsec:cQED_phonon_decoherence}
%
There is a wide range of examples how scattering processes limit the performance of semiconductor nanostructures \cite{michler2017quantum, woggon1997optical,kavokin2017microcavities,jahnke2012quantum}.
To take into account the underlying non-Markovian physics of the phonon environment, simulations must rely either on higher-order perturbative Markovian approaches, time-convolutionless techniques, highly numerically expensive non-equilibrium Green's function models or exact diagonalization.
Here, the influence of the non-Markovian is calculated using a time-convolutionless approach \cite{breuerbook,richter2010time}.
Given the dynamics of the reduced density matrix in second-order of the electron-phonon interaction Eq.~\eqref{eq:second_order_meq}, the system density matrix is approximated in a timelocal form by  setting $ \rho(t-\tau) \approx \rho(t) $. 
Using the full Hamiltonian $H$ including the cavity-QD interaction $H^{cav}_{e-l}$, the dynamics in the single excitation limit yield the following set of equations of motion, if the photon-assisted ground state is abbreviated with $c^\dg\sig{1}{1}^{QD}c=:\sig{1}{1}$.
:
\begin{align}
\dt \ew{\sig{2}{2}} 
=& -\Gamma\ew{\sig{2}{2}}+2\text{Im}\left[g\ew{\sig{1}{2}}\right], \\ 
\dt \ew{\sig{1}{2}} 
=& -[\Gamma/2+\kappa/2+i\Delta-A(t)]\ew{\sig{1}{2}} \\
&-[B(t)+ig]\ew{\sig{1}{1}}+[C(t)+ig]\ew{\sig{2}{2}}, \notag \\
\dt \ew{\sig{1}{1}} 
=& 
-\kappa\ew{\sig{1}{1}}-2 \text{Im}\left[g\ew{\sig{1}{2}}\right] 
\end{align}
where a Markovian radiative decay and a cavity loss is assumed $\kappa\mathcal{D}[c]\rho$ and a detuning between cavity mode and the QD is assumed with $\Delta=\omega_{21}-\omega_c$.
The influence of the acoustic phonons are included in the time-
dependent coefficients $A,B,C$, obtained after inserting a unity in
the time-local master equation:
\begin{align}
A(t) &=
\int_0^t d\tau 
\left[ 
\phi_I(\tau) \bra{1} \sig{2}{2}(-\tau) \ket{1}
-
\phi_I^*(\tau) \bra{2} \sig{2}{2}(-\tau) \ket{2}
\right], \\
B(t) &=
\int_0^t d\tau 
\phi_I^*(\tau) \bra{1} \sig{2}{2}(-\tau) \ket{2}
, \\
C(t) &=
\int_0^t d\tau 
\phi_I(\tau) \bra{1} \sig{2}{2}(-\tau) \ket{2},
\end{align}
with $\bra{i} \sig{2}{2}(-\tau) \ket{j}= 
\bra{i} U(\tau,0)\ket{2} \bra{2}U(-\tau,0)\ket{j}$ the time-evolution matrix element of the unperturbed system with time evolution operator: 
$U(t,0)=\exp[-igt(\sig{1}{2}+\sig{2}{1})-i\Delta t]$, and $\phi_I(t)$ is given in Eq.~\eqref{eq:ibm_phonon_correlation}.
$A(t)$ includes the polaron shift and the dephasing dynamics
via the imaginary and the real part, respectively.
$B(t)$ and $C(t)$ renormalize the interaction strength between
the cavity mode and the QD via the independent boson phonon-correlation function.
This is an important feature of non-Markovian open quantum system dynamics, as a system-reservoir coupling introduces an increase of entanglement and thermalization at the same time, leading to a loss of information and basically new "dressed" system states.
%
\begin{figure}
\centering
\includegraphics[width=10cm]{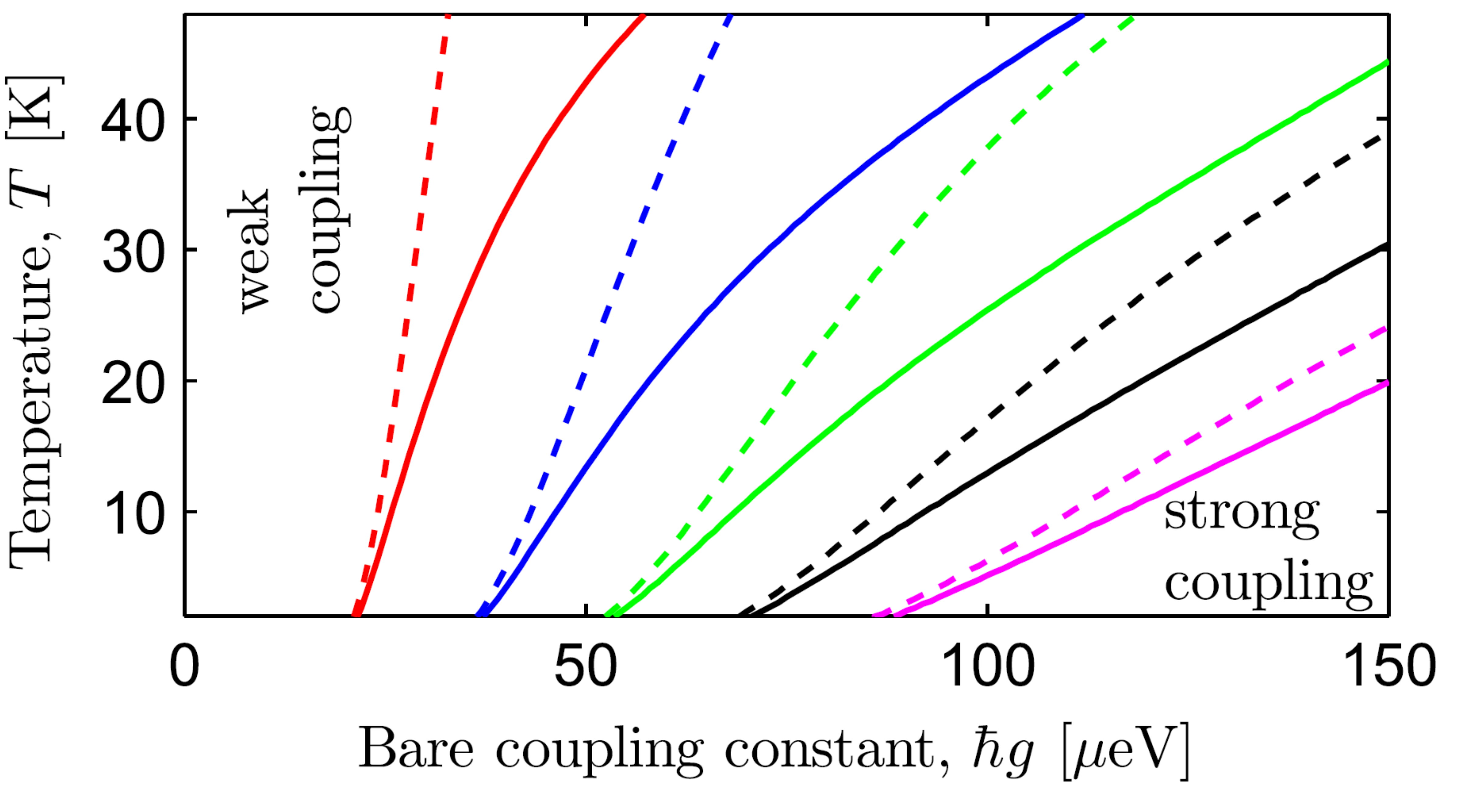}
\caption{Parameter space where strong coupling and weak coupling resides for QD-based cQED depending on the bare coupling constant and the temperature. Non-Markovian phonon interaction lead to a renormalization and affects strongly the transition from weak to strong coupling (solid lines with renormalization, dashed lines without).
Reprinted figure with permission from \cite{mork}.
\copyright2010 by the American Physical
Society.
}
\label{fig:cqed_strong_coupling}
\end{figure}
%
In the Markovian limit $\phi(t) \rightarrow \gamma_p \delta(t)$, we 
recover a Lindblad-type of interaction, in which $B(t)$ and $C(t)$
are vanishing due to $\bra{i} j \rangle=\delta_{ij}$ and
$A(t)$ reduces to $-\gamma_M$ in which all phonon characteristics
are gone and a phenomenological pure dephasing constant remains.
The formulation with finite time-kernel, however, allows one to investigate the phonon impact on the strong coupling regime of the QD-cQED without overestimating the dephasing rate which is the case for a Markovian decay.
A figure of merit for the strong coupling regime, is a non-monotonic decrease of the initial QD occupation $\ew{\sig{2}{2}(0)}=1$ with $\ew{\sig{2}{2}(t_1)}\ge \ew{\sig{2}{2}(t_2)}$ for $t_2>t_1$.
Fig.~\ref{fig:cqed_strong_coupling} shows the parameter space where strong coupling and weak coupling resides depending on the bare coupling constant
$g$ and the temperature.
%
\begin{figure}
\centering
\includegraphics[width=7cm]{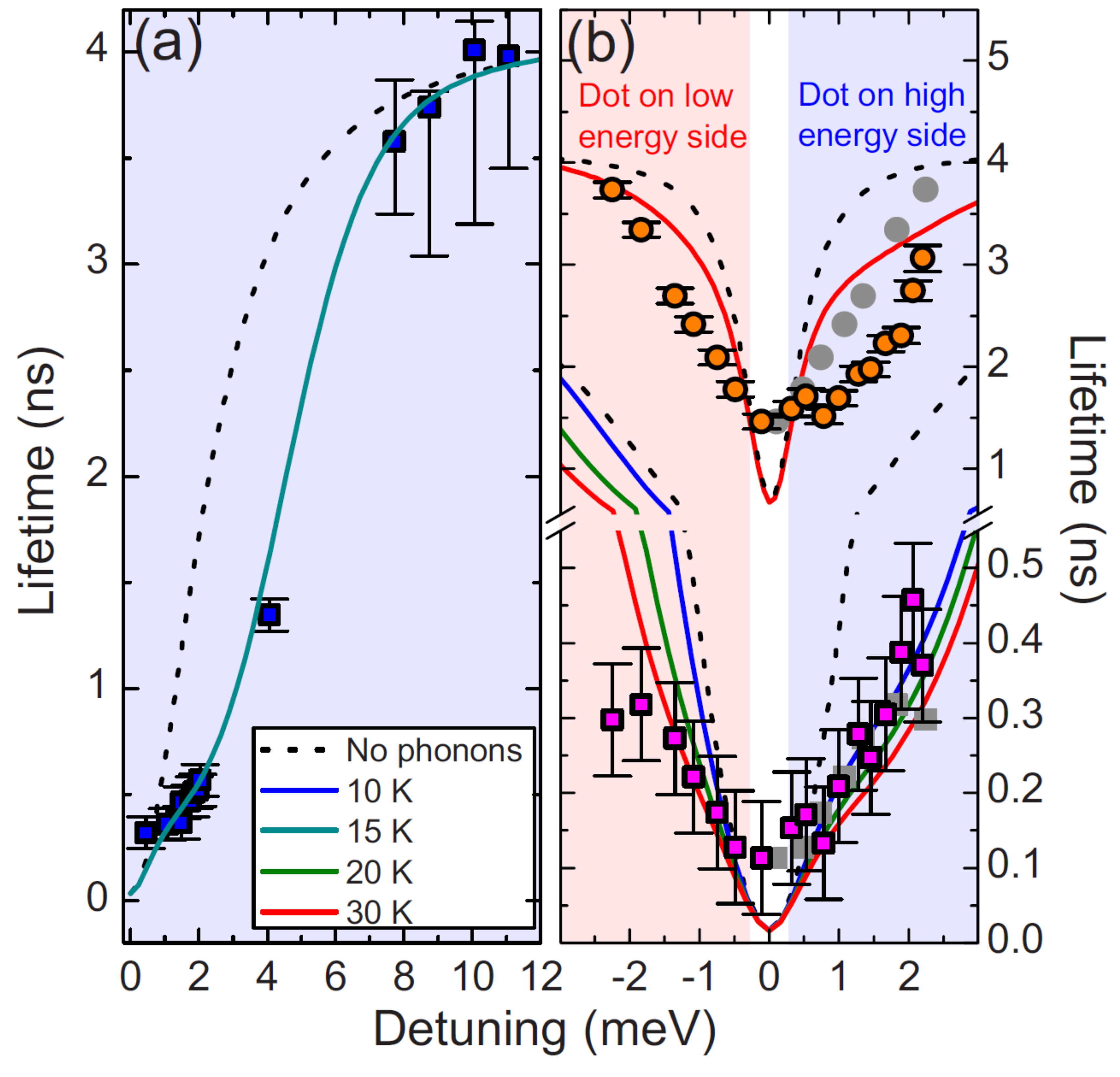}
\caption{Extracted decay times of a QD-cavity system as a function of detuning. Experimental data (dots) are in good agreement with non-Markovian phonon theory (solid line) in comparison without phonons (dashed lines).
In (b), there are two sets of data. Squares refer to the fast and circles to the slow components of the luminescence decay.
Reprinted figure with permission from \cite{hohenester}.
\copyright2009 by the American Physical
Society.
}
\label{fig:cqed_cavity_feeding}
\end{figure}
%
The temperature enters in the phonon correlation function via
$\phi(t)\equiv\phi(t,T)$.
From evaluating the dynamics in the single-excitation regime, it becomes apparent that the higher the temperature the stronger must be the cavity-QD coupling to yield non-monotonous decay dynamics of the QD excitation \cite{madsen2011observation}.
Remarkably, the coupling-renormalization has a strong effect (without renormalization dashed lines, with renormalization solid lines). 
This energy renormalization is also important in explaining asymmetries in life times with respect to the QD-cavity detuning in cavity feeding scenarious.
A pronounced increase in a non-resonant QD-cavity coupling towards elevated temperatures is a strong indication of phonon-mediated relaxation processes \cite{hohenesterFeeding,hughes2011influence,roy2011phonon,PhysRevB.83.165313}. 
The spectral mismatch between the cavity mode and the emitter is then bridged by either phonon emission if the emitter is blue detuned from the cavity mode, or phonon-absorption if the emitter field is red detuned.
This asymmetry is included in the phonon correlation function $\phi(t)$ and not included in the Markovian limit, where $\phi(t)\rightarrow \gamma_M \delta(t)$.
As for low temperatures $ n_q\rightarrow0 $ processes accompanied by phonon emission $ (n_q +1) $ are favored over those obtained by phonon absorption $ (n_q)$, spectral asymmetries become visible and pronounce in this temperature regime.
In Fig.~\ref{fig:cqed_cavity_feeding}, the non-Markovian relaxation dynamics of a detuned QD-cavity QED platform is investigated \cite{hohenester} and given in the single excitation as:
\begin{align}
\dt \ew{\sig{1}{1}}
\approx 2g^2 \text{Re}
\left[ 
\int_0^\infty
d\tau
e^{i\Delta\tau-\phi(\tau)}
\right]
\ew{\sig{2}{2}},
\end{align}
if instead of the phonon-interaction the cavity-QD interaction is taken as a perturbation up to second-order in the reduced but timelocal density matrix, cf.~Eq.~\eqref{eq:second_order_meq}.
This transition strength from excited to ground state can be understood as a generalization of Fermi's golden rule \cite{breuerbook,hohenesterFeeding}.
A model without phonons (dashed line) cannot reproduce the effective life times of the coupled QD-cavity system as a function of the detuning $\Delta$ (experimental data, dots).
The non-Markovian simulations (solid) were computed with a measured quality factor $\mathcal{Q}=2900$ and a cavity-QD coupling strength $\hbar g=45\mu$eV.
Also, the asymmetric dip of the lifetime with respect to the detuning is not reproduced without phonon dynamics, showing that phonon-mediated feeding cannot be captured with a broadened zero-phonon line width.
For higher excitations manifolds, it can furthermore be shown that the cavity feeding effect depends non-trivially on the detunings, e.g. maximal efficiency is obtained for detunings corresponding to transition energies between cavity-dressed states with excitation numbers larger than one \cite{cygorek2017nonlinear}. 
Quantum correlations, such as entanglement and indistinguishability, however still decohere due to phonon-induced noise, as will be discussed next.
%

\subsection{Hong-Ou-Mandel effect: photon indistinguishability}
\label{subsec:HOM}
%
A good figure of merit to characterize the noise robustness of a quantum emitter is the two-photon coherence such as in Hong-Ou-Mandel (HOM) type of experiments \cite{hom}.
If the photon-photon correlation function of two photons subsequently emitted from a single emitter \cite{santori2002indistinguishable} are strongly antibunched after interfering in a Hanbury Brown and Twiss setup, the quantum emitter is time-translational robust and shows therefore no noise.
Such quantum optical experiments probe therefore the indistinguishability of emitted photons and are the basis for entanglement distribution via Bell-state measurements in long-distance quantum communication networks 
\cite{briegel_quantum_repeater,kimble_quantum_internet}. 
Since the visibility on HOM interference experiments is influence by the dephasing of the quantum emitters they are well suited to explore decoherence on the nanoseconds timescale under variation of for instance the temperature with high sensitivity.   
In the theoretical description of HOM experiments we extend the single-mode cavity Hamiltonian in Eq.~\eqref{eq:H_cav} to a multi-mode description $H_{e-l}^{con}$ in Eq.~\eqref{eq:H_con}.
Furthermore, we model the phonon contribution with a stochastic force $F(t)$ via $H_{e-p}^{st}$ in Eq.~\eqref{eq:H_st}.
To probe decoherence on a nanoseconds timescale the QD is excited with a two-pulse sequence with variable pulse separation $\delta t$.
The first pulse creates an excitation via resonant p-shell excitation which radiatively decays under emission of a photon from the s-shell of the QD.
After a time delay $ \delta t $ in which the QD may experience a reconfigured charge environment slightly shifting its energy levels, a second pulse creates another excited state within the QD and another photon is subsequently emitted. 
The corresponding two-photon wave function reads:
\begin{align} \notag
\ket{\Psi} = & E^{(-)}_L(\infty,0)E^{(-)}_S(\infty,\delta t) \ket{\text{vac}} \\
E^{(-)}_n(t,\tau) = &
-i g \
\int_0^\infty d\omega_n 
\int_{\tau}^{t} dt^\prime e^{i(\omega_n-\omega_{21})t^\prime-i\Phi_{\tau}(t^\prime)-\Gamma t^\prime} 
\ c^\dg_{\omega_n} .
\end{align}
We distinguish channels with the labels: $\omega_L$ for long $\omega_S$ for short, i.e. the photons are distinguishable via their spatial traveling mode in the detection path until they superpose at the HOM beam splitter. 
Note the difference in the lower limit of the integrals ($0,\delta t$) and in the integrated noise signals
$\Phi_{\tau}(t) = \int_{\tau}^t dt^\prime F(t^\prime)$, 
and the radiative decay constant is given with $ \Gamma=g^2\pi $. 
If the first photon now takes a longer way to the $50:50$ HOM beam splitter with single-photon detectors $A$ and $B$ at its two output ports and the second photon travels the shorter route, both photons can interfere 
\begin{align}
\sqrt{2} E^{(\pm)}_A &:= E^{(\pm)}_S + E^{(\pm)}_L, \\   \sqrt{2} E^{(\pm)}_B &:= E^{(\pm)}_L - E^{(\pm)}_S   .
\end{align}
In this case the detected photon-photon correlation gives a direct measure for the degree of indistinguishability in terms of the HOM visibility as discussed in the following. 
The unnormalized HOM two-photon correlation function reads:
\begin{align}
G^{(2)}(t_D,t_D+\tau) =& 
\avg{
\left| E^{(+)}_B(t_D+\tau) E^{(+)}_A(t_D)\ket{\Psi(t)} \right|^2  
}
\\ \notag
=& 
\frac{2\Gamma^2}{(2/\pi)^2}
e^{-\Gamma(2t_D+\tau)} 
\avg{1-\text{Re}\left[
e^{-i\xi(t_D,\tau,\delta t)} 
\right]}, \label{eq:corr_two_photon} 
\end{align}
with the noise contribution $ \xi(t_D,\tau,\delta t) = \Phi_0(t_D+\tau)+\Phi_{\delta t}(t_D)-\Phi_{\delta t}(t_D+\tau)-\Phi_0(t_D)$.
The two-photon correlation function vanishes in case of indistinguishable photon emission events, i.e. for vanishing noise $\xi \equiv 0$ or infinite correlated noise: $\Phi_{\delta t}(t)=\Phi_{0}(t)$.
This is the expected result which implies that if the emitter is not subjected to a varying environment or negligible environment influence at all, the emitted photons are indistinguishable as the emission event is time-translational invariant.
However, in the typical experimental setting with finite noise contribution, e.g. due to charge noise by access electron and holes under non-resonant excitation, fluctuating surface charges and/or electron-phonon scattering, decoherence occurs and reduces the HOM visibility.
The figure of merit and experimentally accessible quantity is again the HOM visibility: 
\begin{align}
V(\delta t)=& 
1 
- 
\int_0^\infty\int_0^\infty 
\text{d}\tau \text{d}t 
\frac{G^{(2)}(t,\tau)}{(\pi/2)^2}.
\end{align}
Note, the pulse-delay and noise contribution are still present in the two-photon correlation.
If a delta-correlated white noise is assumed, i.e. in the situation when the noise correlation of different emission events vanishes identically, we obtain:
\begin{align}
V(\delta t,\tau_c)|_\text{Markovian} 
=&  1-\frac{\gamma}{\Gamma+\gamma}
= \frac{\Gamma}{\gamma+\Gamma}
\label{eq:visibility_markovian}.
\end{align}
This is an interesting result which deserves further explanation. 
Firstly we note that here the photon-indistinguishability is independent from the radiative decay constant (as a Markovian radiative decay process has been assumed) but depends strongly on the pure dephasing processes within the emission processes.
However, for a Markovian decoherence process the reduced visibility does not depend on the delay between the two pulses in a two-pulse sequence. 
Thus, a non-Markovian treatment needs to established to fully capture the complexity of the two-pulse experiment
and its underlying physics.
%
\begin{figure}[t!]
\centering
\includegraphics[width=10cm]{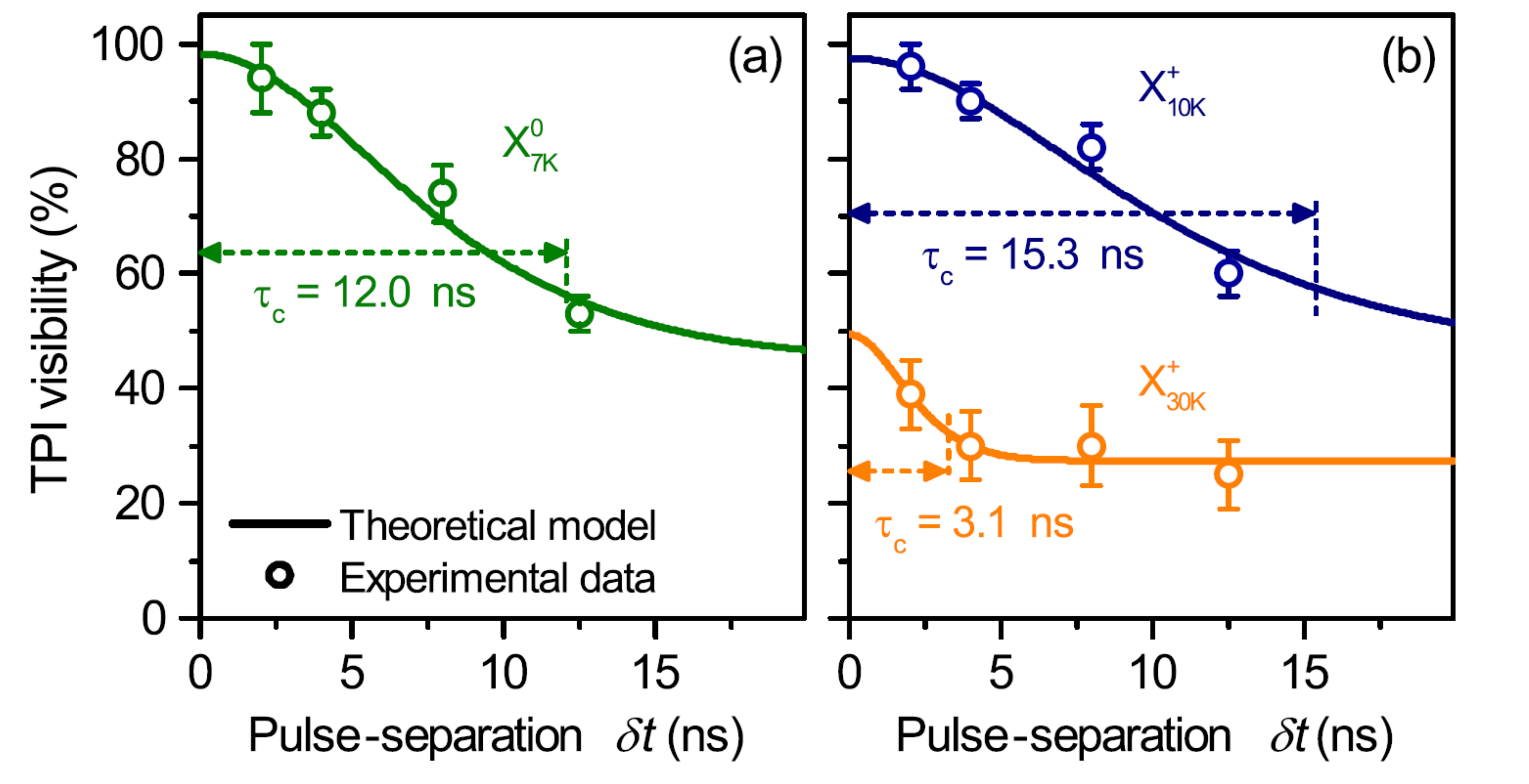}
\caption{Two-photon interference visibilities of consecutively emitted single photons in dependence on the pulse separation $\delta t$. Experimental data for (a) the neutral exciton state and (b) the charged exciton state are quantitatively described by a theoretical model assuming a non-Markovian noise correlation leading to spectral diffusion on a nanosecond timescale.
Reprinted  figure  with  permission  from \cite{thoma}.
\copyright2016  by  the
American Physical Society.}
\label{fig:visibility_thoma}
\end{figure}
%
To illustrate and substantiate this statement we consider experimental results on the HOM visibility of a semiconductor QD presented in Fig.~\ref{fig:visibility_thoma}(a) and (b), where te QD is excited in a two-photon sequence with different pulse delays $ \delta t$.
The HOM visibility clearly decreases with increasing pulse-separation, and a Markovian model cannot be applied to model this result.
In contrast, a colored noise model with a phenomenologically-assumed correlation length of the emission events allows us to characterize the emitter for given visibility data.
For this purpose we include this dependence as a finite memory-effect with specific correlation time of $ \tau_c$: 
\begin{align}
\avg{
\Phi_{t_1}(t_2) \Phi_{t_3}(t_4)
}
= \Gamma^\prime \
e^{-\frac{(t_1-t_3)^2}{\tau_c^2}}
\left( \text{min}[t_2,t_4] - \text{max}[t_1,t_3] \right) .
\end{align}
These kind of noise correlation stems from a non-Markovian low-frequency noise \cite{PhysRevLett.96.097009,PhysRevB.31.490,PhysRevA.30.2381} and shows plateau-like behavior for temporal pulse distances sufficiently short in comparison to the experimentally extracted memory depth.
With the correlation length parameter, we derive for the visibility:
\begin{align}
V(\delta t,\tau_c) =& 
\frac{\Gamma}{\gamma(1-\exp[-(\delta t/\tau_c)^2])+\Gamma}
\label{eq:visibility_non_markovian}.
\end{align}
Therefore, for vanishing pure dephasing the visibility is $1$, i.e. subsequently emitted photons are Fourier transform limited and coalesce at the beamsplitter into a perfect coherent two-photon state.
Further dephasing could also be suppressed by strict-resonant excitation \cite{PhysRevLett.116.020401}, electrical charge control \cite{senellart_review} or suitable surface passivation \cite{PhysRevApplied.9.064019}. 
A detailed understanding of effects limiting the HOM visibility in two-pulse experiments is of vital interest for the further development of state-of-the-art QD-based quantum-light sources towards the implementation of quantum circuits e.g. boson sampling and advanced quantum communication protocols such as measurement device independent quantum key distribution and the quantum repeater. 
This challenging goal can only be achieved by engineering and operating the sources to effectively suppress phonon-induced dephasing and spectral diffusion at timescales well beyond the nanosecond-range in the future.   
%

\subsection{Phonon-enhanced coherence}
\label{subsec:cQED_phonon_coherence}
Lattice vibrations in semiconductor QDs give rise to new effects not known in atomic quantum optics such as phonon-mediated off-resonant cavity feeding [Sec.~\ref{subsec:cQED_phonon_decoherence}], formation of phonon-assisted Mollow triplets~\cite{kabuss} and temperature-dependent vacuum Rabi splittings in cavity emission spectra~\cite{axt_hopfmann_prb}.
Expanding on these developments, another example is phonon-induced quantum optical coherence  \cite{iles2016quantum,carmele_cnr}. 
Here, the memory effects of the phonon bath is investigated with respect on its impact on quantum optical pattern formations in the coherent collapse and revival phenomenon.
This expands the non-Markovian investigation beyond the single excitation limit, as higher-order photon manifolds are involved.
The basis of the investigation is the inductive Heisenberg equation of motion approach \cite{kabuss2011inductive,carmele2010antibunching}, presented in Sec.~\ref{subsec:inductive_eom_approach}.
The cavity-QD dynamics $(H_{e-l}^{cav})$ is calculated within a numerically exact approach while the coupling to the longitudinal acoustic phonon reservoir $(H_{e-p}^{LA})$ is treated for every photon manifold at second-order Born level.
The corresponding equations of motion read for resonant interaction between QD and cavity photons $\Delta=-\Delta_p$ to compensate for the polaron shift:
\begin{align}
\dt  \ew{c^{\dg m} c^m}
=& -2m\kappa\ew{c^{\dg m} c^m}-2m\text{Im}
\left[ g \ew{\sig{1}{2}c^{\dg m} c^{m-1}}\right]
\\
\dt  \ew{\sig{2}{2} c^{\dg m} c^m}
=& -2m\kappa\ew{\sig{2}{2} c^{\dg m} c^m}
+2m\text{Im}
\left[ g \ew{\sig{1}{2}c^{\dg m+1} c^m}\right]
\\ 
\dt  \ew{\sig{1}{2} c^{\dg m+1} c^m}
=& 
-[(2m+1)\kappa-i\Delta]\ew{\sig{1}{2} c^{\dg m+1} c^m}
-ig m\ew{\sig{2}{2} c^{\dg m} c^{m}} \\ \notag
&-ig \left(
2\ew{\sig{2}{2} c^{\dg m+1} c^{m+1}}
-\ew{c^{\dg m+1} c^{m+1}}
\right) \\ \notag
&-i\sum_{\bf q} g^{\bf q}_{12} 
\ew{\sig{1}{2} c^{\dg m+1} c^{m} (b^\dg_{\bf q}+b^\ndg_{\bf q})}.
\end{align}
The most important contribution is proportional to the phonon- and photon-assisted transition dynamics \linebreak
$\ew{\sig{1}{2}c^{\dg m+1} c^{m}(b^\dg_{\bf q}+b^\ndg_{\bf q})}$.
This transition facilitates photon number-dependent dephasing
and blocks the mixing between different photon number
manifolds to stabilize the collapse and revival dynamics.
The equation of motion read, for example:
\begin{align}
\dt  \ew{\sig{1}{2} c^{\dg m+1} c^m b^\dg_{\bf q}}
=& 
-[(2m+1)\kappa-i\omega_q-i\Delta]\ew{\sig{1}{2} c^{\dg m+1} c^m b^\dg_{\bf q}} \\ \notag
&-i g^{\bf q}_{12} \ew{b^\dg_{\bf q} b^\ndg_{\bf q}} \ew{\sig{1}{2} c^{\dg m+1} c^{m}},
\end{align}
where the Born factorization has been applied
to second-order, neglecting contributions proportional to $\propto g g^{\bf q}_{12}$.
This set of equations of motion corresponds to 
Sec.~\ref{subsec:wigner_delay} for higher-order photon
correlations.
%
\begin{figure}
\centering
\includegraphics[width=6cm]{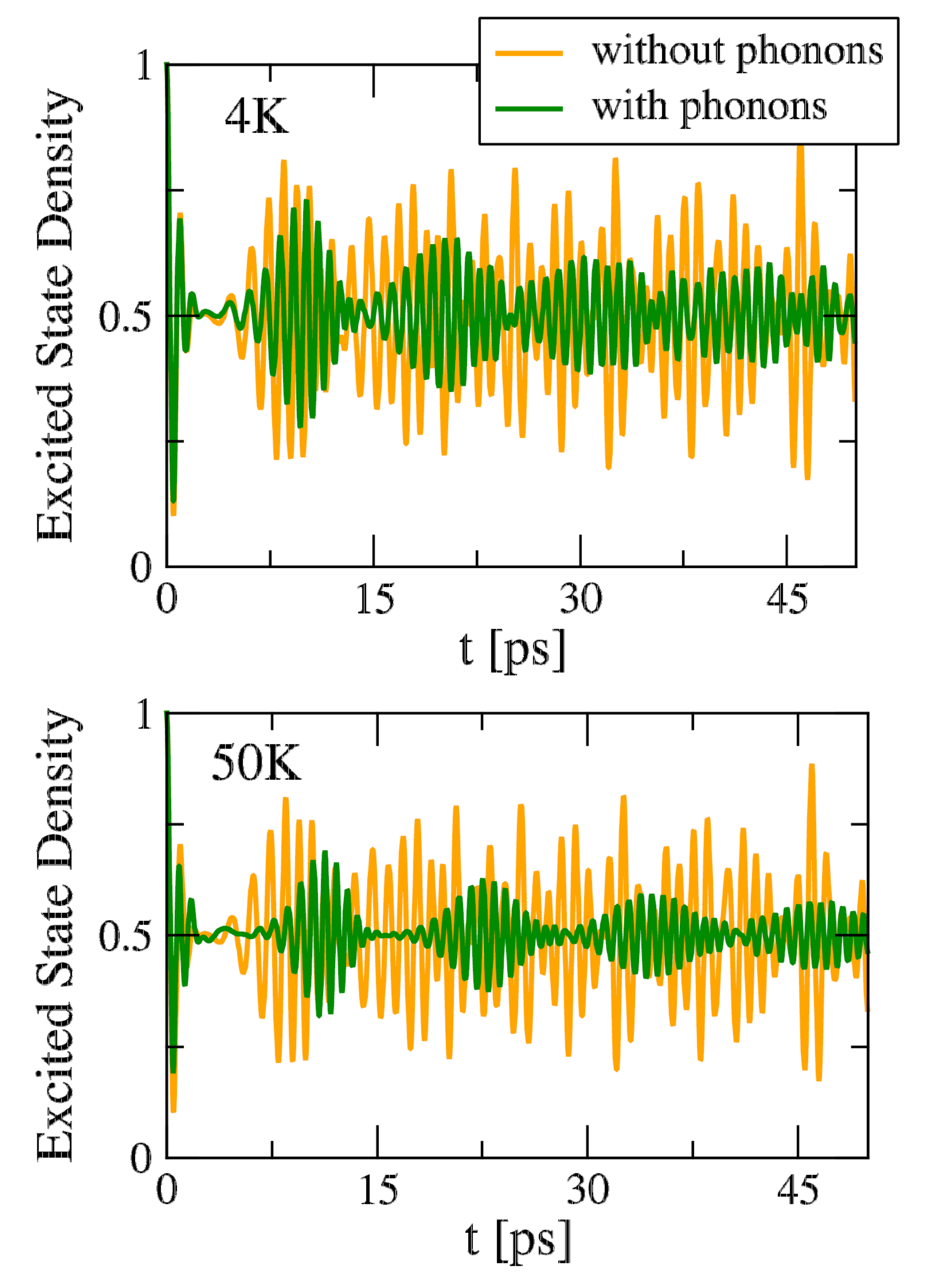}
\caption{The excited state density $\ew{\sig{2}{2}(t)}$ for a cavity initialized with mean coherent photon number of $N=3.5$. The Jaynes-Cummings model solution (gray line) without phonons exhibit only a single collapse with an incomplete revival. Including acoustical phonons with a lattice temperature of $4$K and $50$K many-cycles of collapse and revival are visible}
\label{fig:cnr_phonons}
\end{figure}
%
The set is, however, not closed in terms of the photon number ($m$). 
Therefore, depending on the coupling strength $g$, a sufficient high order in $m$ is taken into account to reach convergence in the calculations. 
Let us focus on a situation that starts with an inverted QD, and a coherent photon state with a mean photon number $N = 3.5$~\cite{rempe}. 
The obtained solution is numerically exact and renders, in the absence of electron-phonon coupling, the well-known Jaynes-Cummings model solutions~\cite{carmele2010antibunching}.  
Due to the small mean photon number, the collapse and revival phenomenon without acoustical phonons known from the Jaynes-Cummings model \cite{eberly1980} is visible only for short times $t<10$ps, cf. Fig.~\ref{fig:cnr_phonons}(orange line in both upper and lower panel).
For longer times $t>15$ps, the pattern of oscillations becomes irregular and the effect of rephasing vanishes.
Excitingly, with phonons for temperatures up to $50$K, the phenomenon is stabilized for much longer times than without phonons (green lines, upper panels for a temperatures of $4$K, lower panel for $50$K).  
The underlying physics is that decoherence manifestly keeps the different photon-assisted Rabi frequencies within the cavity from mixing and allows much longer rephasing dynamics in phase space.
This effect is purely due to the frequency-dependent coupling from the QD to the phonon bath and not included in any Markovian simulations, showing hereby the exciting possibilities provided by non-Markovian, quantum kinetics-induced decoherence. %
Interestingly, it is even stable against cavity loss but strongly depends on the strong coupling limit between cavity and QD, i.e. the cooperativity must be high enough to allow for Rabi oscillations in the microcavity after initialization of the coherent photon field \cite{PhysRevB.86.035319,carmele_cnr}.
The experimental challenge lies, however, in the preparation of the initial state, as a perfect coherent superposition initially is essential, however subjected to losses and transient effect during the preparation stage \cite{rempe}.
These problems have rendered the collapse and revival phenomenon difficult to observe on any platforms.
However, if these difficulties can be overcomed, QDs with acoustical phonon coupling are ideal candidates to observe long and lasting rephasing, collapse and revival dynamics, as standard GaAs parameter have been used.
%

\section{Conclusion}
\label{sec:conclusion}
In this review, we discussed non-Markovian features in QD-based optical experiments.
To understand limits and features of semiconductor nanostructures it is necessary to unravel the microscopic system-environment dynamics.
Electron-phonon interaction is a prominent coupling mechanism and several examples have been given in which the phonon dynamics cannot be reduced to a Lindblad-type of interaction.
Phenomenological modeling overestimates typically the dephasing due to the finite memory kernel of phonons and interesting intrinsical features such as polaron-dressed states and phonon-mediated coherence become
inaccessible.
Several theoretical models, however, have been presented which allow to a high-degree of precision to understand quantitatively and qualitatively the physics in semiconductor quantum optics which in the opinion of the authors have just been started due to advances in experiments and theory to flourish and will continue in doing so.
Of particular interest is the investigation of the electron-phonon dynamics beyond the harmonic approximation.
As non-equilibrium phonon dynamics is already interesting and partially coherence preserving in the harmonic limit, new effects are to be expected beyond lifetime broadening mechanisms if the electron-phonon interaction is treated beyond second-order in its lattice vibrations.
Also, detailed investigation on phononic reservoir engineering is on the verge to be realizable in experiments.
For example, the impact of acoustic cavities on electronic coherences, phonon lasing, and even on surface-reflected acoustic quantum feedback enables the investigation of exotic light-matter interaction.
Due to the high degree of nanotechnological control, collective effects become reachable and superphononance as an acoustic equivalent to superradiance will maybe be observed soon. 
Of fundamental interest is the design of new coupling mechanism and the impact on non-equilibrium steady-states and the question arises whether electron-phonon interaction guarantees ergodicity on a microscopical level or not.
Here, semiconductor quantum optics may shed light due to non-equilibrium electron-phonon dynamics on the very principles of quantum thermodynamics, heat transport, and possible quantum memory platforms. 
The fast progress of nanodesign has just begun to address long-standing questions in many-body physics, and, in the belief of the authors, QDs will play their role in this pursuit of understanding fundamental limits of light-matter interaction due to their universal applicability.

\begin{acknowledgments}
The research leading to the presented results has received funding from the European Research Council under the European Union’s Seventh Framework ERC Grant Agreement No. 615613. We also express gratitude to the DFG (German Science Foundation) for continuous funding of our research via the projects RE2974/5-1, RE2974/9-1, RE2974/12-1, and CRC 787. AC gratefully acknowledges support from the Deutsche Forschungsgemeinschaft (DFG) through the project B1 of the SFB 910 and from the European Unions Horizon 2020 research and innovation program under the SONAR grant Agreement No. 734690 and Andreas Knorr for fruitful discussions.  
\end{acknowledgments}

\bibliographystyle{apsrev4-1}

\begin{thebibliography}{177}%
\makeatletter
\providecommand \@ifxundefined [1]{%
 \@ifx{#1\undefined}
}%
\providecommand \@ifnum [1]{%
 \ifnum #1\expandafter \@firstoftwo
 \else \expandafter \@secondoftwo
 \fi
}%
\providecommand \@ifx [1]{%
 \ifx #1\expandafter \@firstoftwo
 \else \expandafter \@secondoftwo
 \fi
}%
\providecommand \natexlab [1]{#1}%
\providecommand \enquote  [1]{``#1''}%
\providecommand \bibnamefont  [1]{#1}%
\providecommand \bibfnamefont [1]{#1}%
\providecommand \citenamefont [1]{#1}%
\providecommand \href@noop [0]{\@secondoftwo}%
\providecommand \href [0]{\begingroup \@sanitize@url \@href}%
\providecommand \@href[1]{\@@startlink{#1}\@@href}%
\providecommand \@@href[1]{\endgroup#1\@@endlink}%
\providecommand \@sanitize@url [0]{\catcode `\\12\catcode `\$12\catcode
  `\&12\catcode `\#12\catcode `\^12\catcode `\_12\catcode `\%12\relax}%
\providecommand \@@startlink[1]{}%
\providecommand \@@endlink[0]{}%
\providecommand \url  [0]{\begingroup\@sanitize@url \@url }%
\providecommand \@url [1]{\endgroup\@href {#1}{\urlprefix }}%
\providecommand \urlprefix  [0]{URL }%
\providecommand \Eprint [0]{\href }%
\providecommand \doibase [0]{http://dx.doi.org/}%
\providecommand \selectlanguage [0]{\@gobble}%
\providecommand \bibinfo  [0]{\@secondoftwo}%
\providecommand \bibfield  [0]{\@secondoftwo}%
\providecommand \translation [1]{[#1]}%
\providecommand \BibitemOpen [0]{}%
\providecommand \bibitemStop [0]{}%
\providecommand \bibitemNoStop [0]{.\EOS\space}%
\providecommand \EOS [0]{\spacefactor3000\relax}%
\providecommand \BibitemShut  [1]{\csname bibitem#1\endcsname}%
\let\auto@bib@innerbib\@empty
\bibitem [{\citenamefont {Michler}\ \emph {et~al.}(2000)\citenamefont
  {Michler}, \citenamefont {Kiraz}, \citenamefont {Becher}, \citenamefont
  {Schoenfeld}, \citenamefont {Petroff}, \citenamefont {Zhang}, \citenamefont
  {Hu},\ and\ \citenamefont {Imamoglu}}]{michler2000quantum}%
  \BibitemOpen
  \bibfield  {author} {\bibinfo {author} {\bibfnamefont {P.}~\bibnamefont
  {Michler}}, \bibinfo {author} {\bibfnamefont {A.}~\bibnamefont {Kiraz}},
  \bibinfo {author} {\bibfnamefont {C.}~\bibnamefont {Becher}}, \bibinfo
  {author} {\bibfnamefont {W.}~\bibnamefont {Schoenfeld}}, \bibinfo {author}
  {\bibfnamefont {P.}~\bibnamefont {Petroff}}, \bibinfo {author} {\bibfnamefont
  {L.}~\bibnamefont {Zhang}}, \bibinfo {author} {\bibfnamefont
  {E.}~\bibnamefont {Hu}}, \ and\ \bibinfo {author} {\bibfnamefont
  {A.}~\bibnamefont {Imamoglu}},\ }\href@noop {} {\bibfield  {journal}
  {\bibinfo  {journal} {science}\ }\textbf {\bibinfo {volume} {290}},\ \bibinfo
  {pages} {2282} (\bibinfo {year} {2000})}\BibitemShut {NoStop}%
\bibitem [{\citenamefont {Yuan}\ \emph {et~al.}(2002)\citenamefont {Yuan},
  \citenamefont {Kardynal}, \citenamefont {Stevenson}, \citenamefont {Shields},
  \citenamefont {Lobo}, \citenamefont {Cooper}, \citenamefont {Beattie},
  \citenamefont {Ritchie},\ and\ \citenamefont
  {Pepper}}]{yuan2002electrically}%
  \BibitemOpen
  \bibfield  {author} {\bibinfo {author} {\bibfnamefont {Z.}~\bibnamefont
  {Yuan}}, \bibinfo {author} {\bibfnamefont {B.~E.}\ \bibnamefont {Kardynal}},
  \bibinfo {author} {\bibfnamefont {R.~M.}\ \bibnamefont {Stevenson}}, \bibinfo
  {author} {\bibfnamefont {A.~J.}\ \bibnamefont {Shields}}, \bibinfo {author}
  {\bibfnamefont {C.~J.}\ \bibnamefont {Lobo}}, \bibinfo {author}
  {\bibfnamefont {K.}~\bibnamefont {Cooper}}, \bibinfo {author} {\bibfnamefont
  {N.~S.}\ \bibnamefont {Beattie}}, \bibinfo {author} {\bibfnamefont {D.~A.}\
  \bibnamefont {Ritchie}}, \ and\ \bibinfo {author} {\bibfnamefont
  {M.}~\bibnamefont {Pepper}},\ }\href@noop {} {\bibfield  {journal} {\bibinfo
  {journal} {science}\ }\textbf {\bibinfo {volume} {295}},\ \bibinfo {pages}
  {102} (\bibinfo {year} {2002})}\BibitemShut {NoStop}%
\bibitem [{\citenamefont {Akopian}\ \emph {et~al.}(2006)\citenamefont
  {Akopian}, \citenamefont {Lindner}, \citenamefont {Poem}, \citenamefont
  {Berlatzky}, \citenamefont {Avron}, \citenamefont {Gershoni}, \citenamefont
  {Gerardot},\ and\ \citenamefont {Petroff}}]{akopian2006entangled}%
  \BibitemOpen
  \bibfield  {author} {\bibinfo {author} {\bibfnamefont {N.}~\bibnamefont
  {Akopian}}, \bibinfo {author} {\bibfnamefont {N.}~\bibnamefont {Lindner}},
  \bibinfo {author} {\bibfnamefont {E.}~\bibnamefont {Poem}}, \bibinfo {author}
  {\bibfnamefont {Y.}~\bibnamefont {Berlatzky}}, \bibinfo {author}
  {\bibfnamefont {J.}~\bibnamefont {Avron}}, \bibinfo {author} {\bibfnamefont
  {D.}~\bibnamefont {Gershoni}}, \bibinfo {author} {\bibfnamefont
  {B.}~\bibnamefont {Gerardot}}, \ and\ \bibinfo {author} {\bibfnamefont
  {P.}~\bibnamefont {Petroff}},\ }\href@noop {} {\bibfield  {journal} {\bibinfo
   {journal} {Physical Review Letters}\ }\textbf {\bibinfo {volume} {96}},\
  \bibinfo {pages} {130501} (\bibinfo {year} {2006})}\BibitemShut {NoStop}%
\bibitem [{\citenamefont {Hafenbrak}\ \emph {et~al.}(2007)\citenamefont
  {Hafenbrak}, \citenamefont {Ulrich}, \citenamefont {Michler}, \citenamefont
  {Wang}, \citenamefont {Rastelli},\ and\ \citenamefont
  {Schmidt}}]{hafenbrak2007triggered}%
  \BibitemOpen
  \bibfield  {author} {\bibinfo {author} {\bibfnamefont {R.}~\bibnamefont
  {Hafenbrak}}, \bibinfo {author} {\bibfnamefont {S.}~\bibnamefont {Ulrich}},
  \bibinfo {author} {\bibfnamefont {P.}~\bibnamefont {Michler}}, \bibinfo
  {author} {\bibfnamefont {L.}~\bibnamefont {Wang}}, \bibinfo {author}
  {\bibfnamefont {A.}~\bibnamefont {Rastelli}}, \ and\ \bibinfo {author}
  {\bibfnamefont {O.}~\bibnamefont {Schmidt}},\ }\href@noop {} {\bibfield
  {journal} {\bibinfo  {journal} {New Journal of Physics}\ }\textbf {\bibinfo
  {volume} {9}},\ \bibinfo {pages} {315} (\bibinfo {year} {2007})}\BibitemShut
  {NoStop}%
\bibitem [{\citenamefont {Santori}\ \emph {et~al.}(2002)\citenamefont
  {Santori}, \citenamefont {Fattal}, \citenamefont {Vu{\v{c}}kovi{\'c}},
  \citenamefont {Solomon},\ and\ \citenamefont
  {Yamamoto}}]{santori2002indistinguishable}%
  \BibitemOpen
  \bibfield  {author} {\bibinfo {author} {\bibfnamefont {C.}~\bibnamefont
  {Santori}}, \bibinfo {author} {\bibfnamefont {D.}~\bibnamefont {Fattal}},
  \bibinfo {author} {\bibfnamefont {J.}~\bibnamefont {Vu{\v{c}}kovi{\'c}}},
  \bibinfo {author} {\bibfnamefont {G.~S.}\ \bibnamefont {Solomon}}, \ and\
  \bibinfo {author} {\bibfnamefont {Y.}~\bibnamefont {Yamamoto}},\ }\href@noop
  {} {\bibfield  {journal} {\bibinfo  {journal} {Nature}\ }\textbf {\bibinfo
  {volume} {419}},\ \bibinfo {pages} {594} (\bibinfo {year}
  {2002})}\BibitemShut {NoStop}%
\bibitem [{\citenamefont {Reithmaier}\ \emph {et~al.}(2004)\citenamefont
  {Reithmaier}, \citenamefont {Sek}, \citenamefont {L{\"o}ffler}, \citenamefont
  {Hofmann}, \citenamefont {Kuhn}, \citenamefont {Reitzenstein}, \citenamefont
  {Keldysh}, \citenamefont {Kulakovskii}, \citenamefont {Reinecke},\ and\
  \citenamefont {Forchel}}]{reithmaier2004strong}%
  \BibitemOpen
  \bibfield  {author} {\bibinfo {author} {\bibfnamefont {J.~P.}\ \bibnamefont
  {Reithmaier}}, \bibinfo {author} {\bibfnamefont {G.}~\bibnamefont {Sek}},
  \bibinfo {author} {\bibfnamefont {A.}~\bibnamefont {L{\"o}ffler}}, \bibinfo
  {author} {\bibfnamefont {C.}~\bibnamefont {Hofmann}}, \bibinfo {author}
  {\bibfnamefont {S.}~\bibnamefont {Kuhn}}, \bibinfo {author} {\bibfnamefont
  {S.}~\bibnamefont {Reitzenstein}}, \bibinfo {author} {\bibfnamefont
  {L.}~\bibnamefont {Keldysh}}, \bibinfo {author} {\bibfnamefont
  {V.}~\bibnamefont {Kulakovskii}}, \bibinfo {author} {\bibfnamefont
  {T.}~\bibnamefont {Reinecke}}, \ and\ \bibinfo {author} {\bibfnamefont
  {A.}~\bibnamefont {Forchel}},\ }\href@noop {} {\bibfield  {journal} {\bibinfo
   {journal} {Nature}\ }\textbf {\bibinfo {volume} {432}},\ \bibinfo {pages}
  {197} (\bibinfo {year} {2004})}\BibitemShut {NoStop}%
\bibitem [{\citenamefont {Woggon}(1997)}]{woggon1997optical}%
  \BibitemOpen
  \bibfield  {author} {\bibinfo {author} {\bibfnamefont {U.}~\bibnamefont
  {Woggon}},\ }\href@noop {} {\emph {\bibinfo {title} {Optical properties of
  semiconductor quantum dots}}}\ (\bibinfo  {publisher} {Springer},\ \bibinfo
  {year} {1997})\BibitemShut {NoStop}%
\bibitem [{\citenamefont {Bimberg}\ \emph {et~al.}(1999)\citenamefont
  {Bimberg}, \citenamefont {Grundmann},\ and\ \citenamefont
  {Ledentsov}}]{bimberg1999quantum}%
  \BibitemOpen
  \bibfield  {author} {\bibinfo {author} {\bibfnamefont {D.}~\bibnamefont
  {Bimberg}}, \bibinfo {author} {\bibfnamefont {M.}~\bibnamefont {Grundmann}},
  \ and\ \bibinfo {author} {\bibfnamefont {N.~N.}\ \bibnamefont {Ledentsov}},\
  }\href@noop {} {\emph {\bibinfo {title} {Quantum dot heterostructures}}}\
  (\bibinfo  {publisher} {John Wiley \& Sons},\ \bibinfo {year}
  {1999})\BibitemShut {NoStop}%
\bibitem [{\citenamefont {Kiraz}\ \emph {et~al.}(2004)\citenamefont {Kiraz},
  \citenamefont {Atat{\"u}re},\ and\ \citenamefont
  {Imamo{\u{g}}lu}}]{kiraz2004quantum}%
  \BibitemOpen
  \bibfield  {author} {\bibinfo {author} {\bibfnamefont {A.}~\bibnamefont
  {Kiraz}}, \bibinfo {author} {\bibfnamefont {M.}~\bibnamefont {Atat{\"u}re}},
  \ and\ \bibinfo {author} {\bibfnamefont {A.}~\bibnamefont {Imamo{\u{g}}lu}},\
  }\href@noop {} {\bibfield  {journal} {\bibinfo  {journal} {Physical Review
  A}\ }\textbf {\bibinfo {volume} {69}},\ \bibinfo {pages} {032305} (\bibinfo
  {year} {2004})}\BibitemShut {NoStop}%
\bibitem [{\citenamefont {Michler}(2009)}]{michler2009single}%
  \BibitemOpen
  \bibfield  {author} {\bibinfo {author} {\bibfnamefont {P.}~\bibnamefont
  {Michler}},\ }\href@noop {} {\emph {\bibinfo {title} {Single semiconductor
  quantum dots}}},\ Vol.\ \bibinfo {volume} {231}\ (\bibinfo  {publisher}
  {Springer},\ \bibinfo {year} {2009})\BibitemShut {NoStop}%
\bibitem [{\citenamefont {Michler}(2017)}]{michler2017quantum}%
  \BibitemOpen
  \bibfield  {author} {\bibinfo {author} {\bibfnamefont {P.}~\bibnamefont
  {Michler}},\ }\href@noop {} {\emph {\bibinfo {title} {Quantum Dots for
  Quantum Information Technologies}}},\ Vol.\ \bibinfo {volume} {237}\
  (\bibinfo  {publisher} {Springer},\ \bibinfo {year} {2017})\BibitemShut
  {NoStop}%
\bibitem [{\citenamefont {Jahnke}(2012)}]{jahnke2012quantum}%
  \BibitemOpen
  \bibfield  {author} {\bibinfo {author} {\bibfnamefont {F.}~\bibnamefont
  {Jahnke}},\ }\href@noop {} {\emph {\bibinfo {title} {Quantum optics with
  semiconductor nanostructures}}}\ (\bibinfo  {publisher} {Elsevier},\ \bibinfo
  {year} {2012})\BibitemShut {NoStop}%
\bibitem [{\citenamefont {Shields}(2010)}]{shields2010semiconductor}%
  \BibitemOpen
  \bibfield  {author} {\bibinfo {author} {\bibfnamefont {A.~J.}\ \bibnamefont
  {Shields}},\ }in\ \href@noop {} {\emph {\bibinfo {booktitle} {Nanoscience And
  Technology: A Collection of Reviews from Nature Journals}}}\ (\bibinfo
  {publisher} {World Scientific},\ \bibinfo {year} {2010})\ pp.\ \bibinfo
  {pages} {221--229}\BibitemShut {NoStop}%
\bibitem [{\citenamefont {Yu-Ming}\ \emph {et~al.}(2013)\citenamefont
  {Yu-Ming}, \citenamefont {Yu}, \citenamefont {Yu-Jia}, \citenamefont {Dian},
  \citenamefont {Mete}, \citenamefont {Christian}, \citenamefont {Sven},
  \citenamefont {Martin}, \citenamefont {Chao-Yang},\ and\ \citenamefont
  {Jian-Wei}}]{pan}%
  \BibitemOpen
  \bibfield  {author} {\bibinfo {author} {\bibfnamefont {H.}~\bibnamefont
  {Yu-Ming}}, \bibinfo {author} {\bibfnamefont {H.}~\bibnamefont {Yu}},
  \bibinfo {author} {\bibfnamefont {W.}~\bibnamefont {Yu-Jia}}, \bibinfo
  {author} {\bibfnamefont {W.}~\bibnamefont {Dian}}, \bibinfo {author}
  {\bibfnamefont {A.}~\bibnamefont {Mete}}, \bibinfo {author} {\bibfnamefont
  {S.}~\bibnamefont {Christian}}, \bibinfo {author} {\bibfnamefont
  {H.}~\bibnamefont {Sven}}, \bibinfo {author} {\bibfnamefont {K.}~\bibnamefont
  {Martin}}, \bibinfo {author} {\bibfnamefont {L.}~\bibnamefont {Chao-Yang}}, \
  and\ \bibinfo {author} {\bibfnamefont {P.}~\bibnamefont {Jian-Wei}},\
  }\href@noop {} {\bibfield  {journal} {\bibinfo  {journal} {Nature
  Nanotechnology}\ }\textbf {\bibinfo {volume} {8}},\ \bibinfo {pages} {213}
  (\bibinfo {year} {2013})}\BibitemShut {NoStop}%
\bibitem [{\citenamefont {Peter}\ \emph {et~al.}(2004)\citenamefont {Peter},
  \citenamefont {Hours}, \citenamefont {Senellart}, \citenamefont {Vasanelli},
  \citenamefont {Cavanna}, \citenamefont {Bloch},\ and\ \citenamefont
  {G\'erard}}]{senellart}%
  \BibitemOpen
  \bibfield  {author} {\bibinfo {author} {\bibfnamefont {E.}~\bibnamefont
  {Peter}}, \bibinfo {author} {\bibfnamefont {J.}~\bibnamefont {Hours}},
  \bibinfo {author} {\bibfnamefont {P.}~\bibnamefont {Senellart}}, \bibinfo
  {author} {\bibfnamefont {A.}~\bibnamefont {Vasanelli}}, \bibinfo {author}
  {\bibfnamefont {A.}~\bibnamefont {Cavanna}}, \bibinfo {author} {\bibfnamefont
  {J.}~\bibnamefont {Bloch}}, \ and\ \bibinfo {author} {\bibfnamefont {J.~M.}\
  \bibnamefont {G\'erard}},\ }\href {\doibase 10.1103/PhysRevB.69.041307}
  {\bibfield  {journal} {\bibinfo  {journal} {Phys. Rev. B}\ }\textbf {\bibinfo
  {volume} {69}},\ \bibinfo {pages} {041307} (\bibinfo {year}
  {2004})}\BibitemShut {NoStop}%
\bibitem [{\citenamefont {Mendach}\ \emph {et~al.}(2006)\citenamefont
  {Mendach}, \citenamefont {Songmuang}, \citenamefont {Kiravittaya},
  \citenamefont {Rastelli}, \citenamefont {Benyoucef},\ and\ \citenamefont
  {Schmidt}}]{rastelli}%
  \BibitemOpen
  \bibfield  {author} {\bibinfo {author} {\bibfnamefont {S.}~\bibnamefont
  {Mendach}}, \bibinfo {author} {\bibfnamefont {R.}~\bibnamefont {Songmuang}},
  \bibinfo {author} {\bibfnamefont {S.}~\bibnamefont {Kiravittaya}}, \bibinfo
  {author} {\bibfnamefont {A.}~\bibnamefont {Rastelli}}, \bibinfo {author}
  {\bibfnamefont {M.}~\bibnamefont {Benyoucef}}, \ and\ \bibinfo {author}
  {\bibfnamefont {O.~G.}\ \bibnamefont {Schmidt}},\ }\href {\doibase
  10.1063/1.2186509} {\bibfield  {journal} {\bibinfo  {journal} {Applied
  Physics Letters}\ }\textbf {\bibinfo {volume} {88}},\ \bibinfo {pages}
  {111120} (\bibinfo {year} {2006})}\BibitemShut {NoStop}%
\bibitem [{\citenamefont {Schmidt}(2007)}]{schmidt2007lateral}%
  \BibitemOpen
  \bibfield  {author} {\bibinfo {author} {\bibfnamefont {O.~G.}\ \bibnamefont
  {Schmidt}},\ }\href@noop {} {\emph {\bibinfo {title} {Lateral alignment of
  epitaxial quantum dots}}}\ (\bibinfo  {publisher} {Springer Science \&
  Business Media},\ \bibinfo {year} {2007})\BibitemShut {NoStop}%
\bibitem [{\citenamefont {Zhang}\ \emph {et~al.}(2015)\citenamefont {Zhang},
  \citenamefont {Wildmann}, \citenamefont {Ding}, \citenamefont {Trotta},
  \citenamefont {Huo}, \citenamefont {Zallo}, \citenamefont {Huber},
  \citenamefont {Rastelli},\ and\ \citenamefont {Schmidt}}]{ding}%
  \BibitemOpen
  \bibfield  {author} {\bibinfo {author} {\bibfnamefont {J.}~\bibnamefont
  {Zhang}}, \bibinfo {author} {\bibfnamefont {J.~S.}\ \bibnamefont {Wildmann}},
  \bibinfo {author} {\bibfnamefont {F.}~\bibnamefont {Ding}}, \bibinfo {author}
  {\bibfnamefont {R.}~\bibnamefont {Trotta}}, \bibinfo {author} {\bibfnamefont
  {Y.}~\bibnamefont {Huo}}, \bibinfo {author} {\bibfnamefont {E.}~\bibnamefont
  {Zallo}}, \bibinfo {author} {\bibfnamefont {D.}~\bibnamefont {Huber}},
  \bibinfo {author} {\bibfnamefont {A.}~\bibnamefont {Rastelli}}, \ and\
  \bibinfo {author} {\bibfnamefont {O.~G.}\ \bibnamefont {Schmidt}},\
  }\href@noop {} {\bibfield  {journal} {\bibinfo  {journal} {Nature
  communications}\ }\textbf {\bibinfo {volume} {6}},\ \bibinfo {pages} {10067}
  (\bibinfo {year} {2015})}\BibitemShut {NoStop}%
\bibitem [{\citenamefont {Schlehahn}\ \emph {et~al.}(2018)\citenamefont
  {Schlehahn}, \citenamefont {Fischbach}, \citenamefont {Schmidt},
  \citenamefont {Kaganskiy}, \citenamefont {Strittmatter}, \citenamefont
  {Rodt}, \citenamefont {Heindel},\ and\ \citenamefont
  {Reitzenstein}}]{schlehahn2018}%
  \BibitemOpen
  \bibfield  {author} {\bibinfo {author} {\bibfnamefont {A.}~\bibnamefont
  {Schlehahn}}, \bibinfo {author} {\bibfnamefont {S.}~\bibnamefont
  {Fischbach}}, \bibinfo {author} {\bibfnamefont {R.}~\bibnamefont {Schmidt}},
  \bibinfo {author} {\bibfnamefont {A.}~\bibnamefont {Kaganskiy}}, \bibinfo
  {author} {\bibfnamefont {A.}~\bibnamefont {Strittmatter}}, \bibinfo {author}
  {\bibfnamefont {S.}~\bibnamefont {Rodt}}, \bibinfo {author} {\bibfnamefont
  {T.}~\bibnamefont {Heindel}}, \ and\ \bibinfo {author} {\bibfnamefont
  {S.}~\bibnamefont {Reitzenstein}},\ }\href@noop {} {\bibfield  {journal}
  {\bibinfo  {journal} {Scientific reports}\ }\textbf {\bibinfo {volume} {8}},\
  \bibinfo {pages} {1340} (\bibinfo {year} {2018})}\BibitemShut {NoStop}%
\bibitem [{\citenamefont {Yoshie}\ \emph {et~al.}(2004)\citenamefont {Yoshie},
  \citenamefont {Scherer}, \citenamefont {Hendrickson}, \citenamefont
  {Khitrova}, \citenamefont {Gibbs}, \citenamefont {Rupper}, \citenamefont
  {Ell}, \citenamefont {Shchekin},\ and\ \citenamefont
  {Deppe}}]{yoshie2004vacuum}%
  \BibitemOpen
  \bibfield  {author} {\bibinfo {author} {\bibfnamefont {T.}~\bibnamefont
  {Yoshie}}, \bibinfo {author} {\bibfnamefont {A.}~\bibnamefont {Scherer}},
  \bibinfo {author} {\bibfnamefont {J.}~\bibnamefont {Hendrickson}}, \bibinfo
  {author} {\bibfnamefont {G.}~\bibnamefont {Khitrova}}, \bibinfo {author}
  {\bibfnamefont {H.}~\bibnamefont {Gibbs}}, \bibinfo {author} {\bibfnamefont
  {G.}~\bibnamefont {Rupper}}, \bibinfo {author} {\bibfnamefont
  {C.}~\bibnamefont {Ell}}, \bibinfo {author} {\bibfnamefont {O.}~\bibnamefont
  {Shchekin}}, \ and\ \bibinfo {author} {\bibfnamefont {D.}~\bibnamefont
  {Deppe}},\ }\href@noop {} {\bibfield  {journal} {\bibinfo  {journal}
  {Nature}\ }\textbf {\bibinfo {volume} {432}},\ \bibinfo {pages} {200}
  (\bibinfo {year} {2004})}\BibitemShut {NoStop}%
\bibitem [{\citenamefont {Bose}\ \emph {et~al.}(2014)\citenamefont {Bose},
  \citenamefont {Cai}, \citenamefont {Choudhury}, \citenamefont {Solomon},\
  and\ \citenamefont {Waks}}]{bose2014all}%
  \BibitemOpen
  \bibfield  {author} {\bibinfo {author} {\bibfnamefont {R.}~\bibnamefont
  {Bose}}, \bibinfo {author} {\bibfnamefont {T.}~\bibnamefont {Cai}}, \bibinfo
  {author} {\bibfnamefont {K.~R.}\ \bibnamefont {Choudhury}}, \bibinfo {author}
  {\bibfnamefont {G.~S.}\ \bibnamefont {Solomon}}, \ and\ \bibinfo {author}
  {\bibfnamefont {E.}~\bibnamefont {Waks}},\ }\href@noop {} {\bibfield
  {journal} {\bibinfo  {journal} {Nature Photonics}\ }\textbf {\bibinfo
  {volume} {8}},\ \bibinfo {pages} {858} (\bibinfo {year} {2014})}\BibitemShut
  {NoStop}%
\bibitem [{\citenamefont {Senellart}\ \emph {et~al.}(2017)\citenamefont
  {Senellart}, \citenamefont {Solomon},\ and\ \citenamefont
  {White}}]{senellart_review}%
  \BibitemOpen
  \bibfield  {author} {\bibinfo {author} {\bibfnamefont {P.}~\bibnamefont
  {Senellart}}, \bibinfo {author} {\bibfnamefont {G.}~\bibnamefont {Solomon}},
  \ and\ \bibinfo {author} {\bibfnamefont {A.}~\bibnamefont {White}},\
  }\href@noop {} {\bibfield  {journal} {\bibinfo  {journal} {Nature
  nanotechnology}\ }\textbf {\bibinfo {volume} {12}},\ \bibinfo {pages} {1026}
  (\bibinfo {year} {2017})}\BibitemShut {NoStop}%
\bibitem [{\citenamefont {F{\"o}rstner}\ \emph {et~al.}(2003)\citenamefont
  {F{\"o}rstner}, \citenamefont {Weber}, \citenamefont {Danckwerts},\ and\
  \citenamefont {Knorr}}]{forstner2003phonon}%
  \BibitemOpen
  \bibfield  {author} {\bibinfo {author} {\bibfnamefont {J.}~\bibnamefont
  {F{\"o}rstner}}, \bibinfo {author} {\bibfnamefont {C.}~\bibnamefont {Weber}},
  \bibinfo {author} {\bibfnamefont {J.}~\bibnamefont {Danckwerts}}, \ and\
  \bibinfo {author} {\bibfnamefont {A.}~\bibnamefont {Knorr}},\ }\href@noop {}
  {\bibfield  {journal} {\bibinfo  {journal} {physica status solidi (b)}\
  }\textbf {\bibinfo {volume} {238}},\ \bibinfo {pages} {419} (\bibinfo {year}
  {2003})}\BibitemShut {NoStop}%
\bibitem [{\citenamefont {Krummheuer}\ \emph {et~al.}(2005)\citenamefont
  {Krummheuer}, \citenamefont {Axt}, \citenamefont {Kuhn}, \citenamefont
  {D'Amico},\ and\ \citenamefont {Rossi}}]{krummheuer2005}%
  \BibitemOpen
  \bibfield  {author} {\bibinfo {author} {\bibfnamefont {B.}~\bibnamefont
  {Krummheuer}}, \bibinfo {author} {\bibfnamefont {V.~M.}\ \bibnamefont {Axt}},
  \bibinfo {author} {\bibfnamefont {T.}~\bibnamefont {Kuhn}}, \bibinfo {author}
  {\bibfnamefont {I.}~\bibnamefont {D'Amico}}, \ and\ \bibinfo {author}
  {\bibfnamefont {F.}~\bibnamefont {Rossi}},\ }\href {\doibase
  10.1103/PhysRevB.71.235329} {\bibfield  {journal} {\bibinfo  {journal} {Phys.
  Rev. B}\ }\textbf {\bibinfo {volume} {71}},\ \bibinfo {pages} {235329}
  (\bibinfo {year} {2005})}\BibitemShut {NoStop}%
\bibitem [{\citenamefont {Quilter}\ \emph {et~al.}(2015)\citenamefont
  {Quilter}, \citenamefont {Brash}, \citenamefont {Liu}, \citenamefont
  {Gl\"assl}, \citenamefont {Barth}, \citenamefont {Axt}, \citenamefont
  {Ramsay}, \citenamefont {Skolnick},\ and\ \citenamefont
  {Fox}}]{quilter2015phonon}%
  \BibitemOpen
  \bibfield  {author} {\bibinfo {author} {\bibfnamefont {J.~H.}\ \bibnamefont
  {Quilter}}, \bibinfo {author} {\bibfnamefont {A.~J.}\ \bibnamefont {Brash}},
  \bibinfo {author} {\bibfnamefont {F.}~\bibnamefont {Liu}}, \bibinfo {author}
  {\bibfnamefont {M.}~\bibnamefont {Gl\"assl}}, \bibinfo {author}
  {\bibfnamefont {A.~M.}\ \bibnamefont {Barth}}, \bibinfo {author}
  {\bibfnamefont {V.~M.}\ \bibnamefont {Axt}}, \bibinfo {author} {\bibfnamefont
  {A.~J.}\ \bibnamefont {Ramsay}}, \bibinfo {author} {\bibfnamefont {M.~S.}\
  \bibnamefont {Skolnick}}, \ and\ \bibinfo {author} {\bibfnamefont {A.~M.}\
  \bibnamefont {Fox}},\ }\href {\doibase 10.1103/PhysRevLett.114.137401}
  {\bibfield  {journal} {\bibinfo  {journal} {Phys. Rev. Lett.}\ }\textbf
  {\bibinfo {volume} {114}},\ \bibinfo {pages} {137401} (\bibinfo {year}
  {2015})}\BibitemShut {NoStop}%
\bibitem [{\citenamefont {M{\"u}ller}\ \emph {et~al.}(2014)\citenamefont
  {M{\"u}ller}, \citenamefont {Bounouar}, \citenamefont {J{\"o}ns},
  \citenamefont {Gl{\"a}ssl},\ and\ \citenamefont
  {Michler}}]{state_preparation_michler}%
  \BibitemOpen
  \bibfield  {author} {\bibinfo {author} {\bibfnamefont {M.}~\bibnamefont
  {M{\"u}ller}}, \bibinfo {author} {\bibfnamefont {S.}~\bibnamefont
  {Bounouar}}, \bibinfo {author} {\bibfnamefont {K.~D.}\ \bibnamefont
  {J{\"o}ns}}, \bibinfo {author} {\bibfnamefont {M.}~\bibnamefont
  {Gl{\"a}ssl}}, \ and\ \bibinfo {author} {\bibfnamefont {P.}~\bibnamefont
  {Michler}},\ }\href@noop {} {\bibfield  {journal} {\bibinfo  {journal}
  {Nature Photonics}\ }\textbf {\bibinfo {volume} {8}},\ \bibinfo {pages} {224}
  (\bibinfo {year} {2014})}\BibitemShut {NoStop}%
\bibitem [{\citenamefont {Reiter}\ \emph {et~al.}(2014)\citenamefont {Reiter},
  \citenamefont {Kuhn}, \citenamefont {Gl{\"a}ssl},\ and\ \citenamefont
  {Axt}}]{reiter2014role}%
  \BibitemOpen
  \bibfield  {author} {\bibinfo {author} {\bibfnamefont {D.}~\bibnamefont
  {Reiter}}, \bibinfo {author} {\bibfnamefont {T.}~\bibnamefont {Kuhn}},
  \bibinfo {author} {\bibfnamefont {M.}~\bibnamefont {Gl{\"a}ssl}}, \ and\
  \bibinfo {author} {\bibfnamefont {V.}~\bibnamefont {Axt}},\ }\href@noop {}
  {\bibfield  {journal} {\bibinfo  {journal} {Journal of Physics: Condensed
  Matter}\ }\textbf {\bibinfo {volume} {26}},\ \bibinfo {pages} {423203}
  (\bibinfo {year} {2014})}\BibitemShut {NoStop}%
\bibitem [{\citenamefont {Carmele}\ \emph {et~al.}(2010)\citenamefont
  {Carmele}, \citenamefont {Richter}, \citenamefont {Chow},\ and\ \citenamefont
  {Knorr}}]{carmele2010antibunching}%
  \BibitemOpen
  \bibfield  {author} {\bibinfo {author} {\bibfnamefont {A.}~\bibnamefont
  {Carmele}}, \bibinfo {author} {\bibfnamefont {M.}~\bibnamefont {Richter}},
  \bibinfo {author} {\bibfnamefont {W.~W.}\ \bibnamefont {Chow}}, \ and\
  \bibinfo {author} {\bibfnamefont {A.}~\bibnamefont {Knorr}},\ }\href
  {\doibase 10.1103/PhysRevLett.104.156801} {\bibfield  {journal} {\bibinfo
  {journal} {Phys. Rev. Lett.}\ }\textbf {\bibinfo {volume} {104}},\ \bibinfo
  {pages} {156801} (\bibinfo {year} {2010})}\BibitemShut {NoStop}%
\bibitem [{\citenamefont {Carmele}\ \emph
  {et~al.}(2013{\natexlab{a}})\citenamefont {Carmele}, \citenamefont {Knorr},\
  and\ \citenamefont {Milde}}]{carmele_cnr}%
  \BibitemOpen
  \bibfield  {author} {\bibinfo {author} {\bibfnamefont {A.}~\bibnamefont
  {Carmele}}, \bibinfo {author} {\bibfnamefont {A.}~\bibnamefont {Knorr}}, \
  and\ \bibinfo {author} {\bibfnamefont {F.}~\bibnamefont {Milde}},\ }\href
  {http://stacks.iop.org/1367-2630/15/i=10/a=105024} {\bibfield  {journal}
  {\bibinfo  {journal} {New Journal of Physics}\ }\textbf {\bibinfo {volume}
  {15}},\ \bibinfo {pages} {105024} (\bibinfo {year}
  {2013}{\natexlab{a}})}\BibitemShut {NoStop}%
\bibitem [{\citenamefont {Besombes}\ \emph
  {et~al.}(2001{\natexlab{a}})\citenamefont {Besombes}, \citenamefont {Kheng},
  \citenamefont {Marsal},\ and\ \citenamefont
  {Mariette}}]{besombes2001acoustic}%
  \BibitemOpen
  \bibfield  {author} {\bibinfo {author} {\bibfnamefont {L.}~\bibnamefont
  {Besombes}}, \bibinfo {author} {\bibfnamefont {K.}~\bibnamefont {Kheng}},
  \bibinfo {author} {\bibfnamefont {L.}~\bibnamefont {Marsal}}, \ and\ \bibinfo
  {author} {\bibfnamefont {H.}~\bibnamefont {Mariette}},\ }\href@noop {}
  {\bibfield  {journal} {\bibinfo  {journal} {Physical Review B}\ }\textbf
  {\bibinfo {volume} {63}},\ \bibinfo {pages} {155307} (\bibinfo {year}
  {2001}{\natexlab{a}})}\BibitemShut {NoStop}%
\bibitem [{\citenamefont {Borri}\ \emph {et~al.}(2001)\citenamefont {Borri},
  \citenamefont {Langbein}, \citenamefont {Schneider}, \citenamefont {Woggon},
  \citenamefont {Sellin}, \citenamefont {Ouyang},\ and\ \citenamefont
  {Bimberg}}]{borri2001ultralong}%
  \BibitemOpen
  \bibfield  {author} {\bibinfo {author} {\bibfnamefont {P.}~\bibnamefont
  {Borri}}, \bibinfo {author} {\bibfnamefont {W.}~\bibnamefont {Langbein}},
  \bibinfo {author} {\bibfnamefont {S.}~\bibnamefont {Schneider}}, \bibinfo
  {author} {\bibfnamefont {U.}~\bibnamefont {Woggon}}, \bibinfo {author}
  {\bibfnamefont {R.~L.}\ \bibnamefont {Sellin}}, \bibinfo {author}
  {\bibfnamefont {D.}~\bibnamefont {Ouyang}}, \ and\ \bibinfo {author}
  {\bibfnamefont {D.}~\bibnamefont {Bimberg}},\ }\href@noop {} {\bibfield
  {journal} {\bibinfo  {journal} {Physical Review Letters}\ }\textbf {\bibinfo
  {volume} {87}},\ \bibinfo {pages} {157401} (\bibinfo {year}
  {2001})}\BibitemShut {NoStop}%
\bibitem [{\citenamefont {Galland}\ \emph {et~al.}(2008)\citenamefont
  {Galland}, \citenamefont {H{\"o}gele}, \citenamefont {T{\"u}reci},\ and\
  \citenamefont {Imamo{\u{g}}lu}}]{galland2008non}%
  \BibitemOpen
  \bibfield  {author} {\bibinfo {author} {\bibfnamefont {C.}~\bibnamefont
  {Galland}}, \bibinfo {author} {\bibfnamefont {A.}~\bibnamefont {H{\"o}gele}},
  \bibinfo {author} {\bibfnamefont {H.~E.}\ \bibnamefont {T{\"u}reci}}, \ and\
  \bibinfo {author} {\bibfnamefont {A.}~\bibnamefont {Imamo{\u{g}}lu}},\
  }\href@noop {} {\bibfield  {journal} {\bibinfo  {journal} {Physical review
  letters}\ }\textbf {\bibinfo {volume} {101}},\ \bibinfo {pages} {067402}
  (\bibinfo {year} {2008})}\BibitemShut {NoStop}%
\bibitem [{\citenamefont {Vagov}\ \emph {et~al.}(2007)\citenamefont {Vagov},
  \citenamefont {Croitoru}, \citenamefont {Axt}, \citenamefont {Kuhn},\ and\
  \citenamefont {Peeters}}]{vagov2007nonmonotonic}%
  \BibitemOpen
  \bibfield  {author} {\bibinfo {author} {\bibfnamefont {A.}~\bibnamefont
  {Vagov}}, \bibinfo {author} {\bibfnamefont {M.}~\bibnamefont {Croitoru}},
  \bibinfo {author} {\bibfnamefont {V.~M.}\ \bibnamefont {Axt}}, \bibinfo
  {author} {\bibfnamefont {T.}~\bibnamefont {Kuhn}}, \ and\ \bibinfo {author}
  {\bibfnamefont {F.}~\bibnamefont {Peeters}},\ }\href@noop {} {\bibfield
  {journal} {\bibinfo  {journal} {Physical review letters}\ }\textbf {\bibinfo
  {volume} {98}},\ \bibinfo {pages} {227403} (\bibinfo {year}
  {2007})}\BibitemShut {NoStop}%
\bibitem [{\citenamefont {Madsen}\ \emph {et~al.}(2011)\citenamefont {Madsen},
  \citenamefont {Ates}, \citenamefont {Lund-Hansen}, \citenamefont
  {L{\"o}ffler}, \citenamefont {Reitzenstein}, \citenamefont {Forchel},\ and\
  \citenamefont {Lodahl}}]{madsen2011observation}%
  \BibitemOpen
  \bibfield  {author} {\bibinfo {author} {\bibfnamefont {K.~H.}\ \bibnamefont
  {Madsen}}, \bibinfo {author} {\bibfnamefont {S.}~\bibnamefont {Ates}},
  \bibinfo {author} {\bibfnamefont {T.}~\bibnamefont {Lund-Hansen}}, \bibinfo
  {author} {\bibfnamefont {A.}~\bibnamefont {L{\"o}ffler}}, \bibinfo {author}
  {\bibfnamefont {S.}~\bibnamefont {Reitzenstein}}, \bibinfo {author}
  {\bibfnamefont {A.}~\bibnamefont {Forchel}}, \ and\ \bibinfo {author}
  {\bibfnamefont {P.}~\bibnamefont {Lodahl}},\ }\href@noop {} {\bibfield
  {journal} {\bibinfo  {journal} {Physical review letters}\ }\textbf {\bibinfo
  {volume} {106}},\ \bibinfo {pages} {233601} (\bibinfo {year}
  {2011})}\BibitemShut {NoStop}%
\bibitem [{\citenamefont {Carmichael}(2009)}]{carmichael2009statistical}%
  \BibitemOpen
  \bibfield  {author} {\bibinfo {author} {\bibfnamefont {H.~J.}\ \bibnamefont
  {Carmichael}},\ }\href@noop {} {\emph {\bibinfo {title} {Statistical Methods
  in Quantum Optics 1}}}\ (\bibinfo  {publisher} {Springer Science \& Business
  Media},\ \bibinfo {year} {2009})\BibitemShut {NoStop}%
\bibitem [{\citenamefont {Scully}\ and\ \citenamefont
  {Zubairy}(1999)}]{scully1999quantum}%
  \BibitemOpen
  \bibfield  {author} {\bibinfo {author} {\bibfnamefont {M.~O.}\ \bibnamefont
  {Scully}}\ and\ \bibinfo {author} {\bibfnamefont {M.~S.}\ \bibnamefont
  {Zubairy}},\ }\href@noop {} {\enquote {\bibinfo {title} {Quantum optics},}\ }
  (\bibinfo {year} {1999})\BibitemShut {NoStop}%
\bibitem [{\citenamefont {Breuer}\ \emph {et~al.}(2002)\citenamefont {Breuer},
  \citenamefont {Petruccione} \emph {et~al.}}]{breuerbook}%
  \BibitemOpen
  \bibfield  {author} {\bibinfo {author} {\bibfnamefont {H.-P.}\ \bibnamefont
  {Breuer}}, \bibinfo {author} {\bibfnamefont {F.}~\bibnamefont {Petruccione}},
   \emph {et~al.},\ }\href@noop {} {\emph {\bibinfo {title} {The theory of open
  quantum systems}}}\ (\bibinfo  {publisher} {Oxford University Press on
  Demand},\ \bibinfo {year} {2002})\BibitemShut {NoStop}%
\bibitem [{\citenamefont {Gardiner}\ \emph {et~al.}(2004)\citenamefont
  {Gardiner}, \citenamefont {Zoller},\ and\ \citenamefont
  {Zoller}}]{gardinerzollerbook}%
  \BibitemOpen
  \bibfield  {author} {\bibinfo {author} {\bibfnamefont {C.}~\bibnamefont
  {Gardiner}}, \bibinfo {author} {\bibfnamefont {P.}~\bibnamefont {Zoller}}, \
  and\ \bibinfo {author} {\bibfnamefont {P.}~\bibnamefont {Zoller}},\
  }\href@noop {} {\emph {\bibinfo {title} {Quantum noise: a handbook of
  Markovian and non-Markovian quantum stochastic methods with applications to
  quantum optics}}},\ Vol.~\bibinfo {volume} {56}\ (\bibinfo  {publisher}
  {Springer Science \& Business Media},\ \bibinfo {year} {2004})\BibitemShut
  {NoStop}%
\bibitem [{\citenamefont {Breuer}\ \emph {et~al.}(2016)\citenamefont {Breuer},
  \citenamefont {Laine}, \citenamefont {Piilo},\ and\ \citenamefont
  {Vacchini}}]{RevModPhys.88.021002}%
  \BibitemOpen
  \bibfield  {author} {\bibinfo {author} {\bibfnamefont {H.-P.}\ \bibnamefont
  {Breuer}}, \bibinfo {author} {\bibfnamefont {E.-M.}\ \bibnamefont {Laine}},
  \bibinfo {author} {\bibfnamefont {J.}~\bibnamefont {Piilo}}, \ and\ \bibinfo
  {author} {\bibfnamefont {B.}~\bibnamefont {Vacchini}},\ }\href {\doibase
  10.1103/RevModPhys.88.021002} {\bibfield  {journal} {\bibinfo  {journal}
  {Rev. Mod. Phys.}\ }\textbf {\bibinfo {volume} {88}},\ \bibinfo {pages}
  {021002} (\bibinfo {year} {2016})}\BibitemShut {NoStop}%
\bibitem [{\citenamefont {Cerrillo}\ and\ \citenamefont
  {Cao}(2014)}]{PhysRevLett.112.110401}%
  \BibitemOpen
  \bibfield  {author} {\bibinfo {author} {\bibfnamefont {J.}~\bibnamefont
  {Cerrillo}}\ and\ \bibinfo {author} {\bibfnamefont {J.}~\bibnamefont {Cao}},\
  }\href {\doibase 10.1103/PhysRevLett.112.110401} {\bibfield  {journal}
  {\bibinfo  {journal} {Phys. Rev. Lett.}\ }\textbf {\bibinfo {volume} {112}},\
  \bibinfo {pages} {110401} (\bibinfo {year} {2014})}\BibitemShut {NoStop}%
\bibitem [{\citenamefont {de~Vega}\ and\ \citenamefont
  {Alonso}(2017)}]{RevModPhys.89.015001}%
  \BibitemOpen
  \bibfield  {author} {\bibinfo {author} {\bibfnamefont {I.}~\bibnamefont
  {de~Vega}}\ and\ \bibinfo {author} {\bibfnamefont {D.}~\bibnamefont
  {Alonso}},\ }\href {\doibase 10.1103/RevModPhys.89.015001} {\bibfield
  {journal} {\bibinfo  {journal} {Rev. Mod. Phys.}\ }\textbf {\bibinfo {volume}
  {89}},\ \bibinfo {pages} {015001} (\bibinfo {year} {2017})}\BibitemShut
  {NoStop}%
\bibitem [{\citenamefont {Michler}(2003)}]{michler2003single}%
  \BibitemOpen
  \bibfield  {author} {\bibinfo {author} {\bibfnamefont {P.}~\bibnamefont
  {Michler}},\ }\href@noop {} {\emph {\bibinfo {title} {Single quantum dots:
  Fundamentals, applications and new concepts}}},\ Vol.~\bibinfo {volume} {90}\
  (\bibinfo  {publisher} {Springer Science \& Business Media},\ \bibinfo {year}
  {2003})\BibitemShut {NoStop}%
\bibitem [{\citenamefont {Stock}\ \emph {et~al.}(2011)\citenamefont {Stock},
  \citenamefont {Dachner}, \citenamefont {Warming}, \citenamefont {Schliwa},
  \citenamefont {Lochmann}, \citenamefont {Hoffmann}, \citenamefont {Toropov},
  \citenamefont {Bakarov}, \citenamefont {Derebezov}, \citenamefont {Richter},
  \citenamefont {Haisler}, \citenamefont {Knorr},\ and\ \citenamefont
  {Bimberg}}]{stock}%
  \BibitemOpen
  \bibfield  {author} {\bibinfo {author} {\bibfnamefont {E.}~\bibnamefont
  {Stock}}, \bibinfo {author} {\bibfnamefont {M.-R.}\ \bibnamefont {Dachner}},
  \bibinfo {author} {\bibfnamefont {T.}~\bibnamefont {Warming}}, \bibinfo
  {author} {\bibfnamefont {A.}~\bibnamefont {Schliwa}}, \bibinfo {author}
  {\bibfnamefont {A.}~\bibnamefont {Lochmann}}, \bibinfo {author}
  {\bibfnamefont {A.}~\bibnamefont {Hoffmann}}, \bibinfo {author}
  {\bibfnamefont {A.~I.}\ \bibnamefont {Toropov}}, \bibinfo {author}
  {\bibfnamefont {A.~K.}\ \bibnamefont {Bakarov}}, \bibinfo {author}
  {\bibfnamefont {I.~A.}\ \bibnamefont {Derebezov}}, \bibinfo {author}
  {\bibfnamefont {M.}~\bibnamefont {Richter}}, \bibinfo {author} {\bibfnamefont
  {V.~A.}\ \bibnamefont {Haisler}}, \bibinfo {author} {\bibfnamefont
  {A.}~\bibnamefont {Knorr}}, \ and\ \bibinfo {author} {\bibfnamefont
  {D.}~\bibnamefont {Bimberg}},\ }\href {\doibase 10.1103/PhysRevB.83.041304}
  {\bibfield  {journal} {\bibinfo  {journal} {Phys. Rev. B}\ }\textbf {\bibinfo
  {volume} {83}},\ \bibinfo {pages} {041304} (\bibinfo {year}
  {2011})}\BibitemShut {NoStop}%
\bibitem [{\citenamefont {Strau{\ss}}\ \emph {et~al.}(2018)\citenamefont
  {Strau{\ss}}, \citenamefont {Hohn}, \citenamefont {Schneider}, \citenamefont
  {H{\"o}fling}, \citenamefont {Wolters},\ and\ \citenamefont
  {Reitzenstein}}]{strauss}%
  \BibitemOpen
  \bibfield  {author} {\bibinfo {author} {\bibfnamefont {M.}~\bibnamefont
  {Strau{\ss}}}, \bibinfo {author} {\bibfnamefont {M.}~\bibnamefont {Hohn}},
  \bibinfo {author} {\bibfnamefont {C.}~\bibnamefont {Schneider}}, \bibinfo
  {author} {\bibfnamefont {S.}~\bibnamefont {H{\"o}fling}}, \bibinfo {author}
  {\bibfnamefont {J.}~\bibnamefont {Wolters}}, \ and\ \bibinfo {author}
  {\bibfnamefont {S.}~\bibnamefont {Reitzenstein}},\ }\href@noop {} {\bibfield
  {journal} {\bibinfo  {journal} {arXiv preprint arXiv:1805.06357}\ } (\bibinfo
  {year} {2018})}\BibitemShut {NoStop}%
\bibitem [{\citenamefont {Ramsay}\ \emph
  {et~al.}(2010{\natexlab{a}})\citenamefont {Ramsay}, \citenamefont {Godden},
  \citenamefont {Boyle}, \citenamefont {Gauger}, \citenamefont {Nazir},
  \citenamefont {Lovett}, \citenamefont {Fox},\ and\ \citenamefont
  {Skolnick}}]{ramsay2010phonon}%
  \BibitemOpen
  \bibfield  {author} {\bibinfo {author} {\bibfnamefont {A.}~\bibnamefont
  {Ramsay}}, \bibinfo {author} {\bibfnamefont {T.}~\bibnamefont {Godden}},
  \bibinfo {author} {\bibfnamefont {S.}~\bibnamefont {Boyle}}, \bibinfo
  {author} {\bibfnamefont {E.~M.}\ \bibnamefont {Gauger}}, \bibinfo {author}
  {\bibfnamefont {A.}~\bibnamefont {Nazir}}, \bibinfo {author} {\bibfnamefont
  {B.~W.}\ \bibnamefont {Lovett}}, \bibinfo {author} {\bibfnamefont
  {A.}~\bibnamefont {Fox}}, \ and\ \bibinfo {author} {\bibfnamefont
  {M.}~\bibnamefont {Skolnick}},\ }\href@noop {} {\bibfield  {journal}
  {\bibinfo  {journal} {Physical Review Letters}\ }\textbf {\bibinfo {volume}
  {105}},\ \bibinfo {pages} {177402} (\bibinfo {year}
  {2010}{\natexlab{a}})}\BibitemShut {NoStop}%
\bibitem [{\citenamefont {Kaldewey}\ \emph {et~al.}(2017)\citenamefont
  {Kaldewey}, \citenamefont {L{\"u}ker}, \citenamefont {Kuhlmann},
  \citenamefont {Valentin}, \citenamefont {Chauveau}, \citenamefont {Ludwig},
  \citenamefont {Wieck}, \citenamefont {Reiter}, \citenamefont {Kuhn},\ and\
  \citenamefont {Warburton}}]{kaldewey2017demonstrating}%
  \BibitemOpen
  \bibfield  {author} {\bibinfo {author} {\bibfnamefont {T.}~\bibnamefont
  {Kaldewey}}, \bibinfo {author} {\bibfnamefont {S.}~\bibnamefont {L{\"u}ker}},
  \bibinfo {author} {\bibfnamefont {A.~V.}\ \bibnamefont {Kuhlmann}}, \bibinfo
  {author} {\bibfnamefont {S.~R.}\ \bibnamefont {Valentin}}, \bibinfo {author}
  {\bibfnamefont {J.-M.}\ \bibnamefont {Chauveau}}, \bibinfo {author}
  {\bibfnamefont {A.}~\bibnamefont {Ludwig}}, \bibinfo {author} {\bibfnamefont
  {A.~D.}\ \bibnamefont {Wieck}}, \bibinfo {author} {\bibfnamefont {D.~E.}\
  \bibnamefont {Reiter}}, \bibinfo {author} {\bibfnamefont {T.}~\bibnamefont
  {Kuhn}}, \ and\ \bibinfo {author} {\bibfnamefont {R.~J.}\ \bibnamefont
  {Warburton}},\ }\href@noop {} {\bibfield  {journal} {\bibinfo  {journal}
  {Physical Review B}\ }\textbf {\bibinfo {volume} {95}},\ \bibinfo {pages}
  {241306} (\bibinfo {year} {2017})}\BibitemShut {NoStop}%
\bibitem [{\citenamefont {Bounouar}\ \emph {et~al.}(2015)\citenamefont
  {Bounouar}, \citenamefont {M{\"u}ller}, \citenamefont {Barth}, \citenamefont
  {Gl{\"a}ssl}, \citenamefont {Axt},\ and\ \citenamefont
  {Michler}}]{bounouar2015phonon}%
  \BibitemOpen
  \bibfield  {author} {\bibinfo {author} {\bibfnamefont {S.}~\bibnamefont
  {Bounouar}}, \bibinfo {author} {\bibfnamefont {M.}~\bibnamefont
  {M{\"u}ller}}, \bibinfo {author} {\bibfnamefont {A.}~\bibnamefont {Barth}},
  \bibinfo {author} {\bibfnamefont {M.}~\bibnamefont {Gl{\"a}ssl}}, \bibinfo
  {author} {\bibfnamefont {V.}~\bibnamefont {Axt}}, \ and\ \bibinfo {author}
  {\bibfnamefont {P.}~\bibnamefont {Michler}},\ }\href@noop {} {\bibfield
  {journal} {\bibinfo  {journal} {Physical Review B}\ }\textbf {\bibinfo
  {volume} {91}},\ \bibinfo {pages} {161302} (\bibinfo {year}
  {2015})}\BibitemShut {NoStop}%
\bibitem [{\citenamefont {Ardelt}\ \emph {et~al.}(2014)\citenamefont {Ardelt},
  \citenamefont {Hanschke}, \citenamefont {Fischer}, \citenamefont
  {M{\"u}ller}, \citenamefont {Kleinkauf}, \citenamefont {Koller},
  \citenamefont {Bechtold}, \citenamefont {Simmet}, \citenamefont
  {Wierzbowski}, \citenamefont {Riedl} \emph {et~al.}}]{ardelt2014dissipative}%
  \BibitemOpen
  \bibfield  {author} {\bibinfo {author} {\bibfnamefont {P.-L.}\ \bibnamefont
  {Ardelt}}, \bibinfo {author} {\bibfnamefont {L.}~\bibnamefont {Hanschke}},
  \bibinfo {author} {\bibfnamefont {K.~A.}\ \bibnamefont {Fischer}}, \bibinfo
  {author} {\bibfnamefont {K.}~\bibnamefont {M{\"u}ller}}, \bibinfo {author}
  {\bibfnamefont {A.}~\bibnamefont {Kleinkauf}}, \bibinfo {author}
  {\bibfnamefont {M.}~\bibnamefont {Koller}}, \bibinfo {author} {\bibfnamefont
  {A.}~\bibnamefont {Bechtold}}, \bibinfo {author} {\bibfnamefont
  {T.}~\bibnamefont {Simmet}}, \bibinfo {author} {\bibfnamefont
  {J.}~\bibnamefont {Wierzbowski}}, \bibinfo {author} {\bibfnamefont
  {H.}~\bibnamefont {Riedl}},  \emph {et~al.},\ }\href@noop {} {\bibfield
  {journal} {\bibinfo  {journal} {Physical Review B}\ }\textbf {\bibinfo
  {volume} {90}},\ \bibinfo {pages} {241404} (\bibinfo {year}
  {2014})}\BibitemShut {NoStop}%
\bibitem [{\citenamefont {Iles-Smith}\ and\ \citenamefont
  {Nazir}(2016)}]{iles2016quantum}%
  \BibitemOpen
  \bibfield  {author} {\bibinfo {author} {\bibfnamefont {J.}~\bibnamefont
  {Iles-Smith}}\ and\ \bibinfo {author} {\bibfnamefont {A.}~\bibnamefont
  {Nazir}},\ }\href@noop {} {\bibfield  {journal} {\bibinfo  {journal}
  {Optica}\ }\textbf {\bibinfo {volume} {3}},\ \bibinfo {pages} {207} (\bibinfo
  {year} {2016})}\BibitemShut {NoStop}%
\bibitem [{\citenamefont {Weiler}\ \emph {et~al.}(2012)\citenamefont {Weiler},
  \citenamefont {Ulhaq}, \citenamefont {Ulrich}, \citenamefont {Richter},
  \citenamefont {Jetter}, \citenamefont {Michler}, \citenamefont {Roy},\ and\
  \citenamefont {Hughes}}]{hughes_incoherent}%
  \BibitemOpen
  \bibfield  {author} {\bibinfo {author} {\bibfnamefont {S.}~\bibnamefont
  {Weiler}}, \bibinfo {author} {\bibfnamefont {A.}~\bibnamefont {Ulhaq}},
  \bibinfo {author} {\bibfnamefont {S.~M.}\ \bibnamefont {Ulrich}}, \bibinfo
  {author} {\bibfnamefont {D.}~\bibnamefont {Richter}}, \bibinfo {author}
  {\bibfnamefont {M.}~\bibnamefont {Jetter}}, \bibinfo {author} {\bibfnamefont
  {P.}~\bibnamefont {Michler}}, \bibinfo {author} {\bibfnamefont
  {C.}~\bibnamefont {Roy}}, \ and\ \bibinfo {author} {\bibfnamefont
  {S.}~\bibnamefont {Hughes}},\ }\href {\doibase 10.1103/PhysRevB.86.241304}
  {\bibfield  {journal} {\bibinfo  {journal} {Phys. Rev. B}\ }\textbf {\bibinfo
  {volume} {86}},\ \bibinfo {pages} {241304} (\bibinfo {year}
  {2012})}\BibitemShut {NoStop}%
\bibitem [{\citenamefont {Hohenester}(2010)}]{hohenesterFeeding}%
  \BibitemOpen
  \bibfield  {author} {\bibinfo {author} {\bibfnamefont {U.}~\bibnamefont
  {Hohenester}},\ }\href {\doibase 10.1103/PhysRevB.81.155303} {\bibfield
  {journal} {\bibinfo  {journal} {Phys. Rev. B}\ }\textbf {\bibinfo {volume}
  {81}},\ \bibinfo {pages} {155303} (\bibinfo {year} {2010})}\BibitemShut
  {NoStop}%
\bibitem [{\citenamefont {Monniello}\ \emph {et~al.}(2013)\citenamefont
  {Monniello}, \citenamefont {Tonin}, \citenamefont {Hostein}, \citenamefont
  {Lemaitre}, \citenamefont {Martinez}, \citenamefont {Voliotis},\ and\
  \citenamefont {Grousson}}]{monniello2013excitation}%
  \BibitemOpen
  \bibfield  {author} {\bibinfo {author} {\bibfnamefont {L.}~\bibnamefont
  {Monniello}}, \bibinfo {author} {\bibfnamefont {C.}~\bibnamefont {Tonin}},
  \bibinfo {author} {\bibfnamefont {R.}~\bibnamefont {Hostein}}, \bibinfo
  {author} {\bibfnamefont {A.}~\bibnamefont {Lemaitre}}, \bibinfo {author}
  {\bibfnamefont {A.}~\bibnamefont {Martinez}}, \bibinfo {author}
  {\bibfnamefont {V.}~\bibnamefont {Voliotis}}, \ and\ \bibinfo {author}
  {\bibfnamefont {R.}~\bibnamefont {Grousson}},\ }\href@noop {} {\bibfield
  {journal} {\bibinfo  {journal} {Physical Review Letters}\ }\textbf {\bibinfo
  {volume} {111}},\ \bibinfo {pages} {026403} (\bibinfo {year}
  {2013})}\BibitemShut {NoStop}%
\bibitem [{\citenamefont {Ulrich}\ \emph {et~al.}(2011)\citenamefont {Ulrich},
  \citenamefont {Ates}, \citenamefont {Reitzenstein}, \citenamefont
  {L{\"o}ffler}, \citenamefont {Forchel},\ and\ \citenamefont
  {Michler}}]{ulrich2011dephasing}%
  \BibitemOpen
  \bibfield  {author} {\bibinfo {author} {\bibfnamefont {S.}~\bibnamefont
  {Ulrich}}, \bibinfo {author} {\bibfnamefont {S.}~\bibnamefont {Ates}},
  \bibinfo {author} {\bibfnamefont {S.}~\bibnamefont {Reitzenstein}}, \bibinfo
  {author} {\bibfnamefont {A.}~\bibnamefont {L{\"o}ffler}}, \bibinfo {author}
  {\bibfnamefont {A.}~\bibnamefont {Forchel}}, \ and\ \bibinfo {author}
  {\bibfnamefont {P.}~\bibnamefont {Michler}},\ }\href@noop {} {\bibfield
  {journal} {\bibinfo  {journal} {Physical review letters}\ }\textbf {\bibinfo
  {volume} {106}},\ \bibinfo {pages} {247402} (\bibinfo {year}
  {2011})}\BibitemShut {NoStop}%
\bibitem [{\citenamefont {Roy}\ and\ \citenamefont
  {Hughes}(2011)}]{roy2011phonon}%
  \BibitemOpen
  \bibfield  {author} {\bibinfo {author} {\bibfnamefont {C.}~\bibnamefont
  {Roy}}\ and\ \bibinfo {author} {\bibfnamefont {S.}~\bibnamefont {Hughes}},\
  }\href@noop {} {\bibfield  {journal} {\bibinfo  {journal} {Physical review
  letters}\ }\textbf {\bibinfo {volume} {106}},\ \bibinfo {pages} {247403}
  (\bibinfo {year} {2011})}\BibitemShut {NoStop}%
\bibitem [{\citenamefont {Gardiner}\ and\ \citenamefont
  {Zoller}(1991)}]{gardinerzoller}%
  \BibitemOpen
  \bibfield  {author} {\bibinfo {author} {\bibfnamefont {C.}~\bibnamefont
  {Gardiner}}\ and\ \bibinfo {author} {\bibfnamefont {P.}~\bibnamefont
  {Zoller}},\ }\href@noop {} {\emph {\bibinfo {title} {Quantum Noise}}}\
  (\bibinfo  {publisher} {Springer, Berlin Heidelberg New York},\ \bibinfo
  {year} {1991})\BibitemShut {NoStop}%
\bibitem [{\citenamefont {Koch}\ \emph {et~al.}(1993)\citenamefont {Koch} \emph
  {et~al.}}]{koch1993QDs}%
  \BibitemOpen
  \bibfield  {author} {\bibinfo {author} {\bibfnamefont {S.~W.}\ \bibnamefont
  {Koch}} \emph {et~al.},\ }\href@noop {} {\emph {\bibinfo {title}
  {Semiconductor quantum dots}}},\ Vol.~\bibinfo {volume} {2}\ (\bibinfo
  {publisher} {World Scientific},\ \bibinfo {year} {1993})\BibitemShut
  {NoStop}%
\bibitem [{\citenamefont {Grundmann}\ \emph {et~al.}(1995)\citenamefont
  {Grundmann}, \citenamefont {Stier},\ and\ \citenamefont
  {Bimberg}}]{Stier1995QDs}%
  \BibitemOpen
  \bibfield  {author} {\bibinfo {author} {\bibfnamefont {M.}~\bibnamefont
  {Grundmann}}, \bibinfo {author} {\bibfnamefont {O.}~\bibnamefont {Stier}}, \
  and\ \bibinfo {author} {\bibfnamefont {D.}~\bibnamefont {Bimberg}},\ }\href
  {\doibase 10.1103/PhysRevB.52.11969} {\bibfield  {journal} {\bibinfo
  {journal} {Phys. Rev. B}\ }\textbf {\bibinfo {volume} {52}},\ \bibinfo
  {pages} {11969} (\bibinfo {year} {1995})}\BibitemShut {NoStop}%
\bibitem [{\citenamefont {Stier}\ \emph {et~al.}(1999)\citenamefont {Stier},
  \citenamefont {Grundmann},\ and\ \citenamefont {Bimberg}}]{Stier1999QDs}%
  \BibitemOpen
  \bibfield  {author} {\bibinfo {author} {\bibfnamefont {O.}~\bibnamefont
  {Stier}}, \bibinfo {author} {\bibfnamefont {M.}~\bibnamefont {Grundmann}}, \
  and\ \bibinfo {author} {\bibfnamefont {D.}~\bibnamefont {Bimberg}},\ }\href
  {\doibase 10.1103/PhysRevB.59.5688} {\bibfield  {journal} {\bibinfo
  {journal} {Phys. Rev. B}\ }\textbf {\bibinfo {volume} {59}},\ \bibinfo
  {pages} {5688} (\bibinfo {year} {1999})}\BibitemShut {NoStop}%
\bibitem [{\citenamefont {Schliwa}\ \emph {et~al.}(2007)\citenamefont
  {Schliwa}, \citenamefont {Winkelnkemper},\ and\ \citenamefont
  {Bimberg}}]{Schliwa2007QDs}%
  \BibitemOpen
  \bibfield  {author} {\bibinfo {author} {\bibfnamefont {A.}~\bibnamefont
  {Schliwa}}, \bibinfo {author} {\bibfnamefont {M.}~\bibnamefont
  {Winkelnkemper}}, \ and\ \bibinfo {author} {\bibfnamefont {D.}~\bibnamefont
  {Bimberg}},\ }\href {\doibase 10.1103/PhysRevB.76.205324} {\bibfield
  {journal} {\bibinfo  {journal} {Phys. Rev. B}\ }\textbf {\bibinfo {volume}
  {76}},\ \bibinfo {pages} {205324} (\bibinfo {year} {2007})}\BibitemShut
  {NoStop}%
\bibitem [{\citenamefont {Chow}\ and\ \citenamefont {Jahnke}(2013)}]{chow2013}%
  \BibitemOpen
  \bibfield  {author} {\bibinfo {author} {\bibfnamefont {W.~W.}\ \bibnamefont
  {Chow}}\ and\ \bibinfo {author} {\bibfnamefont {F.}~\bibnamefont {Jahnke}},\
  }\href@noop {} {\bibfield  {journal} {\bibinfo  {journal} {Progress in
  quantum electronics}\ }\textbf {\bibinfo {volume} {37}},\ \bibinfo {pages}
  {109} (\bibinfo {year} {2013})}\BibitemShut {NoStop}%
\bibitem [{\citenamefont {Haug}\ and\ \citenamefont {Koch}(2009)}]{haugkoch}%
  \BibitemOpen
  \bibfield  {author} {\bibinfo {author} {\bibfnamefont {H.}~\bibnamefont
  {Haug}}\ and\ \bibinfo {author} {\bibfnamefont {S.~W.}\ \bibnamefont
  {Koch}},\ }\href@noop {} {\emph {\bibinfo {title} {Quantum Theory of the
  Optical and Electronic Properties of Semiconductors: Fivth Edition}}}\
  (\bibinfo  {publisher} {World Scientific Publishing Company},\ \bibinfo
  {year} {2009})\BibitemShut {NoStop}%
\bibitem [{\citenamefont {Brandes}(2005)}]{brandes2005coherent_mesoscopic}%
  \BibitemOpen
  \bibfield  {author} {\bibinfo {author} {\bibfnamefont {T.}~\bibnamefont
  {Brandes}},\ }\href@noop {} {\bibfield  {journal} {\bibinfo  {journal}
  {physics reports}\ }\textbf {\bibinfo {volume} {408}},\ \bibinfo {pages}
  {315} (\bibinfo {year} {2005})}\BibitemShut {NoStop}%
\bibitem [{\citenamefont {Sakaki}(1992)}]{quantumwires}%
  \BibitemOpen
  \bibfield  {author} {\bibinfo {author} {\bibfnamefont {H.}~\bibnamefont
  {Sakaki}},\ }\href@noop {} {\bibfield  {journal} {\bibinfo  {journal}
  {Surface Science}\ }\textbf {\bibinfo {volume} {267}},\ \bibinfo {pages}
  {623} (\bibinfo {year} {1992})}\BibitemShut {NoStop}%
\bibitem [{\citenamefont {Schwab}\ \emph {et~al.}(2000)\citenamefont {Schwab},
  \citenamefont {Henriksen}, \citenamefont {Worlock},\ and\ \citenamefont
  {Roukes}}]{schwab2000measurement_wire}%
  \BibitemOpen
  \bibfield  {author} {\bibinfo {author} {\bibfnamefont {K.}~\bibnamefont
  {Schwab}}, \bibinfo {author} {\bibfnamefont {E.}~\bibnamefont {Henriksen}},
  \bibinfo {author} {\bibfnamefont {J.}~\bibnamefont {Worlock}}, \ and\
  \bibinfo {author} {\bibfnamefont {M.~L.}\ \bibnamefont {Roukes}},\
  }\href@noop {} {\bibfield  {journal} {\bibinfo  {journal} {Nature}\ }\textbf
  {\bibinfo {volume} {404}},\ \bibinfo {pages} {974} (\bibinfo {year}
  {2000})}\BibitemShut {NoStop}%
\bibitem [{\citenamefont {Gornyi}\ \emph {et~al.}(2005)\citenamefont {Gornyi},
  \citenamefont {Mirlin},\ and\ \citenamefont
  {Polyakov}}]{gornyi2005interacting_wires}%
  \BibitemOpen
  \bibfield  {author} {\bibinfo {author} {\bibfnamefont {I.}~\bibnamefont
  {Gornyi}}, \bibinfo {author} {\bibfnamefont {A.}~\bibnamefont {Mirlin}}, \
  and\ \bibinfo {author} {\bibfnamefont {D.}~\bibnamefont {Polyakov}},\
  }\href@noop {} {\bibfield  {journal} {\bibinfo  {journal} {Physical review
  letters}\ }\textbf {\bibinfo {volume} {95}},\ \bibinfo {pages} {206603}
  (\bibinfo {year} {2005})}\BibitemShut {NoStop}%
\bibitem [{\citenamefont {Sun}\ \emph {et~al.}(2000)\citenamefont {Sun},
  \citenamefont {Liang},\ and\ \citenamefont {Yu}}]{sun2000coherent_wells}%
  \BibitemOpen
  \bibfield  {author} {\bibinfo {author} {\bibfnamefont {C.-K.}\ \bibnamefont
  {Sun}}, \bibinfo {author} {\bibfnamefont {J.-C.}\ \bibnamefont {Liang}}, \
  and\ \bibinfo {author} {\bibfnamefont {X.-Y.}\ \bibnamefont {Yu}},\
  }\href@noop {} {\bibfield  {journal} {\bibinfo  {journal} {Physical review
  letters}\ }\textbf {\bibinfo {volume} {84}},\ \bibinfo {pages} {179}
  (\bibinfo {year} {2000})}\BibitemShut {NoStop}%
\bibitem [{\citenamefont {Lugli}\ and\ \citenamefont
  {Goodnick}(1987)}]{lugli1987nonequilibrium_cooling_wells}%
  \BibitemOpen
  \bibfield  {author} {\bibinfo {author} {\bibfnamefont {P.}~\bibnamefont
  {Lugli}}\ and\ \bibinfo {author} {\bibfnamefont {S.}~\bibnamefont
  {Goodnick}},\ }\href@noop {} {\bibfield  {journal} {\bibinfo  {journal}
  {Physical review letters}\ }\textbf {\bibinfo {volume} {59}},\ \bibinfo
  {pages} {716} (\bibinfo {year} {1987})}\BibitemShut {NoStop}%
\bibitem [{\citenamefont {Kim}\ \emph {et~al.}(1992)\citenamefont {Kim},
  \citenamefont {Shah}, \citenamefont {Cunningham}, \citenamefont {Damen},
  \citenamefont {Sch{\"a}fer}, \citenamefont {Hartmann},\ and\ \citenamefont
  {Schmitt-Rink}}]{kim1992giant_excitonic_resonances_wells}%
  \BibitemOpen
  \bibfield  {author} {\bibinfo {author} {\bibfnamefont {D.-S.}\ \bibnamefont
  {Kim}}, \bibinfo {author} {\bibfnamefont {J.}~\bibnamefont {Shah}}, \bibinfo
  {author} {\bibfnamefont {J.}~\bibnamefont {Cunningham}}, \bibinfo {author}
  {\bibfnamefont {T.}~\bibnamefont {Damen}}, \bibinfo {author} {\bibfnamefont
  {W.}~\bibnamefont {Sch{\"a}fer}}, \bibinfo {author} {\bibfnamefont
  {M.}~\bibnamefont {Hartmann}}, \ and\ \bibinfo {author} {\bibfnamefont
  {S.}~\bibnamefont {Schmitt-Rink}},\ }\href@noop {} {\bibfield  {journal}
  {\bibinfo  {journal} {Physical review letters}\ }\textbf {\bibinfo {volume}
  {68}},\ \bibinfo {pages} {1006} (\bibinfo {year} {1992})}\BibitemShut
  {NoStop}%
\bibitem [{\citenamefont {Gammon}\ \emph {et~al.}(1995)\citenamefont {Gammon},
  \citenamefont {Rudin}, \citenamefont {Reinecke}, \citenamefont {Katzer},\
  and\ \citenamefont {Kyono}}]{gammon1995phonon_broadening_wells}%
  \BibitemOpen
  \bibfield  {author} {\bibinfo {author} {\bibfnamefont {D.}~\bibnamefont
  {Gammon}}, \bibinfo {author} {\bibfnamefont {S.}~\bibnamefont {Rudin}},
  \bibinfo {author} {\bibfnamefont {T.}~\bibnamefont {Reinecke}}, \bibinfo
  {author} {\bibfnamefont {D.}~\bibnamefont {Katzer}}, \ and\ \bibinfo {author}
  {\bibfnamefont {C.}~\bibnamefont {Kyono}},\ }\href@noop {} {\bibfield
  {journal} {\bibinfo  {journal} {Physical Review B}\ }\textbf {\bibinfo
  {volume} {51}},\ \bibinfo {pages} {16785} (\bibinfo {year}
  {1995})}\BibitemShut {NoStop}%
\bibitem [{\citenamefont {Richter}\ \emph {et~al.}(2018)\citenamefont
  {Richter}, \citenamefont {Singh}, \citenamefont {Siemens},\ and\
  \citenamefont {Cundiff}}]{richter2018deconvolution}%
  \BibitemOpen
  \bibfield  {author} {\bibinfo {author} {\bibfnamefont {M.}~\bibnamefont
  {Richter}}, \bibinfo {author} {\bibfnamefont {R.}~\bibnamefont {Singh}},
  \bibinfo {author} {\bibfnamefont {M.}~\bibnamefont {Siemens}}, \ and\
  \bibinfo {author} {\bibfnamefont {S.~T.}\ \bibnamefont {Cundiff}},\
  }\href@noop {} {\bibfield  {journal} {\bibinfo  {journal} {Science advances}\
  }\textbf {\bibinfo {volume} {4}},\ \bibinfo {pages} {eaar7697} (\bibinfo
  {year} {2018})}\BibitemShut {NoStop}%
\bibitem [{\citenamefont {Singh}\ \emph {et~al.}(2017)\citenamefont {Singh},
  \citenamefont {Richter}, \citenamefont {Moody}, \citenamefont {Siemens},
  \citenamefont {Li},\ and\ \citenamefont {Cundiff}}]{richter_localization}%
  \BibitemOpen
  \bibfield  {author} {\bibinfo {author} {\bibfnamefont {R.}~\bibnamefont
  {Singh}}, \bibinfo {author} {\bibfnamefont {M.}~\bibnamefont {Richter}},
  \bibinfo {author} {\bibfnamefont {G.}~\bibnamefont {Moody}}, \bibinfo
  {author} {\bibfnamefont {M.~E.}\ \bibnamefont {Siemens}}, \bibinfo {author}
  {\bibfnamefont {H.}~\bibnamefont {Li}}, \ and\ \bibinfo {author}
  {\bibfnamefont {S.~T.}\ \bibnamefont {Cundiff}},\ }\href {\doibase
  10.1103/PhysRevB.95.235307} {\bibfield  {journal} {\bibinfo  {journal} {Phys.
  Rev. B}\ }\textbf {\bibinfo {volume} {95}},\ \bibinfo {pages} {235307}
  (\bibinfo {year} {2017})}\BibitemShut {NoStop}%
\bibitem [{\citenamefont {Lim}\ \emph {et~al.}(2017)\citenamefont {Lim},
  \citenamefont {Ing}, \citenamefont {Rosskopf}, \citenamefont {Jeske},
  \citenamefont {Cole}, \citenamefont {Huelga},\ and\ \citenamefont
  {Plenio}}]{2D_plenio_lim2017signatures}%
  \BibitemOpen
  \bibfield  {author} {\bibinfo {author} {\bibfnamefont {J.}~\bibnamefont
  {Lim}}, \bibinfo {author} {\bibfnamefont {D.~J.}\ \bibnamefont {Ing}},
  \bibinfo {author} {\bibfnamefont {J.}~\bibnamefont {Rosskopf}}, \bibinfo
  {author} {\bibfnamefont {J.}~\bibnamefont {Jeske}}, \bibinfo {author}
  {\bibfnamefont {J.~H.}\ \bibnamefont {Cole}}, \bibinfo {author}
  {\bibfnamefont {S.~F.}\ \bibnamefont {Huelga}}, \ and\ \bibinfo {author}
  {\bibfnamefont {M.~B.}\ \bibnamefont {Plenio}},\ }\href@noop {} {\bibfield
  {journal} {\bibinfo  {journal} {The Journal of chemical physics}\ }\textbf
  {\bibinfo {volume} {146}},\ \bibinfo {pages} {024109} (\bibinfo {year}
  {2017})}\BibitemShut {NoStop}%
\bibitem [{\citenamefont {Wigger}\ \emph {et~al.}(2018)\citenamefont {Wigger},
  \citenamefont {Schneider}, \citenamefont {Gerhardt}, \citenamefont {Kamp},
  \citenamefont {H{\"o}fling}, \citenamefont {Kuhn},\ and\ \citenamefont
  {Kasprzak}}]{2D_wigger2018rabi}%
  \BibitemOpen
  \bibfield  {author} {\bibinfo {author} {\bibfnamefont {D.}~\bibnamefont
  {Wigger}}, \bibinfo {author} {\bibfnamefont {C.}~\bibnamefont {Schneider}},
  \bibinfo {author} {\bibfnamefont {S.}~\bibnamefont {Gerhardt}}, \bibinfo
  {author} {\bibfnamefont {M.}~\bibnamefont {Kamp}}, \bibinfo {author}
  {\bibfnamefont {S.}~\bibnamefont {H{\"o}fling}}, \bibinfo {author}
  {\bibfnamefont {T.}~\bibnamefont {Kuhn}}, \ and\ \bibinfo {author}
  {\bibfnamefont {J.}~\bibnamefont {Kasprzak}},\ }\href@noop {} {\bibfield
  {journal} {\bibinfo  {journal} {Optica}\ }\textbf {\bibinfo {volume} {5}},\
  \bibinfo {pages} {1442} (\bibinfo {year} {2018})}\BibitemShut {NoStop}%
\bibitem [{\citenamefont {Cassette}\ \emph {et~al.}(2015)\citenamefont
  {Cassette}, \citenamefont {Pensack}, \citenamefont {Mahler},\ and\
  \citenamefont {Scholes}}]{2D_electronic_cassette2015room}%
  \BibitemOpen
  \bibfield  {author} {\bibinfo {author} {\bibfnamefont {E.}~\bibnamefont
  {Cassette}}, \bibinfo {author} {\bibfnamefont {R.~D.}\ \bibnamefont
  {Pensack}}, \bibinfo {author} {\bibfnamefont {B.}~\bibnamefont {Mahler}}, \
  and\ \bibinfo {author} {\bibfnamefont {G.~D.}\ \bibnamefont {Scholes}},\
  }\href@noop {} {\bibfield  {journal} {\bibinfo  {journal} {Nature
  communications}\ }\textbf {\bibinfo {volume} {6}},\ \bibinfo {pages} {6086}
  (\bibinfo {year} {2015})}\BibitemShut {NoStop}%
\bibitem [{\citenamefont {Liu}\ \emph {et~al.}(2018{\natexlab{a}})\citenamefont
  {Liu}, \citenamefont {Almeida}, \citenamefont {Bae}, \citenamefont
  {Padilha},\ and\ \citenamefont {Cundiff}}]{2D_spec_liu2018vibrational}%
  \BibitemOpen
  \bibfield  {author} {\bibinfo {author} {\bibfnamefont {A.}~\bibnamefont
  {Liu}}, \bibinfo {author} {\bibfnamefont {D.~B.}\ \bibnamefont {Almeida}},
  \bibinfo {author} {\bibfnamefont {W.~K.}\ \bibnamefont {Bae}}, \bibinfo
  {author} {\bibfnamefont {L.~A.}\ \bibnamefont {Padilha}}, \ and\ \bibinfo
  {author} {\bibfnamefont {S.~T.}\ \bibnamefont {Cundiff}},\ }\href@noop {}
  {\bibfield  {journal} {\bibinfo  {journal} {arXiv preprint arXiv:1806.06112}\
  } (\bibinfo {year} {2018}{\natexlab{a}})}\BibitemShut {NoStop}%
\bibitem [{\citenamefont {Suzuki}\ \emph {et~al.}(2018)\citenamefont {Suzuki},
  \citenamefont {Singh}, \citenamefont {Moody}, \citenamefont {A\ss{}mann},
  \citenamefont {Bayer}, \citenamefont {Ludwig}, \citenamefont {Wieck},\ and\
  \citenamefont {Cundiff}}]{2D_cundiff_PhysRevB.98.195304}%
  \BibitemOpen
  \bibfield  {author} {\bibinfo {author} {\bibfnamefont {T.}~\bibnamefont
  {Suzuki}}, \bibinfo {author} {\bibfnamefont {R.}~\bibnamefont {Singh}},
  \bibinfo {author} {\bibfnamefont {G.}~\bibnamefont {Moody}}, \bibinfo
  {author} {\bibfnamefont {M.}~\bibnamefont {A\ss{}mann}}, \bibinfo {author}
  {\bibfnamefont {M.}~\bibnamefont {Bayer}}, \bibinfo {author} {\bibfnamefont
  {A.}~\bibnamefont {Ludwig}}, \bibinfo {author} {\bibfnamefont {A.~D.}\
  \bibnamefont {Wieck}}, \ and\ \bibinfo {author} {\bibfnamefont {S.~T.}\
  \bibnamefont {Cundiff}},\ }\href {\doibase 10.1103/PhysRevB.98.195304}
  {\bibfield  {journal} {\bibinfo  {journal} {Phys. Rev. B}\ }\textbf {\bibinfo
  {volume} {98}},\ \bibinfo {pages} {195304} (\bibinfo {year}
  {2018})}\BibitemShut {NoStop}%
\bibitem [{\citenamefont {Christiansen}\ \emph {et~al.}(2017)\citenamefont
  {Christiansen}, \citenamefont {Selig}, \citenamefont {Bergh\"auser},
  \citenamefont {Schmidt}, \citenamefont {Niehues}, \citenamefont {Schneider},
  \citenamefont {Arora}, \citenamefont {de~Vasconcellos}, \citenamefont
  {Bratschitsch}, \citenamefont {Malic},\ and\ \citenamefont
  {Knorr}}]{2D_knorr_PhysRevLett.119.187402}%
  \BibitemOpen
  \bibfield  {author} {\bibinfo {author} {\bibfnamefont {D.}~\bibnamefont
  {Christiansen}}, \bibinfo {author} {\bibfnamefont {M.}~\bibnamefont {Selig}},
  \bibinfo {author} {\bibfnamefont {G.}~\bibnamefont {Bergh\"auser}}, \bibinfo
  {author} {\bibfnamefont {R.}~\bibnamefont {Schmidt}}, \bibinfo {author}
  {\bibfnamefont {I.}~\bibnamefont {Niehues}}, \bibinfo {author} {\bibfnamefont
  {R.}~\bibnamefont {Schneider}}, \bibinfo {author} {\bibfnamefont
  {A.}~\bibnamefont {Arora}}, \bibinfo {author} {\bibfnamefont {S.~M.}\
  \bibnamefont {de~Vasconcellos}}, \bibinfo {author} {\bibfnamefont
  {R.}~\bibnamefont {Bratschitsch}}, \bibinfo {author} {\bibfnamefont
  {E.}~\bibnamefont {Malic}}, \ and\ \bibinfo {author} {\bibfnamefont
  {A.}~\bibnamefont {Knorr}},\ }\href {\doibase 10.1103/PhysRevLett.119.187402}
  {\bibfield  {journal} {\bibinfo  {journal} {Phys. Rev. Lett.}\ }\textbf
  {\bibinfo {volume} {119}},\ \bibinfo {pages} {187402} (\bibinfo {year}
  {2017})}\BibitemShut {NoStop}%
\bibitem [{\citenamefont {Malic}\ and\ \citenamefont
  {Knorr}(2013)}]{malicbook}%
  \BibitemOpen
  \bibfield  {author} {\bibinfo {author} {\bibfnamefont {E.}~\bibnamefont
  {Malic}}\ and\ \bibinfo {author} {\bibfnamefont {A.}~\bibnamefont {Knorr}},\
  }\href@noop {} {\emph {\bibinfo {title} {Graphene and Carbon Nanotubes:
  Ultrafast Optics and Relaxation Dynamics}}}\ (\bibinfo  {publisher} {John
  Wiley \& Sons},\ \bibinfo {year} {2013})\BibitemShut {NoStop}%
\bibitem [{\citenamefont {Kira}\ and\ \citenamefont {Koch}(2011)}]{kirabook}%
  \BibitemOpen
  \bibfield  {author} {\bibinfo {author} {\bibfnamefont {M.}~\bibnamefont
  {Kira}}\ and\ \bibinfo {author} {\bibfnamefont {S.~W.}\ \bibnamefont
  {Koch}},\ }\href@noop {} {\emph {\bibinfo {title} {Semiconductor quantum
  optics}}}\ (\bibinfo  {publisher} {Cambridge University Press},\ \bibinfo
  {year} {2011})\BibitemShut {NoStop}%
\bibitem [{\citenamefont {Hohenester}\ \emph {et~al.}(2009)\citenamefont
  {Hohenester}, \citenamefont {Laucht}, \citenamefont {Kaniber}, \citenamefont
  {Hauke}, \citenamefont {Neumann}, \citenamefont {Mohtashami}, \citenamefont
  {Seliger}, \citenamefont {Bichler},\ and\ \citenamefont
  {Finley}}]{hohenester}%
  \BibitemOpen
  \bibfield  {author} {\bibinfo {author} {\bibfnamefont {U.}~\bibnamefont
  {Hohenester}}, \bibinfo {author} {\bibfnamefont {A.}~\bibnamefont {Laucht}},
  \bibinfo {author} {\bibfnamefont {M.}~\bibnamefont {Kaniber}}, \bibinfo
  {author} {\bibfnamefont {N.}~\bibnamefont {Hauke}}, \bibinfo {author}
  {\bibfnamefont {A.}~\bibnamefont {Neumann}}, \bibinfo {author} {\bibfnamefont
  {A.}~\bibnamefont {Mohtashami}}, \bibinfo {author} {\bibfnamefont
  {M.}~\bibnamefont {Seliger}}, \bibinfo {author} {\bibfnamefont
  {M.}~\bibnamefont {Bichler}}, \ and\ \bibinfo {author} {\bibfnamefont
  {J.~J.}\ \bibnamefont {Finley}},\ }\href@noop {} {\bibfield  {journal}
  {\bibinfo  {journal} {Physical Review B}\ }\textbf {\bibinfo {volume} {80}},\
  \bibinfo {pages} {201311} (\bibinfo {year} {2009})}\BibitemShut {NoStop}%
\bibitem [{\citenamefont {Briegel}\ \emph {et~al.}(1998)\citenamefont
  {Briegel}, \citenamefont {D\"ur}, \citenamefont {Cirac},\ and\ \citenamefont
  {Zoller}}]{briegel_quantum_repeater}%
  \BibitemOpen
  \bibfield  {author} {\bibinfo {author} {\bibfnamefont {H.-J.}\ \bibnamefont
  {Briegel}}, \bibinfo {author} {\bibfnamefont {W.}~\bibnamefont {D\"ur}},
  \bibinfo {author} {\bibfnamefont {J.~I.}\ \bibnamefont {Cirac}}, \ and\
  \bibinfo {author} {\bibfnamefont {P.}~\bibnamefont {Zoller}},\ }\href
  {\doibase 10.1103/PhysRevLett.81.5932} {\bibfield  {journal} {\bibinfo
  {journal} {Phys. Rev. Lett.}\ }\textbf {\bibinfo {volume} {81}},\ \bibinfo
  {pages} {5932} (\bibinfo {year} {1998})}\BibitemShut {NoStop}%
\bibitem [{\citenamefont {{H.J. Kimble}}(2008)}]{kimble_quantum_internet}%
  \BibitemOpen
  \bibfield  {author} {\bibinfo {author} {\bibnamefont {{H.J. Kimble}}},\
  }\href {\doibase https://doi.org/10.1038/nature07127 10.1038/nature07127}
  {\bibfield  {journal} {\bibinfo  {journal} {Nature}\ }\textbf {\bibinfo
  {volume} {453}},\ \bibinfo {pages} {1023} (\bibinfo {year}
  {2008})}\BibitemShut {NoStop}%
\bibitem [{\citenamefont {Mukamel}(1995)}]{mukamelbook}%
  \BibitemOpen
  \bibfield  {author} {\bibinfo {author} {\bibfnamefont {S.}~\bibnamefont
  {Mukamel}},\ }\href@noop {} {\emph {\bibinfo {title} {Principles of nonlinear
  optical spectroscopy}}},\ Vol.~\bibinfo {volume} {29}\ (\bibinfo  {publisher}
  {Oxford university press New York},\ \bibinfo {year} {1995})\BibitemShut
  {NoStop}%
\bibitem [{\citenamefont {Axt}\ and\ \citenamefont
  {Mukamel}(1998)}]{axt_mukamel}%
  \BibitemOpen
  \bibfield  {author} {\bibinfo {author} {\bibfnamefont {V.~M.}\ \bibnamefont
  {Axt}}\ and\ \bibinfo {author} {\bibfnamefont {S.}~\bibnamefont {Mukamel}},\
  }\href {\doibase 10.1103/RevModPhys.70.145} {\bibfield  {journal} {\bibinfo
  {journal} {Rev. Mod. Phys.}\ }\textbf {\bibinfo {volume} {70}},\ \bibinfo
  {pages} {145} (\bibinfo {year} {1998})}\BibitemShut {NoStop}%
\bibitem [{\citenamefont {{R. J. Warburton}}(2013)}]{warburton_review}%
  \BibitemOpen
  \bibfield  {author} {\bibinfo {author} {\bibnamefont {{R. J. Warburton}}},\
  }\href {\doibase https://doi.org/10.1038/nmat3585 10.1038/nmat3585}
  {\bibfield  {journal} {\bibinfo  {journal} {Nature Materials}\ }\textbf
  {\bibinfo {volume} {12}},\ \bibinfo {pages} {483} (\bibinfo {year}
  {2013})}\BibitemShut {NoStop}%
\bibitem [{\citenamefont {Matthiesen}\ \emph {et~al.}(2012)\citenamefont
  {Matthiesen}, \citenamefont {Vamivakas},\ and\ \citenamefont
  {Atat\"ure}}]{matthiesen_subnatural}%
  \BibitemOpen
  \bibfield  {author} {\bibinfo {author} {\bibfnamefont {C.}~\bibnamefont
  {Matthiesen}}, \bibinfo {author} {\bibfnamefont {A.~N.}\ \bibnamefont
  {Vamivakas}}, \ and\ \bibinfo {author} {\bibfnamefont {M.}~\bibnamefont
  {Atat\"ure}},\ }\href {\doibase 10.1103/PhysRevLett.108.093602} {\bibfield
  {journal} {\bibinfo  {journal} {Phys. Rev. Lett.}\ }\textbf {\bibinfo
  {volume} {108}},\ \bibinfo {pages} {093602} (\bibinfo {year}
  {2012})}\BibitemShut {NoStop}%
\bibitem [{\citenamefont {May}\ and\ \citenamefont
  {K{\"u}hn}(2008)}]{may2008charge}%
  \BibitemOpen
  \bibfield  {author} {\bibinfo {author} {\bibfnamefont {V.}~\bibnamefont
  {May}}\ and\ \bibinfo {author} {\bibfnamefont {O.}~\bibnamefont {K{\"u}hn}},\
  }\href@noop {} {\emph {\bibinfo {title} {Charge and energy transfer dynamics
  in molecular systems}}}\ (\bibinfo  {publisher} {John Wiley \& Sons},\
  \bibinfo {year} {2008})\BibitemShut {NoStop}%
\bibitem [{\citenamefont {Bourgain}\ \emph {et~al.}(2013)\citenamefont
  {Bourgain}, \citenamefont {Pellegrino}, \citenamefont {Jennewein},
  \citenamefont {Sortais},\ and\ \citenamefont
  {Browaeys}}]{bourgain2013direct}%
  \BibitemOpen
  \bibfield  {author} {\bibinfo {author} {\bibfnamefont {R.}~\bibnamefont
  {Bourgain}}, \bibinfo {author} {\bibfnamefont {J.}~\bibnamefont
  {Pellegrino}}, \bibinfo {author} {\bibfnamefont {S.}~\bibnamefont
  {Jennewein}}, \bibinfo {author} {\bibfnamefont {Y.~R.}\ \bibnamefont
  {Sortais}}, \ and\ \bibinfo {author} {\bibfnamefont {A.}~\bibnamefont
  {Browaeys}},\ }\href@noop {} {\bibfield  {journal} {\bibinfo  {journal}
  {Optics letters}\ }\textbf {\bibinfo {volume} {38}},\ \bibinfo {pages} {1963}
  (\bibinfo {year} {2013})}\BibitemShut {NoStop}%
\bibitem [{\citenamefont {Sondermann}\ and\ \citenamefont
  {Leuchs}(2013)}]{leuchsWignerDelay}%
  \BibitemOpen
  \bibfield  {author} {\bibinfo {author} {\bibfnamefont {M.}~\bibnamefont
  {Sondermann}}\ and\ \bibinfo {author} {\bibfnamefont {G.}~\bibnamefont
  {Leuchs}},\ }\href@noop {} {\bibfield  {journal} {\bibinfo  {journal}
  {Journal of the European Optical Society - Rapid publications}\ }\textbf
  {\bibinfo {volume} {8}} (\bibinfo {year} {2013})}\BibitemShut {NoStop}%
\bibitem [{\citenamefont {Palatchi}\ \emph {et~al.}(2014)\citenamefont
  {Palatchi}, \citenamefont {Dahlstr�m}, \citenamefont {Kheifets},
  \citenamefont {Ivanov}, \citenamefont {Canaday}, \citenamefont {Agostini},\
  and\ \citenamefont {DiMauro}}]{wignerDelayAtomic}%
  \BibitemOpen
  \bibfield  {author} {\bibinfo {author} {\bibfnamefont {C.}~\bibnamefont
  {Palatchi}}, \bibinfo {author} {\bibfnamefont {J.~M.}\ \bibnamefont
  {Dahlstr�m}}, \bibinfo {author} {\bibfnamefont {A.~S.}\ \bibnamefont
  {Kheifets}}, \bibinfo {author} {\bibfnamefont {I.~A.}\ \bibnamefont
  {Ivanov}}, \bibinfo {author} {\bibfnamefont {D.~M.}\ \bibnamefont {Canaday}},
  \bibinfo {author} {\bibfnamefont {P.}~\bibnamefont {Agostini}}, \ and\
  \bibinfo {author} {\bibfnamefont {L.~F.}\ \bibnamefont {DiMauro}},\ }\href
  {http://stacks.iop.org/0953-4075/47/i=24/a=245003} {\bibfield  {journal}
  {\bibinfo  {journal} {Journal of Physics B: Atomic, Molecular and Optical
  Physics}\ }\textbf {\bibinfo {volume} {47}},\ \bibinfo {pages} {245003}
  (\bibinfo {year} {2014})}\BibitemShut {NoStop}%
\bibitem [{\citenamefont {Kira}\ and\ \citenamefont {Koch}(2006)}]{kirareview}%
  \BibitemOpen
  \bibfield  {author} {\bibinfo {author} {\bibfnamefont {M.}~\bibnamefont
  {Kira}}\ and\ \bibinfo {author} {\bibfnamefont {S.}~\bibnamefont {Koch}},\
  }\href@noop {} {\bibfield  {journal} {\bibinfo  {journal} {Progress in
  quantum electronics}\ }\textbf {\bibinfo {volume} {30}},\ \bibinfo {pages}
  {155} (\bibinfo {year} {2006})}\BibitemShut {NoStop}%
\bibitem [{\citenamefont {{\v{C}}{\'\i}{\v{z}}ek}(1969)}]{cicek1969}%
  \BibitemOpen
  \bibfield  {author} {\bibinfo {author} {\bibfnamefont {J.}~\bibnamefont
  {{\v{C}}{\'\i}{\v{z}}ek}},\ }\href@noop {} {\bibfield  {journal} {\bibinfo
  {journal} {Advances in chemical physics}\ ,\ \bibinfo {pages} {35}} (\bibinfo
  {year} {1969})}\BibitemShut {NoStop}%
\bibitem [{\citenamefont {Krummheuer}\ \emph {et~al.}(2002)\citenamefont
  {Krummheuer}, \citenamefont {Axt},\ and\ \citenamefont
  {Kuhn}}]{krummheuer2002}%
  \BibitemOpen
  \bibfield  {author} {\bibinfo {author} {\bibfnamefont {B.}~\bibnamefont
  {Krummheuer}}, \bibinfo {author} {\bibfnamefont {V.~M.}\ \bibnamefont {Axt}},
  \ and\ \bibinfo {author} {\bibfnamefont {T.}~\bibnamefont {Kuhn}},\ }\href
  {\doibase 10.1103/PhysRevB.65.195313} {\bibfield  {journal} {\bibinfo
  {journal} {Phys. Rev. B}\ }\textbf {\bibinfo {volume} {65}},\ \bibinfo
  {pages} {195313} (\bibinfo {year} {2002})}\BibitemShut {NoStop}%
\bibitem [{\citenamefont {Richter}\ \emph {et~al.}(2009)\citenamefont
  {Richter}, \citenamefont {Carmele}, \citenamefont {Sitek},\ and\
  \citenamefont {Knorr}}]{PhysRevLett.103.087407}%
  \BibitemOpen
  \bibfield  {author} {\bibinfo {author} {\bibfnamefont {M.}~\bibnamefont
  {Richter}}, \bibinfo {author} {\bibfnamefont {A.}~\bibnamefont {Carmele}},
  \bibinfo {author} {\bibfnamefont {A.}~\bibnamefont {Sitek}}, \ and\ \bibinfo
  {author} {\bibfnamefont {A.}~\bibnamefont {Knorr}},\ }\href {\doibase
  10.1103/PhysRevLett.103.087407} {\bibfield  {journal} {\bibinfo  {journal}
  {Phys. Rev. Lett.}\ }\textbf {\bibinfo {volume} {103}},\ \bibinfo {pages}
  {087407} (\bibinfo {year} {2009})}\BibitemShut {NoStop}%
\bibitem [{\citenamefont {Rozbicki}\ and\ \citenamefont
  {Machnikowski}(2008)}]{PhysRevLett.100.027401}%
  \BibitemOpen
  \bibfield  {author} {\bibinfo {author} {\bibfnamefont {E.}~\bibnamefont
  {Rozbicki}}\ and\ \bibinfo {author} {\bibfnamefont {P.}~\bibnamefont
  {Machnikowski}},\ }\href {\doibase 10.1103/PhysRevLett.100.027401} {\bibfield
   {journal} {\bibinfo  {journal} {Phys. Rev. Lett.}\ }\textbf {\bibinfo
  {volume} {100}},\ \bibinfo {pages} {027401} (\bibinfo {year}
  {2008})}\BibitemShut {NoStop}%
\bibitem [{\citenamefont {L\"uker}\ \emph {et~al.}(2017)\citenamefont
  {L\"uker}, \citenamefont {Kuhn},\ and\ \citenamefont {Reiter}}]{reiter}%
  \BibitemOpen
  \bibfield  {author} {\bibinfo {author} {\bibfnamefont {S.}~\bibnamefont
  {L\"uker}}, \bibinfo {author} {\bibfnamefont {T.}~\bibnamefont {Kuhn}}, \
  and\ \bibinfo {author} {\bibfnamefont {D.~E.}\ \bibnamefont {Reiter}},\
  }\href {\doibase 10.1103/PhysRevB.96.245306} {\bibfield  {journal} {\bibinfo
  {journal} {Phys. Rev. B}\ }\textbf {\bibinfo {volume} {96}},\ \bibinfo
  {pages} {245306} (\bibinfo {year} {2017})}\BibitemShut {NoStop}%
\bibitem [{\citenamefont {Wilson-Rae}\ and\ \citenamefont
  {Imamo\ifmmode~\breve{g}\else \u{g}\fi{}lu}(2002)}]{wilsonPMEQ}%
  \BibitemOpen
  \bibfield  {author} {\bibinfo {author} {\bibfnamefont {I.}~\bibnamefont
  {Wilson-Rae}}\ and\ \bibinfo {author} {\bibfnamefont {A.}~\bibnamefont
  {Imamo\ifmmode~\breve{g}\else \u{g}\fi{}lu}},\ }\href {\doibase
  10.1103/PhysRevB.65.235311} {\bibfield  {journal} {\bibinfo  {journal} {Phys.
  Rev. B}\ }\textbf {\bibinfo {volume} {65}},\ \bibinfo {pages} {235311}
  (\bibinfo {year} {2002})}\BibitemShut {NoStop}%
\bibitem [{\citenamefont {Laussy}\ \emph {et~al.}(2012)\citenamefont {Laussy},
  \citenamefont {Valle}, \citenamefont {M�nchen}, \citenamefont {Laucht},
  \citenamefont {Gonzalez-Tudela}, \citenamefont {Kaniber}, \citenamefont
  {Finley},\ and\ \citenamefont {Tejedor}}]{laussy}%
  \BibitemOpen
  \bibfield  {author} {\bibinfo {author} {\bibfnamefont {F.}~\bibnamefont
  {Laussy}}, \bibinfo {author} {\bibfnamefont {E.~D.}\ \bibnamefont {Valle}},
  \bibinfo {author} {\bibfnamefont {T.}~\bibnamefont {M�nchen}}, \bibinfo
  {author} {\bibfnamefont {A.}~\bibnamefont {Laucht}}, \bibinfo {author}
  {\bibfnamefont {A.}~\bibnamefont {Gonzalez-Tudela}}, \bibinfo {author}
  {\bibfnamefont {M.}~\bibnamefont {Kaniber}}, \bibinfo {author} {\bibfnamefont
  {J.}~\bibnamefont {Finley}}, \ and\ \bibinfo {author} {\bibfnamefont
  {C.}~\bibnamefont {Tejedor}},\ }in\ \href {\doibase
  https://doi.org/10.1533/9780857096395.3.293} {\emph {\bibinfo {booktitle}
  {Quantum Optics with Semiconductor Nanostructures}}},\ \bibinfo {series and
  number} {Woodhead Publishing Series in Electronic and Optical Materials},\
  \bibinfo {editor} {edited by\ \bibinfo {editor} {\bibfnamefont
  {F.}~\bibnamefont {Jahnke}}}\ (\bibinfo  {publisher} {Woodhead Publishing},\
  \bibinfo {year} {2012})\ pp.\ \bibinfo {pages} {293 -- 331}\BibitemShut
  {NoStop}%
\bibitem [{\citenamefont {Laussy}\ \emph {et~al.}(2011)\citenamefont {Laussy},
  \citenamefont {Laucht}, \citenamefont {del Valle}, \citenamefont {Finley},\
  and\ \citenamefont {Villas-B\^oas}}]{PhysRevB.84.195313}%
  \BibitemOpen
  \bibfield  {author} {\bibinfo {author} {\bibfnamefont {F.~P.}\ \bibnamefont
  {Laussy}}, \bibinfo {author} {\bibfnamefont {A.}~\bibnamefont {Laucht}},
  \bibinfo {author} {\bibfnamefont {E.}~\bibnamefont {del Valle}}, \bibinfo
  {author} {\bibfnamefont {J.~J.}\ \bibnamefont {Finley}}, \ and\ \bibinfo
  {author} {\bibfnamefont {J.~M.}\ \bibnamefont {Villas-B\^oas}},\ }\href
  {\doibase 10.1103/PhysRevB.84.195313} {\bibfield  {journal} {\bibinfo
  {journal} {Phys. Rev. B}\ }\textbf {\bibinfo {volume} {84}},\ \bibinfo
  {pages} {195313} (\bibinfo {year} {2011})}\BibitemShut {NoStop}%
\bibitem [{\citenamefont {del Valle}(2010)}]{delvalle_phd}%
  \BibitemOpen
  \bibfield  {author} {\bibinfo {author} {\bibfnamefont {E.}~\bibnamefont {del
  Valle}},\ }\href@noop {} {\bibfield  {journal} {\bibinfo  {journal}
  {Saarbrcken: VDM Verlag}\ } (\bibinfo {year} {2010})}\BibitemShut {NoStop}%
\bibitem [{\citenamefont {del Valle}\ and\ \citenamefont
  {Laussy}(2010)}]{PhysRevLett.105.233601}%
  \BibitemOpen
  \bibfield  {author} {\bibinfo {author} {\bibfnamefont {E.}~\bibnamefont {del
  Valle}}\ and\ \bibinfo {author} {\bibfnamefont {F.~P.}\ \bibnamefont
  {Laussy}},\ }\href {\doibase 10.1103/PhysRevLett.105.233601} {\bibfield
  {journal} {\bibinfo  {journal} {Phys. Rev. Lett.}\ }\textbf {\bibinfo
  {volume} {105}},\ \bibinfo {pages} {233601} (\bibinfo {year}
  {2010})}\BibitemShut {NoStop}%
\bibitem [{\citenamefont {Kreinberg}\ \emph {et~al.}(2018)\citenamefont
  {Kreinberg}, \citenamefont {Grbe{\v{s}}i{\'c}}, \citenamefont {Strau{\ss}},
  \citenamefont {Carmele}, \citenamefont {Emmerling}, \citenamefont
  {Schneider}, \citenamefont {H{\"o}fling}, \citenamefont {Porte},\ and\
  \citenamefont {Reitzenstein}}]{kreinberg2018quantum}%
  \BibitemOpen
  \bibfield  {author} {\bibinfo {author} {\bibfnamefont {S.}~\bibnamefont
  {Kreinberg}}, \bibinfo {author} {\bibfnamefont {T.}~\bibnamefont
  {Grbe{\v{s}}i{\'c}}}, \bibinfo {author} {\bibfnamefont {M.}~\bibnamefont
  {Strau{\ss}}}, \bibinfo {author} {\bibfnamefont {A.}~\bibnamefont {Carmele}},
  \bibinfo {author} {\bibfnamefont {M.}~\bibnamefont {Emmerling}}, \bibinfo
  {author} {\bibfnamefont {C.}~\bibnamefont {Schneider}}, \bibinfo {author}
  {\bibfnamefont {S.}~\bibnamefont {H{\"o}fling}}, \bibinfo {author}
  {\bibfnamefont {X.}~\bibnamefont {Porte}}, \ and\ \bibinfo {author}
  {\bibfnamefont {S.}~\bibnamefont {Reitzenstein}},\ }\href@noop {} {\bibfield
  {journal} {\bibinfo  {journal} {Light, Science \& Applications}\ }\textbf
  {\bibinfo {volume} {7}} (\bibinfo {year} {2018})}\BibitemShut {NoStop}%
\bibitem [{\citenamefont {McCutcheon}\ \emph {et~al.}(2011)\citenamefont
  {McCutcheon}, \citenamefont {Dattani}, \citenamefont {Gauger}, \citenamefont
  {Lovett},\ and\ \citenamefont {Nazir}}]{nazirPMEQ}%
  \BibitemOpen
  \bibfield  {author} {\bibinfo {author} {\bibfnamefont {D.~P.~S.}\
  \bibnamefont {McCutcheon}}, \bibinfo {author} {\bibfnamefont {N.~S.}\
  \bibnamefont {Dattani}}, \bibinfo {author} {\bibfnamefont {E.~M.}\
  \bibnamefont {Gauger}}, \bibinfo {author} {\bibfnamefont {B.~W.}\
  \bibnamefont {Lovett}}, \ and\ \bibinfo {author} {\bibfnamefont
  {A.}~\bibnamefont {Nazir}},\ }\href {\doibase 10.1103/PhysRevB.84.081305}
  {\bibfield  {journal} {\bibinfo  {journal} {Phys. Rev. B}\ }\textbf {\bibinfo
  {volume} {84}},\ \bibinfo {pages} {081305} (\bibinfo {year}
  {2011})}\BibitemShut {NoStop}%
\bibitem [{\citenamefont {Manson}\ \emph {et~al.}(2016)\citenamefont {Manson},
  \citenamefont {Roy-Choudhury},\ and\ \citenamefont {Hughes}}]{mansonPMEQ}%
  \BibitemOpen
  \bibfield  {author} {\bibinfo {author} {\bibfnamefont {R.}~\bibnamefont
  {Manson}}, \bibinfo {author} {\bibfnamefont {K.}~\bibnamefont
  {Roy-Choudhury}}, \ and\ \bibinfo {author} {\bibfnamefont {S.}~\bibnamefont
  {Hughes}},\ }\href {\doibase 10.1103/PhysRevB.93.155423} {\bibfield
  {journal} {\bibinfo  {journal} {Phys. Rev. B}\ }\textbf {\bibinfo {volume}
  {93}},\ \bibinfo {pages} {155423} (\bibinfo {year} {2016})}\BibitemShut
  {NoStop}%
\bibitem [{\citenamefont {Kabuss}\ \emph
  {et~al.}(2011{\natexlab{a}})\citenamefont {Kabuss}, \citenamefont {Carmele},
  \citenamefont {Richter}, \citenamefont {Chow},\ and\ \citenamefont
  {Knorr}}]{kabuss2011inductive}%
  \BibitemOpen
  \bibfield  {author} {\bibinfo {author} {\bibfnamefont {J.}~\bibnamefont
  {Kabuss}}, \bibinfo {author} {\bibfnamefont {A.}~\bibnamefont {Carmele}},
  \bibinfo {author} {\bibfnamefont {M.}~\bibnamefont {Richter}}, \bibinfo
  {author} {\bibfnamefont {W.~W.}\ \bibnamefont {Chow}}, \ and\ \bibinfo
  {author} {\bibfnamefont {A.}~\bibnamefont {Knorr}},\ }\href@noop {}
  {\bibfield  {journal} {\bibinfo  {journal} {physica status solidi (b)}\
  }\textbf {\bibinfo {volume} {248}},\ \bibinfo {pages} {872} (\bibinfo {year}
  {2011}{\natexlab{a}})}\BibitemShut {NoStop}%
\bibitem [{\citenamefont {Axt}\ \emph {et~al.}(1999)\citenamefont {Axt},
  \citenamefont {Herbst},\ and\ \citenamefont {Kuhn}}]{axt1999coherent}%
  \BibitemOpen
  \bibfield  {author} {\bibinfo {author} {\bibfnamefont {V.}~\bibnamefont
  {Axt}}, \bibinfo {author} {\bibfnamefont {M.}~\bibnamefont {Herbst}}, \ and\
  \bibinfo {author} {\bibfnamefont {T.}~\bibnamefont {Kuhn}},\ }\href@noop {}
  {\bibfield  {journal} {\bibinfo  {journal} {Superlattices and
  Microstructures}\ }\textbf {\bibinfo {volume} {26}},\ \bibinfo {pages} {117}
  (\bibinfo {year} {1999})}\BibitemShut {NoStop}%
\bibitem [{\citenamefont {Kabuss}\ \emph
  {et~al.}(2011{\natexlab{b}})\citenamefont {Kabuss}, \citenamefont {Carmele},
  \citenamefont {Richter},\ and\ \citenamefont {Knorr}}]{kabuss}%
  \BibitemOpen
  \bibfield  {author} {\bibinfo {author} {\bibfnamefont {J.}~\bibnamefont
  {Kabuss}}, \bibinfo {author} {\bibfnamefont {A.}~\bibnamefont {Carmele}},
  \bibinfo {author} {\bibfnamefont {M.}~\bibnamefont {Richter}}, \ and\
  \bibinfo {author} {\bibfnamefont {A.}~\bibnamefont {Knorr}},\ }\href
  {\doibase 10.1103/PhysRevB.84.125324} {\bibfield  {journal} {\bibinfo
  {journal} {Phys. Rev. B}\ }\textbf {\bibinfo {volume} {84}},\ \bibinfo
  {pages} {125324} (\bibinfo {year} {2011}{\natexlab{b}})}\BibitemShut
  {NoStop}%
\bibitem [{\citenamefont {Gl\"assl}\ \emph
  {et~al.}(2011{\natexlab{a}})\citenamefont {Gl\"assl}, \citenamefont {Vagov},
  \citenamefont {L\"uker}, \citenamefont {Reiter}, \citenamefont {Croitoru},
  \citenamefont {Machnikowski}, \citenamefont {Axt},\ and\ \citenamefont
  {Kuhn}}]{clustervspathintegral}%
  \BibitemOpen
  \bibfield  {author} {\bibinfo {author} {\bibfnamefont {M.}~\bibnamefont
  {Gl\"assl}}, \bibinfo {author} {\bibfnamefont {A.}~\bibnamefont {Vagov}},
  \bibinfo {author} {\bibfnamefont {S.}~\bibnamefont {L\"uker}}, \bibinfo
  {author} {\bibfnamefont {D.~E.}\ \bibnamefont {Reiter}}, \bibinfo {author}
  {\bibfnamefont {M.~D.}\ \bibnamefont {Croitoru}}, \bibinfo {author}
  {\bibfnamefont {P.}~\bibnamefont {Machnikowski}}, \bibinfo {author}
  {\bibfnamefont {V.~M.}\ \bibnamefont {Axt}}, \ and\ \bibinfo {author}
  {\bibfnamefont {T.}~\bibnamefont {Kuhn}},\ }\href {\doibase
  10.1103/PhysRevB.84.195311} {\bibfield  {journal} {\bibinfo  {journal} {Phys.
  Rev. B}\ }\textbf {\bibinfo {volume} {84}},\ \bibinfo {pages} {195311}
  (\bibinfo {year} {2011}{\natexlab{a}})}\BibitemShut {NoStop}%
\bibitem [{\citenamefont {Schollw{\"o}ck}(2011)}]{schollwock2011density}%
  \BibitemOpen
  \bibfield  {author} {\bibinfo {author} {\bibfnamefont {U.}~\bibnamefont
  {Schollw{\"o}ck}},\ }\href@noop {} {\bibfield  {journal} {\bibinfo  {journal}
  {Annals of Physics}\ }\textbf {\bibinfo {volume} {326}},\ \bibinfo {pages}
  {96} (\bibinfo {year} {2011})}\BibitemShut {NoStop}%
\bibitem [{\citenamefont {Caldeira}\ and\ \citenamefont
  {Leggett}(1983)}]{caldeira1983path}%
  \BibitemOpen
  \bibfield  {author} {\bibinfo {author} {\bibfnamefont {A.~O.}\ \bibnamefont
  {Caldeira}}\ and\ \bibinfo {author} {\bibfnamefont {A.~J.}\ \bibnamefont
  {Leggett}},\ }\href@noop {} {\bibfield  {journal} {\bibinfo  {journal}
  {Physica A: Statistical mechanics and its Applications}\ }\textbf {\bibinfo
  {volume} {121}},\ \bibinfo {pages} {587} (\bibinfo {year}
  {1983})}\BibitemShut {NoStop}%
\bibitem [{\citenamefont {Vidal}(2004)}]{vidal_mps}%
  \BibitemOpen
  \bibfield  {author} {\bibinfo {author} {\bibfnamefont {G.}~\bibnamefont
  {Vidal}},\ }\href {\doibase 10.1103/PhysRevLett.93.040502} {\bibfield
  {journal} {\bibinfo  {journal} {Phys. Rev. Lett.}\ }\textbf {\bibinfo
  {volume} {93}},\ \bibinfo {pages} {040502} (\bibinfo {year}
  {2004})}\BibitemShut {NoStop}%
\bibitem [{\citenamefont {del Pino}\ \emph {et~al.}(2018)\citenamefont {del
  Pino}, \citenamefont {Schr\"oder}, \citenamefont {Chin}, \citenamefont
  {Feist},\ and\ \citenamefont {Garcia-Vidal}}]{phonon_vidal}%
  \BibitemOpen
  \bibfield  {author} {\bibinfo {author} {\bibfnamefont {J.}~\bibnamefont {del
  Pino}}, \bibinfo {author} {\bibfnamefont {F.~A. Y.~N.}\ \bibnamefont
  {Schr\"oder}}, \bibinfo {author} {\bibfnamefont {A.~W.}\ \bibnamefont
  {Chin}}, \bibinfo {author} {\bibfnamefont {J.}~\bibnamefont {Feist}}, \ and\
  \bibinfo {author} {\bibfnamefont {F.~J.}\ \bibnamefont {Garcia-Vidal}},\
  }\href {\doibase 10.1103/PhysRevLett.121.227401} {\bibfield  {journal}
  {\bibinfo  {journal} {Phys. Rev. Lett.}\ }\textbf {\bibinfo {volume} {121}},\
  \bibinfo {pages} {227401} (\bibinfo {year} {2018})}\BibitemShut {NoStop}%
\bibitem [{\citenamefont {Strathearn}\ \emph {et~al.}(2018)\citenamefont
  {Strathearn}, \citenamefont {Kirton}, \citenamefont {Kilda}, \citenamefont
  {Keeling},\ and\ \citenamefont {Lovett}}]{strathearn2018efficient}%
  \BibitemOpen
  \bibfield  {author} {\bibinfo {author} {\bibfnamefont {A.}~\bibnamefont
  {Strathearn}}, \bibinfo {author} {\bibfnamefont {P.}~\bibnamefont {Kirton}},
  \bibinfo {author} {\bibfnamefont {D.}~\bibnamefont {Kilda}}, \bibinfo
  {author} {\bibfnamefont {J.}~\bibnamefont {Keeling}}, \ and\ \bibinfo
  {author} {\bibfnamefont {B.~W.}\ \bibnamefont {Lovett}},\ }\href@noop {}
  {\bibfield  {journal} {\bibinfo  {journal} {Nature communications}\ }\textbf
  {\bibinfo {volume} {9}},\ \bibinfo {pages} {3322} (\bibinfo {year}
  {2018})}\BibitemShut {NoStop}%
\bibitem [{\citenamefont {Droenner}\ \emph {et~al.}(2018)\citenamefont
  {Droenner}, \citenamefont {Naumann}, \citenamefont {Knorr},\ and\
  \citenamefont {Carmele}}]{droenner2018two}%
  \BibitemOpen
  \bibfield  {author} {\bibinfo {author} {\bibfnamefont {L.}~\bibnamefont
  {Droenner}}, \bibinfo {author} {\bibfnamefont {N.~L.}\ \bibnamefont
  {Naumann}}, \bibinfo {author} {\bibfnamefont {A.}~\bibnamefont {Knorr}}, \
  and\ \bibinfo {author} {\bibfnamefont {A.}~\bibnamefont {Carmele}},\
  }\href@noop {} {\bibfield  {journal} {\bibinfo  {journal} {arXiv preprint
  arXiv:1801.03342}\ } (\bibinfo {year} {2018})}\BibitemShut {NoStop}%
\bibitem [{\citenamefont {Makri}(1998)}]{makri1998quantum}%
  \BibitemOpen
  \bibfield  {author} {\bibinfo {author} {\bibfnamefont {N.}~\bibnamefont
  {Makri}},\ }\href@noop {} {\bibfield  {journal} {\bibinfo  {journal} {The
  Journal of Physical Chemistry A}\ }\textbf {\bibinfo {volume} {102}},\
  \bibinfo {pages} {4414} (\bibinfo {year} {1998})}\BibitemShut {NoStop}%
\bibitem [{\citenamefont {Cygorek}\ \emph {et~al.}(2017)\citenamefont
  {Cygorek}, \citenamefont {Barth}, \citenamefont {Ungar}, \citenamefont
  {Vagov},\ and\ \citenamefont {Axt}}]{cygorek2017nonlinear}%
  \BibitemOpen
  \bibfield  {author} {\bibinfo {author} {\bibfnamefont {M.}~\bibnamefont
  {Cygorek}}, \bibinfo {author} {\bibfnamefont {A.~M.}\ \bibnamefont {Barth}},
  \bibinfo {author} {\bibfnamefont {F.}~\bibnamefont {Ungar}}, \bibinfo
  {author} {\bibfnamefont {A.}~\bibnamefont {Vagov}}, \ and\ \bibinfo {author}
  {\bibfnamefont {V.~M.}\ \bibnamefont {Axt}},\ }\href@noop {} {\bibfield
  {journal} {\bibinfo  {journal} {Physical Review B}\ }\textbf {\bibinfo
  {volume} {96}},\ \bibinfo {pages} {201201} (\bibinfo {year}
  {2017})}\BibitemShut {NoStop}%
\bibitem [{\citenamefont {Hopfmann}\ \emph {et~al.}(2015)\citenamefont
  {Hopfmann}, \citenamefont {Musia\l{}}, \citenamefont {Strau\ss{}},
  \citenamefont {Barth}, \citenamefont {Gl\"assl}, \citenamefont {Vagov},
  \citenamefont {Strau\ss{}}, \citenamefont {Schneider}, \citenamefont
  {H\"ofling}, \citenamefont {Kamp}, \citenamefont {Axt},\ and\ \citenamefont
  {Reitzenstein}}]{axt_hopfmann_prb}%
  \BibitemOpen
  \bibfield  {author} {\bibinfo {author} {\bibfnamefont {C.}~\bibnamefont
  {Hopfmann}}, \bibinfo {author} {\bibfnamefont {A.}~\bibnamefont {Musia\l{}}},
  \bibinfo {author} {\bibfnamefont {M.}~\bibnamefont {Strau\ss{}}}, \bibinfo
  {author} {\bibfnamefont {A.~M.}\ \bibnamefont {Barth}}, \bibinfo {author}
  {\bibfnamefont {M.}~\bibnamefont {Gl\"assl}}, \bibinfo {author}
  {\bibfnamefont {A.}~\bibnamefont {Vagov}}, \bibinfo {author} {\bibfnamefont
  {M.}~\bibnamefont {Strau\ss{}}}, \bibinfo {author} {\bibfnamefont
  {C.}~\bibnamefont {Schneider}}, \bibinfo {author} {\bibfnamefont
  {S.}~\bibnamefont {H\"ofling}}, \bibinfo {author} {\bibfnamefont
  {M.}~\bibnamefont {Kamp}}, \bibinfo {author} {\bibfnamefont {V.~M.}\
  \bibnamefont {Axt}}, \ and\ \bibinfo {author} {\bibfnamefont
  {S.}~\bibnamefont {Reitzenstein}},\ }\href {\doibase
  10.1103/PhysRevB.92.245403} {\bibfield  {journal} {\bibinfo  {journal} {Phys.
  Rev. B}\ }\textbf {\bibinfo {volume} {92}},\ \bibinfo {pages} {245403}
  (\bibinfo {year} {2015})}\BibitemShut {NoStop}%
\bibitem [{\citenamefont {Jackson}(2012)}]{jackson2012classical}%
  \BibitemOpen
  \bibfield  {author} {\bibinfo {author} {\bibfnamefont {J.~D.}\ \bibnamefont
  {Jackson}},\ }\href@noop {} {\emph {\bibinfo {title} {Classical
  electrodynamics}}}\ (\bibinfo  {publisher} {John Wiley \& Sons},\ \bibinfo
  {year} {2012})\BibitemShut {NoStop}%
\bibitem [{\citenamefont {Allen}\ and\ \citenamefont
  {Eberly}(1987)}]{allen1987optical}%
  \BibitemOpen
  \bibfield  {author} {\bibinfo {author} {\bibfnamefont {L.}~\bibnamefont
  {Allen}}\ and\ \bibinfo {author} {\bibfnamefont {J.~H.}\ \bibnamefont
  {Eberly}},\ }\href@noop {} {\emph {\bibinfo {title} {Optical resonance and
  two-level atoms}}},\ Vol.~\bibinfo {volume} {28}\ (\bibinfo  {publisher}
  {Courier Corporation},\ \bibinfo {year} {1987})\BibitemShut {NoStop}%
\bibitem [{\citenamefont {Gardiner}(2009)}]{gardiner2009stochastic}%
  \BibitemOpen
  \bibfield  {author} {\bibinfo {author} {\bibfnamefont {C.}~\bibnamefont
  {Gardiner}},\ }\href@noop {} {\emph {\bibinfo {title} {Stochastic
  methods}}},\ Vol.~\bibinfo {volume} {4}\ (\bibinfo  {publisher} {springer
  Berlin},\ \bibinfo {year} {2009})\BibitemShut {NoStop}%
\bibitem [{\citenamefont {Besombes}\ \emph
  {et~al.}(2001{\natexlab{b}})\citenamefont {Besombes}, \citenamefont {Kheng},
  \citenamefont {Marsal},\ and\ \citenamefont
  {Mariette}}]{QD_lum_besombes2001acoustic}%
  \BibitemOpen
  \bibfield  {author} {\bibinfo {author} {\bibfnamefont {L.}~\bibnamefont
  {Besombes}}, \bibinfo {author} {\bibfnamefont {K.}~\bibnamefont {Kheng}},
  \bibinfo {author} {\bibfnamefont {L.}~\bibnamefont {Marsal}}, \ and\ \bibinfo
  {author} {\bibfnamefont {H.}~\bibnamefont {Mariette}},\ }\href@noop {}
  {\bibfield  {journal} {\bibinfo  {journal} {Physical Review B}\ }\textbf
  {\bibinfo {volume} {63}},\ \bibinfo {pages} {155307} (\bibinfo {year}
  {2001}{\natexlab{b}})}\BibitemShut {NoStop}%
\bibitem [{\citenamefont {Jakubczyk}\ \emph {et~al.}(2016)\citenamefont
  {Jakubczyk}, \citenamefont {Delmonte}, \citenamefont {Fischbach},
  \citenamefont {Wigger}, \citenamefont {Reiter}, \citenamefont {Mermillod},
  \citenamefont {Schnauber}, \citenamefont {Kaganskiy}, \citenamefont
  {Schulze}, \citenamefont {Strittmatter} \emph
  {et~al.}}]{QD_lum_jakubczyk2016impact}%
  \BibitemOpen
  \bibfield  {author} {\bibinfo {author} {\bibfnamefont {T.}~\bibnamefont
  {Jakubczyk}}, \bibinfo {author} {\bibfnamefont {V.}~\bibnamefont {Delmonte}},
  \bibinfo {author} {\bibfnamefont {S.}~\bibnamefont {Fischbach}}, \bibinfo
  {author} {\bibfnamefont {D.}~\bibnamefont {Wigger}}, \bibinfo {author}
  {\bibfnamefont {D.~E.}\ \bibnamefont {Reiter}}, \bibinfo {author}
  {\bibfnamefont {Q.}~\bibnamefont {Mermillod}}, \bibinfo {author}
  {\bibfnamefont {P.}~\bibnamefont {Schnauber}}, \bibinfo {author}
  {\bibfnamefont {A.}~\bibnamefont {Kaganskiy}}, \bibinfo {author}
  {\bibfnamefont {J.-H.}\ \bibnamefont {Schulze}}, \bibinfo {author}
  {\bibfnamefont {A.}~\bibnamefont {Strittmatter}},  \emph {et~al.},\
  }\href@noop {} {\bibfield  {journal} {\bibinfo  {journal} {ACS photonics}\
  }\textbf {\bibinfo {volume} {3}},\ \bibinfo {pages} {2461} (\bibinfo {year}
  {2016})}\BibitemShut {NoStop}%
\bibitem [{\citenamefont {Fras}\ \emph {et~al.}(2016)\citenamefont {Fras},
  \citenamefont {Mermillod}, \citenamefont {Nogues}, \citenamefont {Hoarau},
  \citenamefont {Schneider}, \citenamefont {Kamp}, \citenamefont {H{\"o}fling},
  \citenamefont {Langbein},\ and\ \citenamefont
  {Kasprzak}}]{langbein_fourwave_mixing}%
  \BibitemOpen
  \bibfield  {author} {\bibinfo {author} {\bibfnamefont {F.}~\bibnamefont
  {Fras}}, \bibinfo {author} {\bibfnamefont {Q.}~\bibnamefont {Mermillod}},
  \bibinfo {author} {\bibfnamefont {G.}~\bibnamefont {Nogues}}, \bibinfo
  {author} {\bibfnamefont {C.}~\bibnamefont {Hoarau}}, \bibinfo {author}
  {\bibfnamefont {C.}~\bibnamefont {Schneider}}, \bibinfo {author}
  {\bibfnamefont {M.}~\bibnamefont {Kamp}}, \bibinfo {author} {\bibfnamefont
  {S.}~\bibnamefont {H{\"o}fling}}, \bibinfo {author} {\bibfnamefont
  {W.}~\bibnamefont {Langbein}}, \ and\ \bibinfo {author} {\bibfnamefont
  {J.}~\bibnamefont {Kasprzak}},\ }\href@noop {} {\bibfield  {journal}
  {\bibinfo  {journal} {Nature Photonics}\ }\textbf {\bibinfo {volume} {10}},\
  \bibinfo {pages} {155} (\bibinfo {year} {2016})}\BibitemShut {NoStop}%
\bibitem [{\citenamefont {Madelung}(2012)}]{madelung2012introduction}%
  \BibitemOpen
  \bibfield  {author} {\bibinfo {author} {\bibfnamefont {O.}~\bibnamefont
  {Madelung}},\ }\href@noop {} {\emph {\bibinfo {title} {Introduction to
  solid-state theory}}},\ Vol.~\bibinfo {volume} {2}\ (\bibinfo  {publisher}
  {Springer Science \& Business Media},\ \bibinfo {year} {2012})\BibitemShut
  {NoStop}%
\bibitem [{\citenamefont {Pullerits}\ \emph {et~al.}(1995)\citenamefont
  {Pullerits}, \citenamefont {Monshouwer}, \citenamefont {van Mourik},\ and\
  \citenamefont {van Grondelle}}]{pullerits1995temperature}%
  \BibitemOpen
  \bibfield  {author} {\bibinfo {author} {\bibfnamefont {T.}~\bibnamefont
  {Pullerits}}, \bibinfo {author} {\bibfnamefont {R.}~\bibnamefont
  {Monshouwer}}, \bibinfo {author} {\bibfnamefont {F.}~\bibnamefont {van
  Mourik}}, \ and\ \bibinfo {author} {\bibfnamefont {R.}~\bibnamefont {van
  Grondelle}},\ }\href@noop {} {\bibfield  {journal} {\bibinfo  {journal}
  {Chemical physics}\ }\textbf {\bibinfo {volume} {194}},\ \bibinfo {pages}
  {395} (\bibinfo {year} {1995})}\BibitemShut {NoStop}%
\bibitem [{\citenamefont {Cohen-Tannoudji}\ and\ \citenamefont
  {Gu{\'e}ry-Odelin}(2011)}]{cohen2011advances}%
  \BibitemOpen
  \bibfield  {author} {\bibinfo {author} {\bibfnamefont {C.}~\bibnamefont
  {Cohen-Tannoudji}}\ and\ \bibinfo {author} {\bibfnamefont {D.}~\bibnamefont
  {Gu{\'e}ry-Odelin}},\ }\href@noop {} {\emph {\bibinfo {title} {Advances in
  atomic physics: an overview}}}\ (\bibinfo  {publisher} {World Scientific},\
  \bibinfo {year} {2011})\BibitemShut {NoStop}%
\bibitem [{\citenamefont {Gardiner}\ and\ \citenamefont
  {Zoller}(2015)}]{gardiner2015quantum}%
  \BibitemOpen
  \bibfield  {author} {\bibinfo {author} {\bibfnamefont {C.}~\bibnamefont
  {Gardiner}}\ and\ \bibinfo {author} {\bibfnamefont {P.}~\bibnamefont
  {Zoller}},\ }in\ \href@noop {} {\emph {\bibinfo {booktitle} {The Quantum
  World of Ultra-Cold Atoms and Light Book II: The Physics of Quantum-Optical
  Devices}}}\ (\bibinfo  {publisher} {World Scientific},\ \bibinfo {year}
  {2015})\ pp.\ \bibinfo {pages} {1--524}\BibitemShut {NoStop}%
\bibitem [{\citenamefont {McCutcheon}\ and\ \citenamefont
  {Nazir}(2010)}]{mccutcheon2010quantum}%
  \BibitemOpen
  \bibfield  {author} {\bibinfo {author} {\bibfnamefont {D.~P.}\ \bibnamefont
  {McCutcheon}}\ and\ \bibinfo {author} {\bibfnamefont {A.}~\bibnamefont
  {Nazir}},\ }\href@noop {} {\bibfield  {journal} {\bibinfo  {journal} {New
  Journal of Physics}\ }\textbf {\bibinfo {volume} {12}},\ \bibinfo {pages}
  {113042} (\bibinfo {year} {2010})}\BibitemShut {NoStop}%
\bibitem [{\citenamefont {Kabuss}\ \emph {et~al.}(2010)\citenamefont {Kabuss},
  \citenamefont {Werner}, \citenamefont {Hoffmann}, \citenamefont
  {Hildebrandt}, \citenamefont {Knorr},\ and\ \citenamefont
  {Richter}}]{julia_raman}%
  \BibitemOpen
  \bibfield  {author} {\bibinfo {author} {\bibfnamefont {J.}~\bibnamefont
  {Kabuss}}, \bibinfo {author} {\bibfnamefont {S.}~\bibnamefont {Werner}},
  \bibinfo {author} {\bibfnamefont {A.}~\bibnamefont {Hoffmann}}, \bibinfo
  {author} {\bibfnamefont {P.}~\bibnamefont {Hildebrandt}}, \bibinfo {author}
  {\bibfnamefont {A.}~\bibnamefont {Knorr}}, \ and\ \bibinfo {author}
  {\bibfnamefont {M.}~\bibnamefont {Richter}},\ }\href {\doibase
  10.1103/PhysRevB.81.075314} {\bibfield  {journal} {\bibinfo  {journal} {Phys.
  Rev. B}\ }\textbf {\bibinfo {volume} {81}},\ \bibinfo {pages} {075314}
  (\bibinfo {year} {2010})}\BibitemShut {NoStop}%
\bibitem [{\citenamefont {Carmele}\ \emph {et~al.}(2009)\citenamefont
  {Carmele}, \citenamefont {Knorr},\ and\ \citenamefont
  {Richter}}]{carmele2009photon}%
  \BibitemOpen
  \bibfield  {author} {\bibinfo {author} {\bibfnamefont {A.}~\bibnamefont
  {Carmele}}, \bibinfo {author} {\bibfnamefont {A.}~\bibnamefont {Knorr}}, \
  and\ \bibinfo {author} {\bibfnamefont {M.}~\bibnamefont {Richter}},\
  }\href@noop {} {\bibfield  {journal} {\bibinfo  {journal} {Physical Review
  B}\ }\textbf {\bibinfo {volume} {79}},\ \bibinfo {pages} {035316} (\bibinfo
  {year} {2009})}\BibitemShut {NoStop}%
\bibitem [{\citenamefont {Ramsay}\ \emph
  {et~al.}(2010{\natexlab{b}})\citenamefont {Ramsay}, \citenamefont {Gopal},
  \citenamefont {Gauger}, \citenamefont {Nazir}, \citenamefont {Lovett},
  \citenamefont {Fox},\ and\ \citenamefont
  {Skolnick}}]{QD_rot_PhysRevLett.104.017402}%
  \BibitemOpen
  \bibfield  {author} {\bibinfo {author} {\bibfnamefont {A.~J.}\ \bibnamefont
  {Ramsay}}, \bibinfo {author} {\bibfnamefont {A.~V.}\ \bibnamefont {Gopal}},
  \bibinfo {author} {\bibfnamefont {E.~M.}\ \bibnamefont {Gauger}}, \bibinfo
  {author} {\bibfnamefont {A.}~\bibnamefont {Nazir}}, \bibinfo {author}
  {\bibfnamefont {B.~W.}\ \bibnamefont {Lovett}}, \bibinfo {author}
  {\bibfnamefont {A.~M.}\ \bibnamefont {Fox}}, \ and\ \bibinfo {author}
  {\bibfnamefont {M.~S.}\ \bibnamefont {Skolnick}},\ }\href {\doibase
  10.1103/PhysRevLett.104.017402} {\bibfield  {journal} {\bibinfo  {journal}
  {Phys. Rev. Lett.}\ }\textbf {\bibinfo {volume} {104}},\ \bibinfo {pages}
  {017402} (\bibinfo {year} {2010}{\natexlab{b}})}\BibitemShut {NoStop}%
\bibitem [{\citenamefont {Gl\"assl}\ \emph
  {et~al.}(2011{\natexlab{b}})\citenamefont {Gl\"assl}, \citenamefont
  {Croitoru}, \citenamefont {Vagov}, \citenamefont {Axt},\ and\ \citenamefont
  {Kuhn}}]{QD_rot_PhysRevB.84.125304}%
  \BibitemOpen
  \bibfield  {author} {\bibinfo {author} {\bibfnamefont {M.}~\bibnamefont
  {Gl\"assl}}, \bibinfo {author} {\bibfnamefont {M.~D.}\ \bibnamefont
  {Croitoru}}, \bibinfo {author} {\bibfnamefont {A.}~\bibnamefont {Vagov}},
  \bibinfo {author} {\bibfnamefont {V.~M.}\ \bibnamefont {Axt}}, \ and\
  \bibinfo {author} {\bibfnamefont {T.}~\bibnamefont {Kuhn}},\ }\href {\doibase
  10.1103/PhysRevB.84.125304} {\bibfield  {journal} {\bibinfo  {journal} {Phys.
  Rev. B}\ }\textbf {\bibinfo {volume} {84}},\ \bibinfo {pages} {125304}
  (\bibinfo {year} {2011}{\natexlab{b}})}\BibitemShut {NoStop}%
\bibitem [{\citenamefont {Machnikowski}\ and\ \citenamefont
  {Jacak}(2004)}]{QD_rot_PhysRevB.69.193302}%
  \BibitemOpen
  \bibfield  {author} {\bibinfo {author} {\bibfnamefont {P.}~\bibnamefont
  {Machnikowski}}\ and\ \bibinfo {author} {\bibfnamefont {L.}~\bibnamefont
  {Jacak}},\ }\href {\doibase 10.1103/PhysRevB.69.193302} {\bibfield  {journal}
  {\bibinfo  {journal} {Phys. Rev. B}\ }\textbf {\bibinfo {volume} {69}},\
  \bibinfo {pages} {193302} (\bibinfo {year} {2004})}\BibitemShut {NoStop}%
\bibitem [{\citenamefont {Kr\"ugel}\ \emph {et~al.}(2006)\citenamefont
  {Kr\"ugel}, \citenamefont {Axt},\ and\ \citenamefont
  {Kuhn}}]{QD_rot_PhysRevB.73.035302}%
  \BibitemOpen
  \bibfield  {author} {\bibinfo {author} {\bibfnamefont {A.}~\bibnamefont
  {Kr\"ugel}}, \bibinfo {author} {\bibfnamefont {V.~M.}\ \bibnamefont {Axt}}, \
  and\ \bibinfo {author} {\bibfnamefont {T.}~\bibnamefont {Kuhn}},\ }\href
  {\doibase 10.1103/PhysRevB.73.035302} {\bibfield  {journal} {\bibinfo
  {journal} {Phys. Rev. B}\ }\textbf {\bibinfo {volume} {73}},\ \bibinfo
  {pages} {035302} (\bibinfo {year} {2006})}\BibitemShut {NoStop}%
\bibitem [{\citenamefont {Reiter}(2017)}]{PhysRevB.95.125308}%
  \BibitemOpen
  \bibfield  {author} {\bibinfo {author} {\bibfnamefont {D.~E.}\ \bibnamefont
  {Reiter}},\ }\href {\doibase 10.1103/PhysRevB.95.125308} {\bibfield
  {journal} {\bibinfo  {journal} {Phys. Rev. B}\ }\textbf {\bibinfo {volume}
  {95}},\ \bibinfo {pages} {125308} (\bibinfo {year} {2017})}\BibitemShut
  {NoStop}%
\bibitem [{\citenamefont {Naumann}\ \emph {et~al.}(2016)\citenamefont
  {Naumann}, \citenamefont {Droenner}, \citenamefont {Chow}, \citenamefont
  {Kabuss},\ and\ \citenamefont {Carmele}}]{naumann2016solid}%
  \BibitemOpen
  \bibfield  {author} {\bibinfo {author} {\bibfnamefont {N.~L.}\ \bibnamefont
  {Naumann}}, \bibinfo {author} {\bibfnamefont {L.}~\bibnamefont {Droenner}},
  \bibinfo {author} {\bibfnamefont {W.~W.}\ \bibnamefont {Chow}}, \bibinfo
  {author} {\bibfnamefont {J.}~\bibnamefont {Kabuss}}, \ and\ \bibinfo {author}
  {\bibfnamefont {A.}~\bibnamefont {Carmele}},\ }\href@noop {} {\bibfield
  {journal} {\bibinfo  {journal} {JOSA B}\ }\textbf {\bibinfo {volume} {33}},\
  \bibinfo {pages} {1492} (\bibinfo {year} {2016})}\BibitemShut {NoStop}%
\bibitem [{\citenamefont {Droenner}\ \emph {et~al.}(2017)\citenamefont
  {Droenner}, \citenamefont {Naumann}, \citenamefont {Kabuss},\ and\
  \citenamefont {Carmele}}]{droenner2017collective}%
  \BibitemOpen
  \bibfield  {author} {\bibinfo {author} {\bibfnamefont {L.}~\bibnamefont
  {Droenner}}, \bibinfo {author} {\bibfnamefont {N.~L.}\ \bibnamefont
  {Naumann}}, \bibinfo {author} {\bibfnamefont {J.}~\bibnamefont {Kabuss}}, \
  and\ \bibinfo {author} {\bibfnamefont {A.}~\bibnamefont {Carmele}},\
  }\href@noop {} {\bibfield  {journal} {\bibinfo  {journal} {Physical Review
  A}\ }\textbf {\bibinfo {volume} {96}},\ \bibinfo {pages} {043805} (\bibinfo
  {year} {2017})}\BibitemShut {NoStop}%
\bibitem [{\citenamefont {Kabuss}\ \emph {et~al.}(2012)\citenamefont {Kabuss},
  \citenamefont {Carmele}, \citenamefont {Brandes},\ and\ \citenamefont
  {Knorr}}]{julia_phonon_laser}%
  \BibitemOpen
  \bibfield  {author} {\bibinfo {author} {\bibfnamefont {J.}~\bibnamefont
  {Kabuss}}, \bibinfo {author} {\bibfnamefont {A.}~\bibnamefont {Carmele}},
  \bibinfo {author} {\bibfnamefont {T.}~\bibnamefont {Brandes}}, \ and\
  \bibinfo {author} {\bibfnamefont {A.}~\bibnamefont {Knorr}},\ }\href
  {\doibase 10.1103/PhysRevLett.109.054301} {\bibfield  {journal} {\bibinfo
  {journal} {Phys. Rev. Lett.}\ }\textbf {\bibinfo {volume} {109}},\ \bibinfo
  {pages} {054301} (\bibinfo {year} {2012})}\BibitemShut {NoStop}%
\bibitem [{\citenamefont {Kepesidis}\ \emph {et~al.}(2013)\citenamefont
  {Kepesidis}, \citenamefont {Bennett}, \citenamefont {Portolan}, \citenamefont
  {Lukin},\ and\ \citenamefont {Rabl}}]{rabl_phonon_laser}%
  \BibitemOpen
  \bibfield  {author} {\bibinfo {author} {\bibfnamefont {K.~V.}\ \bibnamefont
  {Kepesidis}}, \bibinfo {author} {\bibfnamefont {S.~D.}\ \bibnamefont
  {Bennett}}, \bibinfo {author} {\bibfnamefont {S.}~\bibnamefont {Portolan}},
  \bibinfo {author} {\bibfnamefont {M.~D.}\ \bibnamefont {Lukin}}, \ and\
  \bibinfo {author} {\bibfnamefont {P.}~\bibnamefont {Rabl}},\ }\href {\doibase
  10.1103/PhysRevB.88.064105} {\bibfield  {journal} {\bibinfo  {journal} {Phys.
  Rev. B}\ }\textbf {\bibinfo {volume} {88}},\ \bibinfo {pages} {064105}
  (\bibinfo {year} {2013})}\BibitemShut {NoStop}%
\bibitem [{\citenamefont {Czerniuk}\ \emph {et~al.}(2017)\citenamefont
  {Czerniuk}, \citenamefont {Wigger}, \citenamefont {Akimov}, \citenamefont
  {Schneider}, \citenamefont {Kamp}, \citenamefont {H\"ofling}, \citenamefont
  {Yakovlev}, \citenamefont {Kuhn}, \citenamefont {Reiter},\ and\ \citenamefont
  {Bayer}}]{reiter_bayer_phonon_laser}%
  \BibitemOpen
  \bibfield  {author} {\bibinfo {author} {\bibfnamefont {T.}~\bibnamefont
  {Czerniuk}}, \bibinfo {author} {\bibfnamefont {D.}~\bibnamefont {Wigger}},
  \bibinfo {author} {\bibfnamefont {A.~V.}\ \bibnamefont {Akimov}}, \bibinfo
  {author} {\bibfnamefont {C.}~\bibnamefont {Schneider}}, \bibinfo {author}
  {\bibfnamefont {M.}~\bibnamefont {Kamp}}, \bibinfo {author} {\bibfnamefont
  {S.}~\bibnamefont {H\"ofling}}, \bibinfo {author} {\bibfnamefont {D.~R.}\
  \bibnamefont {Yakovlev}}, \bibinfo {author} {\bibfnamefont {T.}~\bibnamefont
  {Kuhn}}, \bibinfo {author} {\bibfnamefont {D.~E.}\ \bibnamefont {Reiter}}, \
  and\ \bibinfo {author} {\bibfnamefont {M.}~\bibnamefont {Bayer}},\ }\href
  {\doibase 10.1103/PhysRevLett.118.133901} {\bibfield  {journal} {\bibinfo
  {journal} {Phys. Rev. Lett.}\ }\textbf {\bibinfo {volume} {118}},\ \bibinfo
  {pages} {133901} (\bibinfo {year} {2017})}\BibitemShut {NoStop}%
\bibitem [{\citenamefont {Richter}\ and\ \citenamefont
  {Knorr}(2010)}]{richter2010time}%
  \BibitemOpen
  \bibfield  {author} {\bibinfo {author} {\bibfnamefont {M.}~\bibnamefont
  {Richter}}\ and\ \bibinfo {author} {\bibfnamefont {A.}~\bibnamefont
  {Knorr}},\ }\href@noop {} {\bibfield  {journal} {\bibinfo  {journal} {Annals
  of Physics}\ }\textbf {\bibinfo {volume} {325}},\ \bibinfo {pages} {711}
  (\bibinfo {year} {2010})}\BibitemShut {NoStop}%
\bibitem [{\citenamefont {Stauber}\ \emph {et~al.}(2000)\citenamefont
  {Stauber}, \citenamefont {Zimmermann},\ and\ \citenamefont
  {Castella}}]{stauber}%
  \BibitemOpen
  \bibfield  {author} {\bibinfo {author} {\bibfnamefont {T.}~\bibnamefont
  {Stauber}}, \bibinfo {author} {\bibfnamefont {R.}~\bibnamefont {Zimmermann}},
  \ and\ \bibinfo {author} {\bibfnamefont {H.}~\bibnamefont {Castella}},\
  }\href {\doibase 10.1103/PhysRevB.62.7336} {\bibfield  {journal} {\bibinfo
  {journal} {Phys. Rev. B}\ }\textbf {\bibinfo {volume} {62}},\ \bibinfo
  {pages} {7336} (\bibinfo {year} {2000})}\BibitemShut {NoStop}%
\bibitem [{\citenamefont {Heitz}\ \emph {et~al.}(1999)\citenamefont {Heitz},
  \citenamefont {Mukhametzhanov}, \citenamefont {Stier}, \citenamefont
  {Madhukar},\ and\ \citenamefont {Bimberg}}]{PhysRevLett.83.4654}%
  \BibitemOpen
  \bibfield  {author} {\bibinfo {author} {\bibfnamefont {R.}~\bibnamefont
  {Heitz}}, \bibinfo {author} {\bibfnamefont {I.}~\bibnamefont
  {Mukhametzhanov}}, \bibinfo {author} {\bibfnamefont {O.}~\bibnamefont
  {Stier}}, \bibinfo {author} {\bibfnamefont {A.}~\bibnamefont {Madhukar}}, \
  and\ \bibinfo {author} {\bibfnamefont {D.}~\bibnamefont {Bimberg}},\ }\href
  {\doibase 10.1103/PhysRevLett.83.4654} {\bibfield  {journal} {\bibinfo
  {journal} {Phys. Rev. Lett.}\ }\textbf {\bibinfo {volume} {83}},\ \bibinfo
  {pages} {4654} (\bibinfo {year} {1999})}\BibitemShut {NoStop}%
\bibitem [{\citenamefont {Heitz}\ \emph {et~al.}(2001)\citenamefont {Heitz},
  \citenamefont {Born}, \citenamefont {Guffarth}, \citenamefont {Stier},
  \citenamefont {Schliwa}, \citenamefont {Hoffmann},\ and\ \citenamefont
  {Bimberg}}]{PhysRevB.64.241305}%
  \BibitemOpen
  \bibfield  {author} {\bibinfo {author} {\bibfnamefont {R.}~\bibnamefont
  {Heitz}}, \bibinfo {author} {\bibfnamefont {H.}~\bibnamefont {Born}},
  \bibinfo {author} {\bibfnamefont {F.}~\bibnamefont {Guffarth}}, \bibinfo
  {author} {\bibfnamefont {O.}~\bibnamefont {Stier}}, \bibinfo {author}
  {\bibfnamefont {A.}~\bibnamefont {Schliwa}}, \bibinfo {author} {\bibfnamefont
  {A.}~\bibnamefont {Hoffmann}}, \ and\ \bibinfo {author} {\bibfnamefont
  {D.}~\bibnamefont {Bimberg}},\ }\href {\doibase 10.1103/PhysRevB.64.241305}
  {\bibfield  {journal} {\bibinfo  {journal} {Phys. Rev. B}\ }\textbf {\bibinfo
  {volume} {64}},\ \bibinfo {pages} {241305} (\bibinfo {year}
  {2001})}\BibitemShut {NoStop}%
\bibitem [{\citenamefont {Droenner}\ \emph {et~al.}(2019)\citenamefont
  {Droenner}, \citenamefont {Finsterh{\"o}lzl}, \citenamefont {Heyl},\ and\
  \citenamefont {Carmele}}]{droenner2019stabilizing}%
  \BibitemOpen
  \bibfield  {author} {\bibinfo {author} {\bibfnamefont {L.}~\bibnamefont
  {Droenner}}, \bibinfo {author} {\bibfnamefont {R.}~\bibnamefont
  {Finsterh{\"o}lzl}}, \bibinfo {author} {\bibfnamefont {M.}~\bibnamefont
  {Heyl}}, \ and\ \bibinfo {author} {\bibfnamefont {A.}~\bibnamefont
  {Carmele}},\ }\href@noop {} {\bibfield  {journal} {\bibinfo  {journal} {arXiv
  preprint arXiv:1902.04986}\ } (\bibinfo {year} {2019})}\BibitemShut {NoStop}%
\bibitem [{\citenamefont {Leggett}\ \emph {et~al.}(1987)\citenamefont
  {Leggett}, \citenamefont {Chakravarty}, \citenamefont {Dorsey}, \citenamefont
  {Fisher}, \citenamefont {Garg},\ and\ \citenamefont
  {Zwerger}}]{RevModPhys.59.1}%
  \BibitemOpen
  \bibfield  {author} {\bibinfo {author} {\bibfnamefont {A.~J.}\ \bibnamefont
  {Leggett}}, \bibinfo {author} {\bibfnamefont {S.}~\bibnamefont
  {Chakravarty}}, \bibinfo {author} {\bibfnamefont {A.~T.}\ \bibnamefont
  {Dorsey}}, \bibinfo {author} {\bibfnamefont {M.~P.~A.}\ \bibnamefont
  {Fisher}}, \bibinfo {author} {\bibfnamefont {A.}~\bibnamefont {Garg}}, \ and\
  \bibinfo {author} {\bibfnamefont {W.}~\bibnamefont {Zwerger}},\ }\href
  {\doibase 10.1103/RevModPhys.59.1} {\bibfield  {journal} {\bibinfo  {journal}
  {Rev. Mod. Phys.}\ }\textbf {\bibinfo {volume} {59}},\ \bibinfo {pages} {1}
  (\bibinfo {year} {1987})}\BibitemShut {NoStop}%
\bibitem [{\citenamefont {Vagov}\ \emph {et~al.}(2006)\citenamefont {Vagov},
  \citenamefont {Croitoru}, \citenamefont {Axt}, \citenamefont {Kuhn},\ and\
  \citenamefont {Peeters}}]{vagov2006high}%
  \BibitemOpen
  \bibfield  {author} {\bibinfo {author} {\bibfnamefont {A.}~\bibnamefont
  {Vagov}}, \bibinfo {author} {\bibfnamefont {M.}~\bibnamefont {Croitoru}},
  \bibinfo {author} {\bibfnamefont {V.}~\bibnamefont {Axt}}, \bibinfo {author}
  {\bibfnamefont {T.}~\bibnamefont {Kuhn}}, \ and\ \bibinfo {author}
  {\bibfnamefont {F.}~\bibnamefont {Peeters}},\ }\href@noop {} {\bibfield
  {journal} {\bibinfo  {journal} {physica status solidi (b)}\ }\textbf
  {\bibinfo {volume} {243}},\ \bibinfo {pages} {2233} (\bibinfo {year}
  {2006})}\BibitemShut {NoStop}%
\bibitem [{\citenamefont {Vagov}\ \emph {et~al.}(2011)\citenamefont {Vagov},
  \citenamefont {Croitoru}, \citenamefont {Axt}, \citenamefont {Machnikowski},\
  and\ \citenamefont {Kuhn}}]{vagov2011dynamics}%
  \BibitemOpen
  \bibfield  {author} {\bibinfo {author} {\bibfnamefont {A.}~\bibnamefont
  {Vagov}}, \bibinfo {author} {\bibfnamefont {M.}~\bibnamefont {Croitoru}},
  \bibinfo {author} {\bibfnamefont {V.}~\bibnamefont {Axt}}, \bibinfo {author}
  {\bibfnamefont {P.}~\bibnamefont {Machnikowski}}, \ and\ \bibinfo {author}
  {\bibfnamefont {T.}~\bibnamefont {Kuhn}},\ }\href@noop {} {\bibfield
  {journal} {\bibinfo  {journal} {physica status solidi (b)}\ }\textbf
  {\bibinfo {volume} {248}},\ \bibinfo {pages} {839} (\bibinfo {year}
  {2011})}\BibitemShut {NoStop}%
\bibitem [{\citenamefont {Dousse}\ \emph {et~al.}(2008)\citenamefont {Dousse},
  \citenamefont {Lanco}, \citenamefont {Suffczy\ifmmode~\acute{n}\else
  \'{n}\fi{}ski}, \citenamefont {Semenova}, \citenamefont {Miard},
  \citenamefont {Lema\^{\i}tre}, \citenamefont {Sagnes}, \citenamefont
  {Roblin}, \citenamefont {Bloch},\ and\ \citenamefont {Senellart}}]{dousse}%
  \BibitemOpen
  \bibfield  {author} {\bibinfo {author} {\bibfnamefont {A.}~\bibnamefont
  {Dousse}}, \bibinfo {author} {\bibfnamefont {L.}~\bibnamefont {Lanco}},
  \bibinfo {author} {\bibfnamefont {J.}~\bibnamefont
  {Suffczy\ifmmode~\acute{n}\else \'{n}\fi{}ski}}, \bibinfo {author}
  {\bibfnamefont {E.}~\bibnamefont {Semenova}}, \bibinfo {author}
  {\bibfnamefont {A.}~\bibnamefont {Miard}}, \bibinfo {author} {\bibfnamefont
  {A.}~\bibnamefont {Lema\^{\i}tre}}, \bibinfo {author} {\bibfnamefont
  {I.}~\bibnamefont {Sagnes}}, \bibinfo {author} {\bibfnamefont
  {C.}~\bibnamefont {Roblin}}, \bibinfo {author} {\bibfnamefont
  {J.}~\bibnamefont {Bloch}}, \ and\ \bibinfo {author} {\bibfnamefont
  {P.}~\bibnamefont {Senellart}},\ }\href {\doibase
  10.1103/PhysRevLett.101.267404} {\bibfield  {journal} {\bibinfo  {journal}
  {Phys. Rev. Lett.}\ }\textbf {\bibinfo {volume} {101}},\ \bibinfo {pages}
  {267404} (\bibinfo {year} {2008})}\BibitemShut {NoStop}%
\bibitem [{\citenamefont {Gschrey}\ \emph {et~al.}(2013)\citenamefont
  {Gschrey}, \citenamefont {Gericke}, \citenamefont {Sch{\"u}{\ss}ler},
  \citenamefont {Schmidt}, \citenamefont {Schulze}, \citenamefont {Heindel},
  \citenamefont {Rodt}, \citenamefont {Strittmatter},\ and\ \citenamefont
  {Reitzenstein}}]{gschrey2013situ}%
  \BibitemOpen
  \bibfield  {author} {\bibinfo {author} {\bibfnamefont {M.}~\bibnamefont
  {Gschrey}}, \bibinfo {author} {\bibfnamefont {F.}~\bibnamefont {Gericke}},
  \bibinfo {author} {\bibfnamefont {A.}~\bibnamefont {Sch{\"u}{\ss}ler}},
  \bibinfo {author} {\bibfnamefont {R.}~\bibnamefont {Schmidt}}, \bibinfo
  {author} {\bibfnamefont {J.-H.}\ \bibnamefont {Schulze}}, \bibinfo {author}
  {\bibfnamefont {T.}~\bibnamefont {Heindel}}, \bibinfo {author} {\bibfnamefont
  {S.}~\bibnamefont {Rodt}}, \bibinfo {author} {\bibfnamefont {A.}~\bibnamefont
  {Strittmatter}}, \ and\ \bibinfo {author} {\bibfnamefont {S.}~\bibnamefont
  {Reitzenstein}},\ }\href@noop {} {\bibfield  {journal} {\bibinfo  {journal}
  {Applied Physics Letters}\ }\textbf {\bibinfo {volume} {102}},\ \bibinfo
  {pages} {251113} (\bibinfo {year} {2013})}\BibitemShut {NoStop}%
\bibitem [{\citenamefont {Sotier}\ \emph {et~al.}(2009)\citenamefont {Sotier},
  \citenamefont {Thomay}, \citenamefont {Hanke}, \citenamefont {Korger},
  \citenamefont {Mahapatra}, \citenamefont {Frey}, \citenamefont {Brunner},
  \citenamefont {Bratschitsch},\ and\ \citenamefont
  {Leitenstorfer}}]{sotier2009femtosecond}%
  \BibitemOpen
  \bibfield  {author} {\bibinfo {author} {\bibfnamefont {F.}~\bibnamefont
  {Sotier}}, \bibinfo {author} {\bibfnamefont {T.}~\bibnamefont {Thomay}},
  \bibinfo {author} {\bibfnamefont {T.}~\bibnamefont {Hanke}}, \bibinfo
  {author} {\bibfnamefont {J.}~\bibnamefont {Korger}}, \bibinfo {author}
  {\bibfnamefont {S.}~\bibnamefont {Mahapatra}}, \bibinfo {author}
  {\bibfnamefont {A.}~\bibnamefont {Frey}}, \bibinfo {author} {\bibfnamefont
  {K.}~\bibnamefont {Brunner}}, \bibinfo {author} {\bibfnamefont
  {R.}~\bibnamefont {Bratschitsch}}, \ and\ \bibinfo {author} {\bibfnamefont
  {A.}~\bibnamefont {Leitenstorfer}},\ }\href@noop {} {\bibfield  {journal}
  {\bibinfo  {journal} {Nature Physics}\ }\textbf {\bibinfo {volume} {5}},\
  \bibinfo {pages} {352} (\bibinfo {year} {2009})}\BibitemShut {NoStop}%
\bibitem [{\citenamefont {Flagg}\ \emph {et~al.}(2009)\citenamefont {Flagg},
  \citenamefont {Muller}, \citenamefont {Robertson}, \citenamefont {Founta},
  \citenamefont {Deppe}, \citenamefont {Xiao}, \citenamefont {Ma},
  \citenamefont {Salamo},\ and\ \citenamefont {Shih}}]{flagg2009resonantly}%
  \BibitemOpen
  \bibfield  {author} {\bibinfo {author} {\bibfnamefont {E.~B.}\ \bibnamefont
  {Flagg}}, \bibinfo {author} {\bibfnamefont {A.}~\bibnamefont {Muller}},
  \bibinfo {author} {\bibfnamefont {J.}~\bibnamefont {Robertson}}, \bibinfo
  {author} {\bibfnamefont {S.}~\bibnamefont {Founta}}, \bibinfo {author}
  {\bibfnamefont {D.}~\bibnamefont {Deppe}}, \bibinfo {author} {\bibfnamefont
  {M.}~\bibnamefont {Xiao}}, \bibinfo {author} {\bibfnamefont {W.}~\bibnamefont
  {Ma}}, \bibinfo {author} {\bibfnamefont {G.}~\bibnamefont {Salamo}}, \ and\
  \bibinfo {author} {\bibfnamefont {C.-K.}\ \bibnamefont {Shih}},\ }\href@noop
  {} {\bibfield  {journal} {\bibinfo  {journal} {Nature Physics}\ }\textbf
  {\bibinfo {volume} {5}},\ \bibinfo {pages} {203} (\bibinfo {year}
  {2009})}\BibitemShut {NoStop}%
\bibitem [{\citenamefont {Laussy}\ \emph {et~al.}(2008)\citenamefont {Laussy},
  \citenamefont {Del~Valle},\ and\ \citenamefont {Tejedor}}]{laussy2008strong}%
  \BibitemOpen
  \bibfield  {author} {\bibinfo {author} {\bibfnamefont {F.~P.}\ \bibnamefont
  {Laussy}}, \bibinfo {author} {\bibfnamefont {E.}~\bibnamefont {Del~Valle}}, \
  and\ \bibinfo {author} {\bibfnamefont {C.}~\bibnamefont {Tejedor}},\
  }\href@noop {} {\bibfield  {journal} {\bibinfo  {journal} {Physical review
  letters}\ }\textbf {\bibinfo {volume} {101}},\ \bibinfo {pages} {083601}
  (\bibinfo {year} {2008})}\BibitemShut {NoStop}%
\bibitem [{\citenamefont {Kasprzak}\ \emph {et~al.}(2010)\citenamefont
  {Kasprzak}, \citenamefont {Reitzenstein}, \citenamefont {Muljarov},
  \citenamefont {Kistner}, \citenamefont {Schneider}, \citenamefont {Strauss},
  \citenamefont {H{\"o}fling}, \citenamefont {Forchel},\ and\ \citenamefont
  {Langbein}}]{kasprzak2010up}%
  \BibitemOpen
  \bibfield  {author} {\bibinfo {author} {\bibfnamefont {J.}~\bibnamefont
  {Kasprzak}}, \bibinfo {author} {\bibfnamefont {S.}~\bibnamefont
  {Reitzenstein}}, \bibinfo {author} {\bibfnamefont {E.~A.}\ \bibnamefont
  {Muljarov}}, \bibinfo {author} {\bibfnamefont {C.}~\bibnamefont {Kistner}},
  \bibinfo {author} {\bibfnamefont {C.}~\bibnamefont {Schneider}}, \bibinfo
  {author} {\bibfnamefont {M.}~\bibnamefont {Strauss}}, \bibinfo {author}
  {\bibfnamefont {S.}~\bibnamefont {H{\"o}fling}}, \bibinfo {author}
  {\bibfnamefont {A.}~\bibnamefont {Forchel}}, \ and\ \bibinfo {author}
  {\bibfnamefont {W.}~\bibnamefont {Langbein}},\ }\href@noop {} {\bibfield
  {journal} {\bibinfo  {journal} {Nature materials}\ }\textbf {\bibinfo
  {volume} {9}},\ \bibinfo {pages} {304} (\bibinfo {year} {2010})}\BibitemShut
  {NoStop}%
\bibitem [{\citenamefont {Kavokin}(2017)}]{kavokin2017microcavities}%
  \BibitemOpen
  \bibfield  {author} {\bibinfo {author} {\bibfnamefont {A.}~\bibnamefont
  {Kavokin}},\ }\href@noop {} {\emph {\bibinfo {title} {Microcavities}}},\
  \bibinfo {number} {21}\ (\bibinfo  {publisher} {Oxford University Press},\
  \bibinfo {year} {2017})\BibitemShut {NoStop}%
\bibitem [{\citenamefont {Lodahl}\ \emph {et~al.}(2015)\citenamefont {Lodahl},
  \citenamefont {Mahmoodian},\ and\ \citenamefont {Stobbe}}]{lodahl2015}%
  \BibitemOpen
  \bibfield  {author} {\bibinfo {author} {\bibfnamefont {P.}~\bibnamefont
  {Lodahl}}, \bibinfo {author} {\bibfnamefont {S.}~\bibnamefont {Mahmoodian}},
  \ and\ \bibinfo {author} {\bibfnamefont {S.}~\bibnamefont {Stobbe}},\ }\href
  {\doibase 10.1103/RevModPhys.87.347} {\bibfield  {journal} {\bibinfo
  {journal} {Rev. Mod. Phys.}\ }\textbf {\bibinfo {volume} {87}},\ \bibinfo
  {pages} {347} (\bibinfo {year} {2015})}\BibitemShut {NoStop}%
\bibitem [{\citenamefont {Schneebeli}\ \emph {et~al.}(2008)\citenamefont
  {Schneebeli}, \citenamefont {Kira},\ and\ \citenamefont
  {Koch}}]{schneebeli2008}%
  \BibitemOpen
  \bibfield  {author} {\bibinfo {author} {\bibfnamefont {L.}~\bibnamefont
  {Schneebeli}}, \bibinfo {author} {\bibfnamefont {M.}~\bibnamefont {Kira}}, \
  and\ \bibinfo {author} {\bibfnamefont {S.~W.}\ \bibnamefont {Koch}},\ }\href
  {\doibase 10.1103/PhysRevLett.101.097401} {\bibfield  {journal} {\bibinfo
  {journal} {Phys. Rev. Lett.}\ }\textbf {\bibinfo {volume} {101}},\ \bibinfo
  {pages} {097401} (\bibinfo {year} {2008})}\BibitemShut {NoStop}%
\bibitem [{\citenamefont {Hopfmann}\ \emph {et~al.}(2017)\citenamefont
  {Hopfmann}, \citenamefont {Carmele}, \citenamefont {Musia{\l}}, \citenamefont
  {Schneider}, \citenamefont {Kamp}, \citenamefont {H{\"o}fling}, \citenamefont
  {Knorr},\ and\ \citenamefont {Reitzenstein}}]{hopfmann2017transition}%
  \BibitemOpen
  \bibfield  {author} {\bibinfo {author} {\bibfnamefont {C.}~\bibnamefont
  {Hopfmann}}, \bibinfo {author} {\bibfnamefont {A.}~\bibnamefont {Carmele}},
  \bibinfo {author} {\bibfnamefont {A.}~\bibnamefont {Musia{\l}}}, \bibinfo
  {author} {\bibfnamefont {C.}~\bibnamefont {Schneider}}, \bibinfo {author}
  {\bibfnamefont {M.}~\bibnamefont {Kamp}}, \bibinfo {author} {\bibfnamefont
  {S.}~\bibnamefont {H{\"o}fling}}, \bibinfo {author} {\bibfnamefont
  {A.}~\bibnamefont {Knorr}}, \ and\ \bibinfo {author} {\bibfnamefont
  {S.}~\bibnamefont {Reitzenstein}},\ }\href@noop {} {\bibfield  {journal}
  {\bibinfo  {journal} {Physical Review B}\ }\textbf {\bibinfo {volume} {95}},\
  \bibinfo {pages} {035302} (\bibinfo {year} {2017})}\BibitemShut {NoStop}%
\bibitem [{\citenamefont {Lounis}\ and\ \citenamefont
  {Orrit}(2005)}]{lounis2005single}%
  \BibitemOpen
  \bibfield  {author} {\bibinfo {author} {\bibfnamefont {B.}~\bibnamefont
  {Lounis}}\ and\ \bibinfo {author} {\bibfnamefont {M.}~\bibnamefont {Orrit}},\
  }\href@noop {} {\bibfield  {journal} {\bibinfo  {journal} {Reports on
  Progress in Physics}\ }\textbf {\bibinfo {volume} {68}},\ \bibinfo {pages}
  {1129} (\bibinfo {year} {2005})}\BibitemShut {NoStop}%
\bibitem [{\citenamefont {Ding}\ \emph {et~al.}(2016)\citenamefont {Ding},
  \citenamefont {He}, \citenamefont {Duan}, \citenamefont {Gregersen},
  \citenamefont {Chen}, \citenamefont {Unsleber}, \citenamefont {Maier},
  \citenamefont {Schneider}, \citenamefont {Kamp}, \citenamefont {H\"ofling},
  \citenamefont {Lu},\ and\ \citenamefont {Pan}}]{PhysRevLett.116.020401}%
  \BibitemOpen
  \bibfield  {author} {\bibinfo {author} {\bibfnamefont {X.}~\bibnamefont
  {Ding}}, \bibinfo {author} {\bibfnamefont {Y.}~\bibnamefont {He}}, \bibinfo
  {author} {\bibfnamefont {Z.-C.}\ \bibnamefont {Duan}}, \bibinfo {author}
  {\bibfnamefont {N.}~\bibnamefont {Gregersen}}, \bibinfo {author}
  {\bibfnamefont {M.-C.}\ \bibnamefont {Chen}}, \bibinfo {author}
  {\bibfnamefont {S.}~\bibnamefont {Unsleber}}, \bibinfo {author}
  {\bibfnamefont {S.}~\bibnamefont {Maier}}, \bibinfo {author} {\bibfnamefont
  {C.}~\bibnamefont {Schneider}}, \bibinfo {author} {\bibfnamefont
  {M.}~\bibnamefont {Kamp}}, \bibinfo {author} {\bibfnamefont {S.}~\bibnamefont
  {H\"ofling}}, \bibinfo {author} {\bibfnamefont {C.-Y.}\ \bibnamefont {Lu}}, \
  and\ \bibinfo {author} {\bibfnamefont {J.-W.}\ \bibnamefont {Pan}},\ }\href
  {\doibase 10.1103/PhysRevLett.116.020401} {\bibfield  {journal} {\bibinfo
  {journal} {Phys. Rev. Lett.}\ }\textbf {\bibinfo {volume} {116}},\ \bibinfo
  {pages} {020401} (\bibinfo {year} {2016})}\BibitemShut {NoStop}%
\bibitem [{\citenamefont {Somaschi}\ \emph {et~al.}(2016)\citenamefont
  {Somaschi}, \citenamefont {Giesz}, \citenamefont {De~Santis}, \citenamefont
  {Loredo}, \citenamefont {Almeida}, \citenamefont {Hornecker}, \citenamefont
  {Portalupi}, \citenamefont {Grange}, \citenamefont {Ant{\'o}n}, \citenamefont
  {Demory} \emph {et~al.}}]{somaschi2016near}%
  \BibitemOpen
  \bibfield  {author} {\bibinfo {author} {\bibfnamefont {N.}~\bibnamefont
  {Somaschi}}, \bibinfo {author} {\bibfnamefont {V.}~\bibnamefont {Giesz}},
  \bibinfo {author} {\bibfnamefont {L.}~\bibnamefont {De~Santis}}, \bibinfo
  {author} {\bibfnamefont {J.}~\bibnamefont {Loredo}}, \bibinfo {author}
  {\bibfnamefont {M.~P.}\ \bibnamefont {Almeida}}, \bibinfo {author}
  {\bibfnamefont {G.}~\bibnamefont {Hornecker}}, \bibinfo {author}
  {\bibfnamefont {S.~L.}\ \bibnamefont {Portalupi}}, \bibinfo {author}
  {\bibfnamefont {T.}~\bibnamefont {Grange}}, \bibinfo {author} {\bibfnamefont
  {C.}~\bibnamefont {Ant{\'o}n}}, \bibinfo {author} {\bibfnamefont
  {J.}~\bibnamefont {Demory}},  \emph {et~al.},\ }\href@noop {} {\bibfield
  {journal} {\bibinfo  {journal} {Nature Photonics}\ }\textbf {\bibinfo
  {volume} {10}},\ \bibinfo {pages} {340} (\bibinfo {year} {2016})}\BibitemShut
  {NoStop}%
\bibitem [{\citenamefont {Iles-Smith}\ \emph {et~al.}(2017)\citenamefont
  {Iles-Smith}, \citenamefont {McCutcheon}, \citenamefont {Nazir},\ and\
  \citenamefont {M{\o}rk}}]{iles2017phonon}%
  \BibitemOpen
  \bibfield  {author} {\bibinfo {author} {\bibfnamefont {J.}~\bibnamefont
  {Iles-Smith}}, \bibinfo {author} {\bibfnamefont {D.~P.}\ \bibnamefont
  {McCutcheon}}, \bibinfo {author} {\bibfnamefont {A.}~\bibnamefont {Nazir}}, \
  and\ \bibinfo {author} {\bibfnamefont {J.}~\bibnamefont {M{\o}rk}},\
  }\href@noop {} {\bibfield  {journal} {\bibinfo  {journal} {Nature Photonics}\
  }\textbf {\bibinfo {volume} {11}},\ \bibinfo {pages} {521} (\bibinfo {year}
  {2017})}\BibitemShut {NoStop}%
\bibitem [{\citenamefont {Calic}\ \emph {et~al.}(2011)\citenamefont {Calic},
  \citenamefont {Gallo}, \citenamefont {Felici}, \citenamefont {Atlasov},
  \citenamefont {Dwir}, \citenamefont {Rudra}, \citenamefont {Biasiol},
  \citenamefont {Sorba}, \citenamefont {Tarel}, \citenamefont {Savona} \emph
  {et~al.}}]{savona}%
  \BibitemOpen
  \bibfield  {author} {\bibinfo {author} {\bibfnamefont {M.}~\bibnamefont
  {Calic}}, \bibinfo {author} {\bibfnamefont {P.}~\bibnamefont {Gallo}},
  \bibinfo {author} {\bibfnamefont {M.}~\bibnamefont {Felici}}, \bibinfo
  {author} {\bibfnamefont {K.}~\bibnamefont {Atlasov}}, \bibinfo {author}
  {\bibfnamefont {B.}~\bibnamefont {Dwir}}, \bibinfo {author} {\bibfnamefont
  {A.}~\bibnamefont {Rudra}}, \bibinfo {author} {\bibfnamefont
  {G.}~\bibnamefont {Biasiol}}, \bibinfo {author} {\bibfnamefont
  {L.}~\bibnamefont {Sorba}}, \bibinfo {author} {\bibfnamefont
  {G.}~\bibnamefont {Tarel}}, \bibinfo {author} {\bibfnamefont
  {V.}~\bibnamefont {Savona}},  \emph {et~al.},\ }\href@noop {} {\bibfield
  {journal} {\bibinfo  {journal} {Physical review letters}\ }\textbf {\bibinfo
  {volume} {106}},\ \bibinfo {pages} {227402} (\bibinfo {year}
  {2011})}\BibitemShut {NoStop}%
\bibitem [{\citenamefont {Kaer}\ \emph {et~al.}(2010)\citenamefont {Kaer},
  \citenamefont {Nielsen}, \citenamefont {Lodahl}, \citenamefont {Jauho},\ and\
  \citenamefont {M\o{}rk}}]{mork}%
  \BibitemOpen
  \bibfield  {author} {\bibinfo {author} {\bibfnamefont {P.}~\bibnamefont
  {Kaer}}, \bibinfo {author} {\bibfnamefont {T.~R.}\ \bibnamefont {Nielsen}},
  \bibinfo {author} {\bibfnamefont {P.}~\bibnamefont {Lodahl}}, \bibinfo
  {author} {\bibfnamefont {A.-P.}\ \bibnamefont {Jauho}}, \ and\ \bibinfo
  {author} {\bibfnamefont {J.}~\bibnamefont {M\o{}rk}},\ }\href {\doibase
  10.1103/PhysRevLett.104.157401} {\bibfield  {journal} {\bibinfo  {journal}
  {Phys. Rev. Lett.}\ }\textbf {\bibinfo {volume} {104}},\ \bibinfo {pages}
  {157401} (\bibinfo {year} {2010})}\BibitemShut {NoStop}%
\bibitem [{\citenamefont {Hughes}\ \emph
  {et~al.}(2011{\natexlab{a}})\citenamefont {Hughes}, \citenamefont {Yao},
  \citenamefont {Milde}, \citenamefont {Knorr}, \citenamefont {Dalacu},
  \citenamefont {Mnaymneh}, \citenamefont {Sazonova}, \citenamefont {Poole},
  \citenamefont {Aers}, \citenamefont {Lapointe}, \citenamefont {Cheriton},\
  and\ \citenamefont {Williams}}]{PhysRevB.83.165313}%
  \BibitemOpen
  \bibfield  {author} {\bibinfo {author} {\bibfnamefont {S.}~\bibnamefont
  {Hughes}}, \bibinfo {author} {\bibfnamefont {P.}~\bibnamefont {Yao}},
  \bibinfo {author} {\bibfnamefont {F.}~\bibnamefont {Milde}}, \bibinfo
  {author} {\bibfnamefont {A.}~\bibnamefont {Knorr}}, \bibinfo {author}
  {\bibfnamefont {D.}~\bibnamefont {Dalacu}}, \bibinfo {author} {\bibfnamefont
  {K.}~\bibnamefont {Mnaymneh}}, \bibinfo {author} {\bibfnamefont
  {V.}~\bibnamefont {Sazonova}}, \bibinfo {author} {\bibfnamefont {P.~J.}\
  \bibnamefont {Poole}}, \bibinfo {author} {\bibfnamefont {G.~C.}\ \bibnamefont
  {Aers}}, \bibinfo {author} {\bibfnamefont {J.}~\bibnamefont {Lapointe}},
  \bibinfo {author} {\bibfnamefont {R.}~\bibnamefont {Cheriton}}, \ and\
  \bibinfo {author} {\bibfnamefont {R.~L.}\ \bibnamefont {Williams}},\ }\href
  {\doibase 10.1103/PhysRevB.83.165313} {\bibfield  {journal} {\bibinfo
  {journal} {Phys. Rev. B}\ }\textbf {\bibinfo {volume} {83}},\ \bibinfo
  {pages} {165313} (\bibinfo {year} {2011}{\natexlab{a}})}\BibitemShut
  {NoStop}%
\bibitem [{\citenamefont {Winger}\ \emph {et~al.}(2009)\citenamefont {Winger},
  \citenamefont {Volz}, \citenamefont {Tarel}, \citenamefont {Portolan},
  \citenamefont {Badolato}, \citenamefont {Hennessy}, \citenamefont {Hu},
  \citenamefont {Beveratos}, \citenamefont {Finley}, \citenamefont {Savona}
  \emph {et~al.}}]{winger2009explanation}%
  \BibitemOpen
  \bibfield  {author} {\bibinfo {author} {\bibfnamefont {M.}~\bibnamefont
  {Winger}}, \bibinfo {author} {\bibfnamefont {T.}~\bibnamefont {Volz}},
  \bibinfo {author} {\bibfnamefont {G.}~\bibnamefont {Tarel}}, \bibinfo
  {author} {\bibfnamefont {S.}~\bibnamefont {Portolan}}, \bibinfo {author}
  {\bibfnamefont {A.}~\bibnamefont {Badolato}}, \bibinfo {author}
  {\bibfnamefont {K.~J.}\ \bibnamefont {Hennessy}}, \bibinfo {author}
  {\bibfnamefont {E.~L.}\ \bibnamefont {Hu}}, \bibinfo {author} {\bibfnamefont
  {A.}~\bibnamefont {Beveratos}}, \bibinfo {author} {\bibfnamefont
  {J.}~\bibnamefont {Finley}}, \bibinfo {author} {\bibfnamefont
  {V.}~\bibnamefont {Savona}},  \emph {et~al.},\ }\href@noop {} {\bibfield
  {journal} {\bibinfo  {journal} {Physical review letters}\ }\textbf {\bibinfo
  {volume} {103}},\ \bibinfo {pages} {207403} (\bibinfo {year}
  {2009})}\BibitemShut {NoStop}%
\bibitem [{\citenamefont {Hughes}\ \emph
  {et~al.}(2011{\natexlab{b}})\citenamefont {Hughes}, \citenamefont {Yao},
  \citenamefont {Milde}, \citenamefont {Knorr}, \citenamefont {Dalacu},
  \citenamefont {Mnaymneh}, \citenamefont {Sazonova}, \citenamefont {Poole},
  \citenamefont {Aers}, \citenamefont {Lapointe} \emph
  {et~al.}}]{hughes2011influence}%
  \BibitemOpen
  \bibfield  {author} {\bibinfo {author} {\bibfnamefont {S.}~\bibnamefont
  {Hughes}}, \bibinfo {author} {\bibfnamefont {P.}~\bibnamefont {Yao}},
  \bibinfo {author} {\bibfnamefont {F.}~\bibnamefont {Milde}}, \bibinfo
  {author} {\bibfnamefont {A.}~\bibnamefont {Knorr}}, \bibinfo {author}
  {\bibfnamefont {D.}~\bibnamefont {Dalacu}}, \bibinfo {author} {\bibfnamefont
  {K.}~\bibnamefont {Mnaymneh}}, \bibinfo {author} {\bibfnamefont
  {V.}~\bibnamefont {Sazonova}}, \bibinfo {author} {\bibfnamefont
  {P.}~\bibnamefont {Poole}}, \bibinfo {author} {\bibfnamefont
  {G.}~\bibnamefont {Aers}}, \bibinfo {author} {\bibfnamefont {J.}~\bibnamefont
  {Lapointe}},  \emph {et~al.},\ }\href@noop {} {\bibfield  {journal} {\bibinfo
   {journal} {Physical Review B}\ }\textbf {\bibinfo {volume} {83}},\ \bibinfo
  {pages} {165313} (\bibinfo {year} {2011}{\natexlab{b}})}\BibitemShut
  {NoStop}%
\bibitem [{\citenamefont {Carmele}\ \emph
  {et~al.}(2013{\natexlab{b}})\citenamefont {Carmele}, \citenamefont {Kabuss},\
  and\ \citenamefont {Chow}}]{carmele2013highly}%
  \BibitemOpen
  \bibfield  {author} {\bibinfo {author} {\bibfnamefont {A.}~\bibnamefont
  {Carmele}}, \bibinfo {author} {\bibfnamefont {J.}~\bibnamefont {Kabuss}}, \
  and\ \bibinfo {author} {\bibfnamefont {W.~W.}\ \bibnamefont {Chow}},\
  }\href@noop {} {\bibfield  {journal} {\bibinfo  {journal} {Physical Review
  B}\ }\textbf {\bibinfo {volume} {87}},\ \bibinfo {pages} {041305} (\bibinfo
  {year} {2013}{\natexlab{b}})}\BibitemShut {NoStop}%
\bibitem [{\citenamefont {Thoma}\ \emph {et~al.}(2016)\citenamefont {Thoma},
  \citenamefont {Schnauber}, \citenamefont {Gschrey}, \citenamefont {Seifried},
  \citenamefont {Wolters}, \citenamefont {Schulze}, \citenamefont
  {Strittmatter}, \citenamefont {Rodt}, \citenamefont {Carmele}, \citenamefont
  {Knorr}, \citenamefont {Heindel},\ and\ \citenamefont
  {Reitzenstein}}]{thoma}%
  \BibitemOpen
  \bibfield  {author} {\bibinfo {author} {\bibfnamefont {A.}~\bibnamefont
  {Thoma}}, \bibinfo {author} {\bibfnamefont {P.}~\bibnamefont {Schnauber}},
  \bibinfo {author} {\bibfnamefont {M.}~\bibnamefont {Gschrey}}, \bibinfo
  {author} {\bibfnamefont {M.}~\bibnamefont {Seifried}}, \bibinfo {author}
  {\bibfnamefont {J.}~\bibnamefont {Wolters}}, \bibinfo {author} {\bibfnamefont
  {J.-H.}\ \bibnamefont {Schulze}}, \bibinfo {author} {\bibfnamefont
  {A.}~\bibnamefont {Strittmatter}}, \bibinfo {author} {\bibfnamefont
  {S.}~\bibnamefont {Rodt}}, \bibinfo {author} {\bibfnamefont {A.}~\bibnamefont
  {Carmele}}, \bibinfo {author} {\bibfnamefont {A.}~\bibnamefont {Knorr}},
  \bibinfo {author} {\bibfnamefont {T.}~\bibnamefont {Heindel}}, \ and\
  \bibinfo {author} {\bibfnamefont {S.}~\bibnamefont {Reitzenstein}},\ }\href
  {\doibase 10.1103/PhysRevLett.116.033601} {\bibfield  {journal} {\bibinfo
  {journal} {Phys. Rev. Lett.}\ }\textbf {\bibinfo {volume} {116}},\ \bibinfo
  {pages} {033601} (\bibinfo {year} {2016})}\BibitemShut {NoStop}%
\bibitem [{\citenamefont {Rempe}\ \emph {et~al.}(1987)\citenamefont {Rempe},
  \citenamefont {Walther},\ and\ \citenamefont {Klein}}]{rempe}%
  \BibitemOpen
  \bibfield  {author} {\bibinfo {author} {\bibfnamefont {G.}~\bibnamefont
  {Rempe}}, \bibinfo {author} {\bibfnamefont {H.}~\bibnamefont {Walther}}, \
  and\ \bibinfo {author} {\bibfnamefont {N.}~\bibnamefont {Klein}},\ }\href
  {\doibase 10.1103/PhysRevLett.58.353} {\bibfield  {journal} {\bibinfo
  {journal} {Phys. Rev. Lett.}\ }\textbf {\bibinfo {volume} {58}},\ \bibinfo
  {pages} {353} (\bibinfo {year} {1987})}\BibitemShut {NoStop}%
\bibitem [{\citenamefont {Eberly}\ \emph {et~al.}(1980)\citenamefont {Eberly},
  \citenamefont {Narozhny},\ and\ \citenamefont
  {Sanchez-Mondragon}}]{eberly1980}%
  \BibitemOpen
  \bibfield  {author} {\bibinfo {author} {\bibfnamefont {J.~H.}\ \bibnamefont
  {Eberly}}, \bibinfo {author} {\bibfnamefont {N.~B.}\ \bibnamefont
  {Narozhny}}, \ and\ \bibinfo {author} {\bibfnamefont {J.~J.}\ \bibnamefont
  {Sanchez-Mondragon}},\ }\href {\doibase 10.1103/PhysRevLett.44.1323}
  {\bibfield  {journal} {\bibinfo  {journal} {Phys. Rev. Lett.}\ }\textbf
  {\bibinfo {volume} {44}},\ \bibinfo {pages} {1323} (\bibinfo {year}
  {1980})}\BibitemShut {NoStop}%
\bibitem [{\citenamefont {Hong}\ \emph {et~al.}(1987)\citenamefont {Hong},
  \citenamefont {Ou},\ and\ \citenamefont {Mandel}}]{hom}%
  \BibitemOpen
  \bibfield  {author} {\bibinfo {author} {\bibfnamefont {C.~K.}\ \bibnamefont
  {Hong}}, \bibinfo {author} {\bibfnamefont {Z.~Y.}\ \bibnamefont {Ou}}, \ and\
  \bibinfo {author} {\bibfnamefont {L.}~\bibnamefont {Mandel}},\ }\href
  {\doibase 10.1103/PhysRevLett.59.2044} {\bibfield  {journal} {\bibinfo
  {journal} {Phys. Rev. Lett.}\ }\textbf {\bibinfo {volume} {59}},\ \bibinfo
  {pages} {2044} (\bibinfo {year} {1987})}\BibitemShut {NoStop}%
\bibitem [{\citenamefont {Galperin}\ \emph {et~al.}(2006)\citenamefont
  {Galperin}, \citenamefont {Altshuler}, \citenamefont {Bergli},\ and\
  \citenamefont {Shantsev}}]{PhysRevLett.96.097009}%
  \BibitemOpen
  \bibfield  {author} {\bibinfo {author} {\bibfnamefont {Y.~M.}\ \bibnamefont
  {Galperin}}, \bibinfo {author} {\bibfnamefont {B.~L.}\ \bibnamefont
  {Altshuler}}, \bibinfo {author} {\bibfnamefont {J.}~\bibnamefont {Bergli}}, \
  and\ \bibinfo {author} {\bibfnamefont {D.~V.}\ \bibnamefont {Shantsev}},\
  }\href {\doibase 10.1103/PhysRevLett.96.097009} {\bibfield  {journal}
  {\bibinfo  {journal} {Phys. Rev. Lett.}\ }\textbf {\bibinfo {volume} {96}},\
  \bibinfo {pages} {097009} (\bibinfo {year} {2006})}\BibitemShut {NoStop}%
\bibitem [{\citenamefont {Laikhtman}(1985)}]{PhysRevB.31.490}%
  \BibitemOpen
  \bibfield  {author} {\bibinfo {author} {\bibfnamefont {B.~D.}\ \bibnamefont
  {Laikhtman}},\ }\href {\doibase 10.1103/PhysRevB.31.490} {\bibfield
  {journal} {\bibinfo  {journal} {Phys. Rev. B}\ }\textbf {\bibinfo {volume}
  {31}},\ \bibinfo {pages} {490} (\bibinfo {year} {1985})}\BibitemShut
  {NoStop}%
\bibitem [{\citenamefont {Eberly}\ \emph {et~al.}(1984)\citenamefont {Eberly},
  \citenamefont {W\'odkiewicz},\ and\ \citenamefont
  {Shore}}]{PhysRevA.30.2381}%
  \BibitemOpen
  \bibfield  {author} {\bibinfo {author} {\bibfnamefont {J.~H.}\ \bibnamefont
  {Eberly}}, \bibinfo {author} {\bibfnamefont {K.}~\bibnamefont
  {W\'odkiewicz}}, \ and\ \bibinfo {author} {\bibfnamefont {B.~W.}\
  \bibnamefont {Shore}},\ }\href {\doibase 10.1103/PhysRevA.30.2381} {\bibfield
   {journal} {\bibinfo  {journal} {Phys. Rev. A}\ }\textbf {\bibinfo {volume}
  {30}},\ \bibinfo {pages} {2381} (\bibinfo {year} {1984})}\BibitemShut
  {NoStop}%
\bibitem [{\citenamefont {Liu}\ \emph {et~al.}(2018{\natexlab{b}})\citenamefont
  {Liu}, \citenamefont {Konthasinghe}, \citenamefont
  {Davan\ifmmode~\mbox{\c{c}}\else \c{c}\fi{}o}, \citenamefont {Lawall},
  \citenamefont {Anant}, \citenamefont {Verma}, \citenamefont {Mirin},
  \citenamefont {Nam}, \citenamefont {Song}, \citenamefont {Ma}, \citenamefont
  {Chen}, \citenamefont {Ni}, \citenamefont {Niu},\ and\ \citenamefont
  {Srinivasan}}]{PhysRevApplied.9.064019}%
  \BibitemOpen
  \bibfield  {author} {\bibinfo {author} {\bibfnamefont {J.}~\bibnamefont
  {Liu}}, \bibinfo {author} {\bibfnamefont {K.}~\bibnamefont {Konthasinghe}},
  \bibinfo {author} {\bibfnamefont {M.}~\bibnamefont
  {Davan\ifmmode~\mbox{\c{c}}\else \c{c}\fi{}o}}, \bibinfo {author}
  {\bibfnamefont {J.}~\bibnamefont {Lawall}}, \bibinfo {author} {\bibfnamefont
  {V.}~\bibnamefont {Anant}}, \bibinfo {author} {\bibfnamefont
  {V.}~\bibnamefont {Verma}}, \bibinfo {author} {\bibfnamefont
  {R.}~\bibnamefont {Mirin}}, \bibinfo {author} {\bibfnamefont {S.~W.}\
  \bibnamefont {Nam}}, \bibinfo {author} {\bibfnamefont {J.~D.}\ \bibnamefont
  {Song}}, \bibinfo {author} {\bibfnamefont {B.}~\bibnamefont {Ma}}, \bibinfo
  {author} {\bibfnamefont {Z.~S.}\ \bibnamefont {Chen}}, \bibinfo {author}
  {\bibfnamefont {H.~Q.}\ \bibnamefont {Ni}}, \bibinfo {author} {\bibfnamefont
  {Z.~C.}\ \bibnamefont {Niu}}, \ and\ \bibinfo {author} {\bibfnamefont
  {K.}~\bibnamefont {Srinivasan}},\ }\href {\doibase
  10.1103/PhysRevApplied.9.064019} {\bibfield  {journal} {\bibinfo  {journal}
  {Phys. Rev. Applied}\ }\textbf {\bibinfo {volume} {9}},\ \bibinfo {pages}
  {064019} (\bibinfo {year} {2018}{\natexlab{b}})}\BibitemShut {NoStop}%
\bibitem [{\citenamefont {Gl\"assl}\ \emph {et~al.}(2012)\citenamefont
  {Gl\"assl}, \citenamefont {S\"orgel}, \citenamefont {Vagov}, \citenamefont
  {Croitoru}, \citenamefont {Kuhn},\ and\ \citenamefont
  {Axt}}]{PhysRevB.86.035319}%
  \BibitemOpen
  \bibfield  {author} {\bibinfo {author} {\bibfnamefont {M.}~\bibnamefont
  {Gl\"assl}}, \bibinfo {author} {\bibfnamefont {L.}~\bibnamefont {S\"orgel}},
  \bibinfo {author} {\bibfnamefont {A.}~\bibnamefont {Vagov}}, \bibinfo
  {author} {\bibfnamefont {M.~D.}\ \bibnamefont {Croitoru}}, \bibinfo {author}
  {\bibfnamefont {T.}~\bibnamefont {Kuhn}}, \ and\ \bibinfo {author}
  {\bibfnamefont {V.~M.}\ \bibnamefont {Axt}},\ }\href {\doibase
  10.1103/PhysRevB.86.035319} {\bibfield  {journal} {\bibinfo  {journal} {Phys.
  Rev. B}\ }\textbf {\bibinfo {volume} {86}},\ \bibinfo {pages} {035319}
  (\bibinfo {year} {2012})}\BibitemShut {NoStop}%
\end{thebibliography}
%

\end{document}